\documentclass[aps,prx,twocolumn,superscriptaddress]{revtex4-2}

\usepackage{graphicx}
\usepackage{xcolor}
\usepackage[normalem]{ulem}
\usepackage{braket}
\usepackage{amsmath}
\usepackage{array}
\usepackage[colorlinks,citecolor={magenta},linkcolor={magenta},urlcolor=blue,hyperindex]{hyperref}

\usepackage{bm}  

\newcommand{\Ba}{$^{137}\mathrm{Ba}^+$}

\newcommand{\slevel}{$\mathrm{S}_{1/2}$}

\newcommand{\dlevel}{$\mathrm{D}_{5/2}$}

\newcommand{\flevel}{$\mathrm{F}_{7/2}$}

\begin{document}


\title{Utility of virtual qubits in trapped-ion quantum computers}



\author{Saumya Shivam}
\thanks{These authors contributed equally.}
\affiliation{Department of Physics, Princeton University, Princeton, New Jersey 08544, USA}

\author{Fabian Pokorny}
\thanks{These authors contributed equally.}

\author{Andres Vazquez-Brennan}

\author{Ana S. Sotirova}

\author{Jamie D. Leppard}

\author{Sophie M. Decoppet}

\author{C. J. Ballance}
\affiliation{Department of Physics, University of Oxford, Clarendon Laboratory, Parks Road, Oxford OX1 3PU, UK}

\author{S.~L. Sondhi}
\affiliation{Rudolf Peierls Centre for Theoretical Physics, University of Oxford, Oxford OX1 3PU, UK}


\date{\today}

\begin{abstract}

We propose encoding multiple qubits inside ions in existing trapped-ion quantum computers to access more qubits and to simplify circuits implementing standard algorithms. By using such `virtual' qubits, some inter-ion gates can be replaced by intra-ion gates, reducing the use of vibrational modes of the ion chain, leading to less noise. We discuss specific examples such as the Bernstein-Vazirani algorithm and random circuit sampling, using a small number of virtual qubits. Additionally, virtual qubits enable using larger number of data qubits for an error correcting code, and we consider the repetition code as an example. We also lay out practical considerations to be made when choosing states to encode virtual qubits in \Ba ions, and for preparing states and performing measurements. 
\end{abstract}

\maketitle
\tableofcontents

\section{Introduction}\label{sec:intro}

Trapped ions are among the most promising platforms for realizing a useful, universal quantum computer, with advantages such as naturally identical qubits, all-to-all connectivity, high-fidelity gates, long coherence times and the ability to create remote entanglement \cite{ballance2017high,stephenson2020high,bruzewicz2019trapped,wang2021single}. Traditionally, two states chosen within each ion serve as a qubit, and individually addressed laser beams can be used to drive transitions in each qubit. Multi-qubit entangling gates use the interaction between the internal levels of an ion and vibrational modes of the ion chain, and form a universal gate set when combined with the single qubit gates \cite{cai2023entangling}. However, additional levels inside ions can also be accessed using the same mechanisms, opening up the possibility to encode multiple qubits in just one ion. These qubits aren't physically distinct like traditional qubits, hence throughout this work we will refer to them as \textit{virtual qubits}. 

In the near term, with the number of individually controllable ions in a chain limited to a few dozens, utilizing a larger Hilbert space with the same number of ions by encoding multiple qubits per ion can enable exploration of problems which would be intractable otherwise. This can also help in optimizing modular approaches to scaling up a trapped-ion quantum computer \cite{pino2021demonstration,moses2023race,malinowski2023wire}.  In this work, we focus on the special case where a small number of virtual qubits are defined using multiple states inside a single ion. This allows for a meaningful comparison to the traditional one-qubit-per-ion approach, in contrast to exploring qudit-based quantum computing, another, increasingly popular approach to make use of the available higher dimensionality \cite{elder_mult_sc_2020,low_practical_qudit_2020,rambach_qudit_tomo_2021,chi2022programmable,ringbauer2022universal,gao2022role,nikolaeva2021efficient,low2023control,de2023navigating,hrmo2023native}.

Note that the idea of using higher dimensional systems as multi-qubit systems has been proposed and implemented in other platforms such as superconducting qubits \cite{multilevel_sc_2015,svetitsky2014hidden,dong2022simulation,rambow2021reduction,jankovic2023noisy,cao2023emulating,litteken2023dancing}, spin systems \cite{kessel1999multiqubit,gun2013construction} and very recently in ion traps  \cite{campbell2022polyqubit,nikolaeva2023compiling}. 

 Our contribution to this research area is to provide explicit examples demonstrating potential use of an ion chain with two virtual qubits per ion for implementing algorithms and error correction. Our results are not dependent on the ability to implement a modified qudit-based inter-ion gate \cite{low_practical_qudit_2020,hrmo2023native}, but just use conventional inter-ion gates. We also provide decompositions for various standard gates for different number of virtual qubits, and comment on the behavior as the number of virtual qubits increases. Finally, we discuss the practical considerations in defining such a system with minimal modifications to existing setups, and propose an explicit implementation using the metastable manifold of \Ba ions to define virtual qubits.

This paper is organized as follows: We start out by defining the notations used and the nature of the setup we are considering. Then, in Sections \ref{sec:native_intra} and \ref{sec:inter_ion} we elaborate on how the native gates transform when using virtual qubits, and comment on the role of connectivity and encoding maps in Section \ref{sec:lim-conn}. 

In Section \ref{sec:applications}, we provide specific examples when using virtual qubits can be effective. In Section \ref{sec:app_ent_ops}, we discuss ways in which entangling operations can be simplified when using virtual qubits.  A circuit implementing the Bernstein-Vazirani algorithm using two virtual qubits per ion is shown in \ref{sec:bv}. Next, in Section \ref{sec:XEB}, we numerically show that random circuits composed out of intra- and inter-ion gates require fewer gates for their linear cross entropy fidelity to converge.  Then, in Section \ref{sec:bit-flip}, we simulate a bit-flip repetition code with two virtual qubits encoded as data qubits, and find that it leads to lower logical error rates. 

Finally, in Section \ref{sec:experiment}, we outline the requirements to realize virtual qubits in trapped ions, and explicitly propose their implementation in the metastable level of \Ba ions. We also provide a brief description of a state preparation and measurement protocol using our proposed states. 

\subsection{Definitions}
\begin{figure}[h]
    \centering
    \includegraphics[width=0.9\linewidth,trim = 0cm 1cm 0cm 0cm]{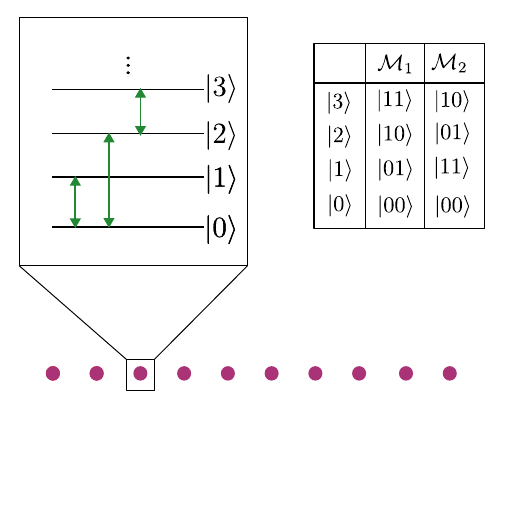}
    \caption{A sketch of a trapped ion chain with four states inside each ion encoded as $n=2$ virtual qubits, with two simple choices of the basis encoding map $\mathcal{M}$ given in the table on the right. Native intra-ion gates allow transitions between two states through application of electromagnetic fields of appropriate frequencies. A universal gate set for a single ion can be achieved as long as the coupled states form a connected graph. An example of allowed transitions which create a universal gate set is shown by green arrows.  }
    \label{fig:intro_fig}
\end{figure}

 Consider a chain of $L$ ions, with $d_{i}=2^{n_i}$ states of the $\mathrm{i^{th}}$ ion encoded as $n_i$ virtual qubits. The total number of effective qubits  in the chain is then $N=\sum_i n_i$. When denoted without a subscript, we will assume a constant number of virtual qubits $n$ in each ion, unless specified otherwise. Within an ion, the states associated with the qubits are denoted as $\ket{\alpha}$, where $\alpha=0,1,\dots ,d_i-1$, in increasing order of their energy. The encoding of the basis states of the virtual qubits is then defined through a map $\mathcal{M}$ from the span of the $d_i$ states to a Hilbert space comprising of $n_i$ qubits. While there are infinite possible $\mathcal{M}$, a reasonable and convenient restriction on the mappings can be imposed by requiring that measurements collapsing into one of the $d_i$ states $\ket{\alpha}$ implies a collapse on one of $Z$ basis states ($\ket{a_1,a_2,...,a_{n_i}}$) in the space of virtual qubits. In other words,  $\mathcal{M}(\ket{(\alpha)}=\ket{a_1,a_2,...,a_{n_i}}$ ($a_j~\epsilon~\{0,1\}~\forall j)$. There are $d_i$! such maps in total. Two simple mapping examples are shown in Figure \ref{fig:intro_fig}.

 In some sections of the paper, we might refer to using a single qubit per ion as using $n=1$ virtual qubits, for convenience. In addition, we avoid using the term ``physical qubits" in the context of error correction, and instead refer to them as ``data qubits", which may be a virtual qubit inside an ion. A logical qubit can also in principle be formed out of virtual qubits.

\section{Native and standard gates with virtual qubits}

\subsection{Intra-ion gates}\label{sec:native_intra}

Inside each ion, driving Rabi oscillations between two states $\ket{\alpha}$ and $\ket{\beta}$ enables implementing the native gates $R_{\alpha \beta}(\theta,\phi)$. For example, the explicit form of the gate for $n=2$ and $\alpha=0,\beta=1$ is given by :
\begin{align}\label{eq:Rij}
R_{01}(\theta,\phi)=
\begin{pmatrix}
    \cos(\theta) & -i e^{-i\phi}\sin(\theta)& 0 & 0 \\
    -i e^{i\phi}\sin(\theta)  &  \cos(\theta)& 0 & 0\\
    0 & 0 & 1 & 0 \\
    0 & 0 & 0 & 1 \\
\end{pmatrix}
\end{align}
More generally, $R_{\alpha \beta}(\theta,\phi)=\cos(\theta)(\ket{\alpha}\bra{\alpha}+\ket{\beta}\bra{\beta})-i\sin(\theta)(e^{-i\phi}\ket{\alpha}\bra{\beta}+e^{i\phi}\ket{\beta}\bra{\alpha})$.
 Such gates can be implemented by direct excitation corresponding to the frequency difference between the two states, or through a Raman beatnote \cite{cirac1995quantum,islam2012quantum}. The angles ($\theta, \phi$) depend on the duration and phase of the incident beam(s) on the ion \cite{islam2011onset}.

Interestingly, $R_{\alpha \beta}(\theta,\phi)$ is typically a multi-qubit gate: it acts on all $n$ qubits simultaneously. In the computational basis, it becomes an increasingly sparse operator as $n$ increases, and can produce a variety of entangled states. For example, for $n=2$, in the mappings where $\mathcal{M}(\ket{0}) = \ket{00}$ and $\mathcal{M}(\ket{1}) = \ket{11}$, $R_{01}(\pi/4,-\pi/2)$ acting on an initial state $\ket{00}$ can produce the two qubit GHZ state~\footnote{Note that since the two qubits inside an ion cannot be separated, applications such as quantum teleportation are not considered here.}. The advantage of creating entangled states in this setting is that they do not require using the vibrational modes of the ion chain, and just utilize the internal $R_{\alpha \beta}$ gates, the latter typically having much lower error rates . On the other hand, if $\mathcal{M}(\ket{0})=\ket{00}$ and $\mathcal{M}(\ket{1})=\ket{01}$, the same gate $R_{01}(\pi/4,-\pi/2)$ acting on $\ket{00}$ leads to another product state, $\frac{1}{\sqrt{2}}(\ket{00}+\ket{01})$. This variability can be used to optimize the decomposition of gates according to the desired application. 

Easier multi-qubit gates such as $R_{\alpha \beta}(\theta,\phi)$ come at a cost: single-qubit gates can now require larger number of native gates. The decomposition of standard single-qubit gates depends on two factors, the number of $R_{\alpha \beta}$ that are allowed in practice for a given ion (connectivity determined by the allowed transitions), and the nature of the map $\mathcal{M}$. 

The minimum number of $R_{\alpha \beta}(\theta,\phi)$ gates required to implement a non-trivial single-qubit gate is $d/2$ (while the upper bound scales as $O(d^2)$, as shown in Ref. \cite{low_practical_qudit_2020}). The lower limit of $d/2$ can be reached when all possible $R_{\alpha \beta}$ are allowed -- which we will assume initially. Additional gates may be required to match the global phase of standard gates \footnote{The native intra-ion $R_{\alpha \beta}$ can only be used to construct gates belonging to the group SU($d$)}. For constructing arbitrary gates inside an ion, the set of allowed $R_{\alpha \beta}$ must also form a universal gate set in the $d$-dimensional Hilbert space. That criterion is satisfied as long as the states that $R_{\alpha \beta}$ connects form a connected graph \cite{low_practical_qudit_2020} -- requiring a minimum of $d-1$ allowed $R_{\alpha \beta}$. 

As $d=2^n$ increases, the number of native gates required to construct an arbitrary single-qubit gate then increases exponentially in $n$, for both limited and all-to-all connectivity of $R_{\alpha \beta}$ inside an ion. As a result, scaling up the number of virtual qubits $n$ to a large value is impractical. 

Besides the number of native gates needed to implement single-qubit gates, other factors such as the nature of inter-ion gates, decoherence, complex state preparation and measurement protocols  impose a limit on $n$  \cite{blume2002climbing,low_practical_qudit_2020}. 

Even though scaling $n$ is impractical, we now argue for the benefits of using a small $n$ $\in \{2,3,4\}$, by providing examples of decompositions of specific gates and estimating the gate errors. We defer a detailed discussion on the choice of states in a given ion for realizing a a universal gate set to Section \ref{sec:guidelines_comp_manifold}, and the role of limited connectivity and encoding map (like the one illustrated in Figure \ref{fig:intro_fig}) to Section \ref{sec:lim-conn}. 

\subsubsection{Single and two qubit gates}

\begin{figure*}[t]
    \centering
    \includegraphics[width=\textwidth,trim = 0cm 0.5cm 0cm 0cm]{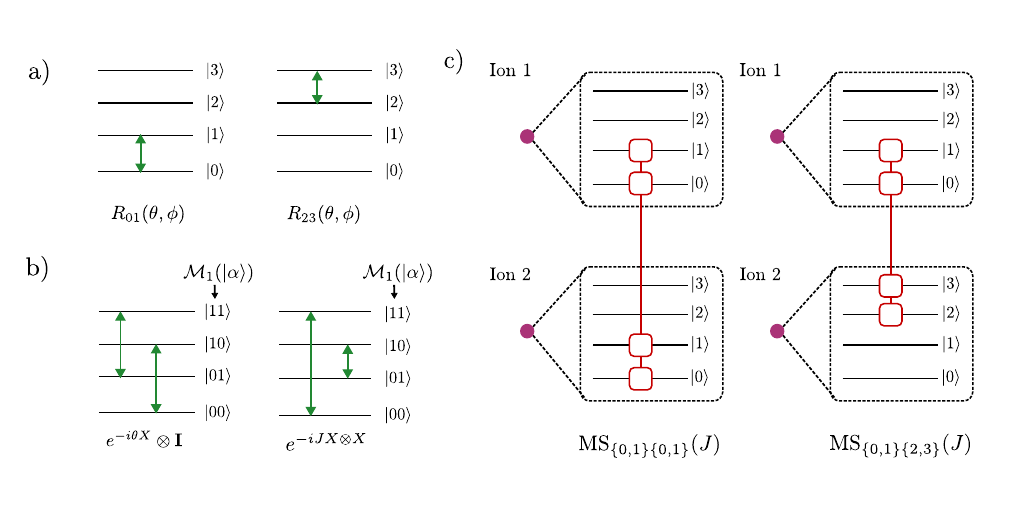}
    \caption{Native and standard gates with $n=2$ virtual qubits defined in an ion. a) Native intra-ion gates $R_{\alpha \beta}$ perform Rabi oscillations between $\ket{\alpha}$ and $\ket{\beta}$, such that $R_{\alpha \beta}(\theta,\phi)=\cos(\theta)(\ket{\alpha}\bra{\alpha}+\ket{\beta}\bra{\beta})-i\sin(\theta)(e^{-i\phi}\ket{\alpha}\bra{\beta}+e^{i\phi}\ket{\beta}\bra{\alpha})$. b) Decompositions of the single-qubit gate $e^{-i\theta X \otimes \mathbf{I}}$ and the two-qubit gate $e^{-i\theta X \otimes X}$. The decomposition depends on the map between the ionic states and the virtual qubit states, and we use the map $\mathcal{M}_1$ -- defined in Figure \ref{fig:intro_fig}. The XX gate, which is an entangling gate, can be implemented using only two $R_{\alpha \beta}$ gates, and without using any vibrational modes, resulting in a lower error rate. c) Examples of native inter-ion gates, assuming a traditional MS interaction between two ions. $\mathrm{MS}_{\{\alpha_i \beta_i\}\{\alpha_j \beta_j\}}(J)$ couples the states $\ket{\alpha_i},\ket{\beta_i}$ in the $i^{\mathrm{th}}$ ion with states $\ket{\alpha_j},\ket{\beta_j}$ in the $j^{\mathrm{th}}$ ion.  For $n=2$, 36 such gates can be constructed, although only one is necessary to obtain a universal gate-set along with intra-ion gates. Interestingly, typical $\mathrm{MS}_{\{\alpha_i \beta_i\}\{\alpha_j \beta_j\}}$ are \textit{four-qubit} gates. For example, using map $\mathcal{M}_1$, the gate $\mathrm{MS}_{\{0,1\}\{0,1\}}$ can be explicitly written as    ${\mathrm{MS}_{\{0,1\}\{0,1\}}}(J)=e^{-iJ H_{12}}$, where $H_{12}=\frac{J}{4}(\mathbf{I}-Z_{1,1})X_{1,2}(\mathbf{I}-Z_{2,1})X_{2,2}$, such that $X_{i,m}(Z_{i,m})$ denotes the $X(Z)$ gate acting on the $m^{th}$ virtual qubit in the $i^{th}$ ion. }
    \label{fig:all_to_all}
\end{figure*}

Consider $n=2$ in a single ion, and the encoding map $\mathcal{M}_1$ from Figure \ref{fig:intro_fig}. The decomposition of the single-qubit gate $e^{-i\theta X_1 }$ is illustrated in Figure \ref{fig:all_to_all}, and requires two native gates. Possible decompositions for other standard single-qubit gates are provided in the Appendix, and typically require $2-3$ $R_{\alpha \beta}$ gates. 

Two-qubit gates between two virtual qubits on the same ion can be implemented through a series of $R_{\alpha \beta}$ gates without using the vibrational modes. For example, the left panel of  Figure \ref{fig:all_to_all} shows the decomposition for the XX gate $U=e^{-i J  X_1X_2}$ in an ion with $n=2$, which under the encoding map $\mathcal{M}_1$ and the connectivity shown in Figure \ref{fig:intro_fig} requires only two native $R_{\alpha \beta}$ gates. The gate parameters are provided in the Appendix.

As an aside, we note that $R_{\alpha \beta}$ gates can only implement gates belonging to SU($d$). Arbitrary gates in U($d$) can be be implemented up to a global phase, but by using transitions to an additional state,  the global phase can be matched.  To implement a CNOT gate between two virtual qubits inside an ion, we need 5 intra-ion gates (see Appendix \ref{sec:app_intra-ion}). However, with a single allowed transition to a fifth state, it can be implemented with two intra-ion gates \cite{multilevel_sc_2015,kiktenko2015single,campbell2022polyqubit}. More generally, the minimum number of gates ($d/2$) remains unchanged upon adding transitions to additional states.

\subsubsection{Comparison of gate errors}
We expect each $R_{\alpha \beta}$ gate to have an error comparable to a typical single-qubit gate error (in one ion with single qubit), which is around $10^{-4}$ in state-of-the-art systems \cite{moses2023race,wright2019benchmarking,harty2014high}. This means that two-qubit gates implemented within an ion will have much lower error rates compared to the standard approach. For example, an XX gate between two ions, each with one qubit, has an error rate around $10^{-3}$ \cite{moses2023race,ballance2017high}.   When the same gate is implemented within an ion with $n=2$ virtual qubits, it only requires 2 $R_{\alpha \beta}$ gates, implying its error rate is approximately $2\times$ $10^{-4}$, which is five times better than the XX gate. It will also be faster than the XX gate since it does not use vibrational modes.

The error rates for single-qubit gates can be estimated similarly. As discussed in the previous section, single-qubit gates implemented in an ion with $n$ virtual qubits require at least $d/2=2^{n-1}$ gates, such that the error rate is approximately $2^{n-1}\times 10^{-4}$. Thus, for $n=2$ and $n=3$, the error rate is comparable to $10^{-4}$, but becomes larger by at least an order of magnitude for higher $n$.

So far, we have ignored the length of the ion chain. For a proper comparison, single/two-qubit gates in a single ion with two virtual qubits should be compared to a single/two-qubit gate in a two-ion chain with one qubit per ion. That's because, in practice, error rates can be optimized for smaller number of ions. That means the error rates in the previous paragraph may be improved further in experimental implementations.

\subsection{Inter-ion gates} \label{sec:inter_ion}

Various types of multi-qubit inter-ion gates can be constructed by building upon the Mølmer–Sørensen (MS) interaction~\cite{molmer_s_1999,molmer_s_2000}, which allows direct implementation of the XX gate ($e^{-i J  X_1X_2}$) between two ions with single qubits, by coupling the ions' internal states with the ion chain's vibrational modes. We first illustrate how the same gate transforms when we consider the higher dimensional space given by $n>1$ (without adding any additional lasers or modifying the setup). Consider the MS gate that couples the states $\ket{0}$ and $\ket{1}$ in each ion (Figure \ref{fig:all_to_all}c), denoted by $\mathrm{MS}_{\{0,1\}\{0,1\}}(J)$.

When $n=2,L=2$, the gate becomes a \textit{four-qubit gate} under encoding map $\mathcal{M}_1$ (Figure \ref{fig:intro_fig})  :

\begin{align}\label{eq:MS_gate_two_qubit}
    {\mathrm{MS}_{\{0,1\}\{0,1\}}}(J)=e^{-iH_{12}}
\end{align}
where 
\begin{align}
H_{12}=\frac{J}{4}(\mathbf{I}-Z_{1,1})X_{1,2}(\mathbf{I}-Z_{2,1})X_{2,2}
\end{align}
and $X_{i,m}(Z_{i,m})$ denotes the $X(Z)$ gate acting on the $m^{th}$ virtual qubit in the $i^{th}$ ion. $H_{ij}$ then denotes the interaction term between ions $i$ and $j$ (here $i=1,j=2$).  We assume that whenever an ion is in a state with no overlap in the space spanned by $\{\ket{0},\ket{1}\}$, i.e. belonging to the space spanned by $\{\ket{2},\ket{3}\}$, the frequencies of the applied MS gates are such that  vibrational modes are not coupled, and the gate effectively acts as an identity operation. 

Using the same physical mechanism, an MS gate can potentially be applied coupling two arbitrary states $\ket{\alpha}$ and $\ket{\beta}$ in one ion with different states in another ion (see Figure \ref{fig:all_to_all}c). We will denote such gates with the shorthand $\mathrm{MS}_{\{\alpha_i \beta_i\}\{\alpha_j \beta_j\}}(J)$, with the states $\alpha_i,\beta_i$ and $\alpha_j,\beta_j$ belonging to  $i^{th}$ and $j^{th}$ ion respectively. The explicit form of the gate independent of the encoding map is :

\begin{multline}
     {\mathrm{MS}_{\{\alpha_i,\beta_i\}\{\alpha_j,\beta_j\}}}(J) = \exp \Big(-iJ \big( \ket{\alpha}_i\otimes \ket{\alpha}_j\bra{\beta}_i\otimes \bra{\beta}_j\\+\ket{\alpha}_i\otimes \ket{\beta}_j\bra{\beta}_i\otimes \bra{\alpha}_j+\mathrm{h.c.} \big) \Big)
\end{multline}

In writing it this way, the assumption about the gate from $\mathrm{MS}_{\{0,1\}\{0,1\}}$ generalizes as follows : whenever the $i^{th}$ ion is in a state with no overlap in the space spanned by $\{\ket{\alpha_i},\ket{\beta_i}\}$, or the $j^{th}$ ion is in a state with no overlap in the space spanned by $\{\ket{\alpha_j},\ket{\beta_j}\}$, the gate acts as an identity operation. The validity of the assumption depends on the manifold of the states and arising selection rules, and corresponding frequencies, as we elaborate in Section  \ref{sec:guidelines_comp_manifold}. 

Since any such  gate is capable of generating entanglement between qubits within different ions (independent of $n$), it forms a universal gate set combined with the native intra-ion gates introduced in the previous section \cite{brylinski2002universal}.

The native gate in Eq. \ref{eq:MS_gate_two_qubit} (and other $\mathrm{MS}_{\{\alpha_i \beta_i\}\{\alpha_j \beta_j\}}(J)$ gates) can be further generalized when the beatnotes are applied on all ions simultaneously. We denote the vibrational modes coupling between ions $i,j$ as $J_{ij}$. Then, the $\mathrm{MS}_{\{0,1\}\{0,1\}}$ gate for $n=2$ using map $\mathcal{M}_1$ is given by

\begin{align}\label{eq:MS_gate}
    {\mathrm{MS}_{\{0,1\}\{0,1\}}}(\{J_{ij}\})=e^{-\sum_{ij}iJ_{ij} H_{ij}}
\end{align}
where $H_{ij}=\frac{J}{4}(\mathbf{I}-Z_{i,1})X_{i,2}(\mathbf{I}-Z_{j,1})X_{j,2}$. While it appears very different than the traditional MS gate with $n=1$ ($e^{-i J  X_1X_2}$), they both share the property that $[H_{ij},H_{kl}]=0, \forall  i\neq j,k\neq l$, independent of the map $\mathcal{M}$ used. This is also true for all other $\mathrm{MS}_{\{\alpha_i \beta_i\}\{\alpha_j \beta_j\}}$ gates. Such Hamiltonians belong to the class of stabilizer Hamiltonians, that can be used to explore topological phenomena in non-equilibrium systems, among other applications \cite{CvK2016phase1,CvK2016phase2,bahri2015localization}.

\begin{figure*}[t]
    \centering
    \includegraphics[width=\linewidth,trim = 0cm 0cm 0cm 0cm]{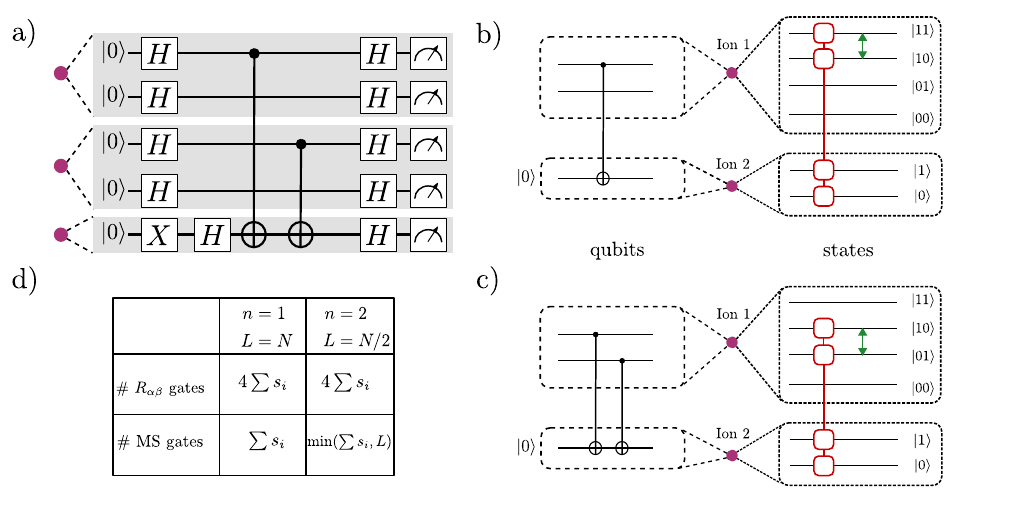}
    \caption{a) A textbook quantum circuit implementation of the  Bernstein Vazirani algorithm, including the oracle which implements $f(x)=x\cdot s$ (mod 2), with the hidden string $s=1010$. In general, $\sum s_i$ CNOT gates are required for the hidden string $s$ (counting the number of 1's in $s$). We propose defining $n=2$ virtual qubits for ions as shown by the shaded regions. b) Implementation of a CNOT gate in terms of native gates (Figure \ref{fig:all_to_all}) in an ion chain with $n=2$ virtual qubits, and $n=1$ for the auxiliary ion. A single intra- and inter-ion gate suffices, whereas for ions with single qubits, four intra-ion gates and one intra-ion gate are needed.  c) Two CNOT gates with control qubits in the same ion can be implemented by a single intra- and inter-ion gate, significantly reducing the number of gates and gate errors, as inter-ion gates have an order of magnitude worse error rates than intra-ion gates. 
    d) Counting the total number of native gates used in the circuit, after optimizing the circuit for each hidden string, and comparing with a similar implementation for the traditional ion chain $n=1$. We see that the number of gates required are lower on average. This implies that for a given number of qubits $N$, the algorithm performs better in a smaller ion chain with $L=N/2$ ions with two virtual qubits each, than with $L=N$ ions with single qubit per ion. }
    \label{fig:BV}
\end{figure*}

What do standard gates between virtual qubits in different ions look like in terms of the native inter- and intra-ion gates? For simplicity, we first consider two ions hosting three qubits, i.e. $n_1=2$, and  $n_2=1$. The top right panel of Figure \ref{fig:BV} shows a decomposition of the $_{1,1}\mathrm{CNOT}_2$ gate, where subscripts denote that the control qubit is the first qubit within the first ion in the chosen map $\mathcal{M}_1$, while the sole qubit in the second ion serves as the target qubit. Gate parameters and decompositions of some other standard two/three-qubit gates are provided in Appendix \ref{sec:app_inter_ion}.

The $_{1,1}\mathrm{CNOT}_2$ gate requires only a single $R_{\alpha \beta}$ gate and a single MS gate. This is also true for the  $_{1,2}\mathrm{CNOT}_2$ gate (Figure \ref{fig:stab_meas}). In contrast, a CNOT gate between two ions having only single qubits requires 4 single-qubit gates and an MS gate \cite{debnath2016demonstration}.  

The errors could be reduced further in practice, since fewer ions would be needed for the same total number of qubits. A smaller ion chain enables better optimization of MS gates due to less complex vibrational mode spectra and lower power requirements, and consequently they have better fidelities.

This advantage of having virtual qubits breaks down as the number of virtual qubits increases. In the intermediate case when $n=2$ in both the ions, at least two MS gates are needed to perform a CNOT or XX gate between the qubits in each ion, with a higher number of gates required to construct inter-ion XX gates (see Appendix \ref{sec:app_inter_ion}). 

This limitation can be partially surpassed by modifying the standard MS interaction.  One line of approach proposed in \cite{campbell2022polyqubit,low_2020_practical,katz2023programmable} is to excite additional pairs of states in each ion simultaneously. For example, Ref \cite{campbell2022polyqubit} proposed exciting the states $\ket{2}$ and $\ket{3}$ at the same time as $\ket{0}$ and $\ket{1}$ (for $n=2$ virtual qubits in two ions), which under $\mathcal{M}_1$ leads directly to an XX gate between two qubits in different ions:
\begin{align}\label{eq:MS_polyqubit}
    U=e^{-iJX_{1,2}X_{2,2}}
\end{align}
The challenge of this approach lies in the practical difficulty of implementing simultaneous coupling of multiple pairs of states with the vibrational modes and fine-tuning the laser pulse intensities \cite{campbell2022polyqubit}. 

Since modifying the MS interaction can be difficult, we don't consider such gates for the remainder of the paper, and instead focus on the applications using $\mathrm{MS}_{\{\alpha_i \beta_i\}\{\alpha_j \beta_j\}}$ for a small number of virtual qubits.

So far, we have focused on the number of native gates required to implement a particular gate for a fixed encoding map $\mathcal{M}_1$. Next, we examine the role of encoding maps.


\begin{center}
\begin{figure}[h]
    \centering
    \includegraphics[scale=0.75,width=\linewidth,trim = 0cm 0.2cm 0cm 0cm]{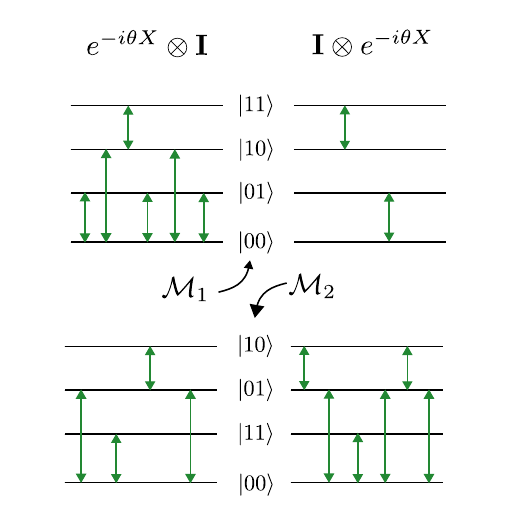}
    \caption{A decomposition of the single qubit gate $e^{-i\theta X}$, using a minimal native gate set under two encoding maps, defined in Figure \ref{fig:intro_fig}. The number of native gates depends on the specific virtual qubit within an ion that is being considered, and the map that is used to encode the states into a two qubit system. This asymmetry between the two virtual qubits vanishes when \textit{all} possible intra-ion transitions are allowed, and two native gates would suffice to construct each of the four gates, for both $\mathcal{M}_1$ and $\mathcal{M}_2$ -- like in Figure \ref{fig:all_to_all}.}
    \label{fig:single_qubit_gate}
\end{figure}
\end{center}

\subsection{The role of connectivity and encoding map}\label{sec:lim-conn}

When all $R_{\alpha \beta}$ and $\mathrm{MS}_{\{\alpha_i \beta_i\}\{\alpha_j \beta_j\}}(J)$ are allowed for a given number of virtual qubits, we expect the minimum number of native gates required to construct a standard gate to be independent of the basis encoding map used. This can be easily seen for an intra-ion gate, as the decomposition of a unitary $U$ under a map $\mathcal{M}_a$ can be used to find the decomposition under another map $\mathcal{M}_b$, by replacing each $R_{\alpha \beta}(\theta,\phi)$ in the former by $R_{\mathcal{M}^{-1}_b (\mathcal{M}_a(\alpha)) , \mathcal{M}^{-1}_b (\mathcal{M}_a(\beta))}(\theta,\phi)$. A similar reasoning can be used to generalize this independence with MS gates.

What happens when the allowed transitions are limited, yet forming a connected graph so that the gate set is universal? In this case, the different virtual qubits inside an ion become asymmetric, in terms of the number of native gates needed to implement a single-qubit gate. Further, different encoding maps no longer remain equivalent. 

As an example, assuming the connectivity inside a single ion with $n=2$ to be as shown in Figure \ref{fig:intro_fig}, consider the gate $U=\exp(-i\theta X)$ acting on a single qubit. In Figure \ref{fig:single_qubit_gate}, we see that the minimum number of native gates needed to implement the unitary depends on \textit{which} virtual qubit it acts on within an ion. Further, the number of native gates is different under encoding maps $\mathcal{M}_1$ and $\mathcal{M}_2$, suggesting that generally one would need to find an optimal encoding map depending on the desired application. A similar conclusion is expected to hold for inter-ion gates.

Since intra-ion gates are easier to implement than inter-ion gates, how important is the connectivity in intra-ion gates vs inter-ion gates? Surprisingly, for the random circuits that we consider in Section \ref{sec:XEB}, a limited connectivity in inter-ion gates can be compensated by an all-to-all connectivity of intra-ion gates. We also comment on practically achievable connectivity in \Ba ions in Section \ref{sec:guidelines_comp_manifold}.

\section{Applications}\label{sec:applications}

Until now, we have argued that error rates for certain single- and multi-qubit gates can be reduced by defining a small number of virtual qubits inside an ion -- using one or two ions. Does that improvement hold for a larger ion-chain implementing a circuit with a variety of gates? 

Here, we provide example applications using $n=2$ and $n=3$ virtual qubits, where fewer number of native gates would be needed for the same total number of qubits. Alternatively, a larger number of qubits can be accessed using the same number of ions.

\subsection{Entangling operations}\label{sec:app_ent_ops}
We noticed in Section \ref{sec:native_intra} that native intra-ion gates are typically multi-qubit gates, and capable of generating entanglement between the qubits inside an ion, without using vibrational modes of the chain. Because intra-ion gates are easier to apply parallelly in different ions, this simplifies application of parallel entangling gates, which is important considering finite coherence times of NISQ devices.

 For example, the following unitary can be easily implemented in an  $L$ ion chain with $n=2$ in each ion
\begin{align}
    U=e^{-i\sum_{i=1}^{L}J_{i} X_{2i-1}X_{2i}}
\end{align}
where each $J_i$ can be independently chosen, without needing to excite motional modes. Thus, this unitary can entangle qubits $2i-1$ and $2i$ for each $i$.

While such a unitary could in principle be applied using inter-ion MS gates in an ion chain with one-qubit per ion, the pulse sequences need to be carefully optimized to only interact with specific vibrational modes \cite{figgatt2019parallel,grzesiak2020efficient}.

Combining the entangling intra-ion operations with native inter-ion $\mathrm{MS}_{\{\alpha_i \beta_i\}\{\alpha_j \beta_j\}}(J)$ then allows access to a variety of entangled states incorporating parallel operations. Counting the maximum number of allowed MS gates -- 36 for $n=2$ -- illustrates the advantage over choosing $n=1$. 

When using the same number of ions to access a larger number of qubits, the variety of native entanglement operations can then allow generation of entanglement over many more qubits than trapped ions can currently accommodate \cite{pogorelov2021compact,moses2023race}. 

All of this can be done without modifying the native inter-ion gate interaction. Modified MS gates -- such as one that couple more than two internal states at the same time \cite{campbell2022polyqubit} -- can be used to create specific entangled states more efficiently. For example, a global GHZ state can be generated by applying such a gate just \textit{once}. To see that, consider $n=2$ and the map  $\mathcal{M}_2$, defined by $\mathcal{M}_2(\ket{0})=\ket{00}$, $\mathcal{M}_2(\ket{1})=\ket{11}$, $\mathcal{M}_2(\ket{2})=\ket{01}$ and $\mathcal{M}_2(\ket{3})=\ket{10}$. If the modified MS gate simultaneously couples $\{\ket{0},\ket{1}\}$ and  $\{\ket{2},\ket{3}\}$ in each ion, and is applied with a uniform strength, then the resulting gate is the $2L$-qubit gate :
\begin{align}
    U=e^{-iJ\prod_{i=1}^{2L}X_{i}}
\end{align}
which leads to a $2L$-qubit GHZ state when acted on $\ket{000...}$ with $J=\pi/4$. For the remaining applications, we won't consider such modified MS gates.

\subsection{Implementing standard algorithms}\label{sec:bv}
 The Bernstein-Vazirani (BV) algorithm can be used to find an unknown string $s$ (size $N$) that an oracle  implements in the function $f(x)=x\cdot s$ (mod 2). The algorithm can find $s$ in a single iteration, as opposed to $N$ iterations required for an efficient classical solution. It has been practically demonstrated on many platforms, including a trapped-ion quantum computer with single qubit per ion \cite{wright2019benchmarking}. 

The typical quantum circuit implementing both the oracle and the algorithm, using standard gates, consists of Hadamard and CNOT gates. Note that it includes an auxiliary qubit. The string $s$ is then found by measuring the output of the $N$ data qubits. For example, Figure \ref{fig:BV}a) shows the circuit for $N=4$, $s=1010$. 

In an ion-chain with single qubit per ion, the Hadamard gate can be implemented directly using a single intra-ion gate, while the CNOT gate uses four intra-ion gates and one inter-ion MS gate \cite{wright2019benchmarking}. The oracle implementation differs for different strings $s$. For an $N$-bit unknown string $s$, $N$ qubits are needed. If the $i^{th}$ bit of $s$ is one ($s_i=1$), a CNOT gate is implemented between the $i^{th}$ qubit and an auxiliary qubit. 

The circuit implementation for all possible strings $s$ can be simplified by using $n=2$ virtual qubits per ion, except for the auxiliary ion, which has a single qubit. In Figure \ref{fig:BV}, we show the decomposition for CNOT gates using the native gate set for an ion-chain with $n=2$, along with the auxiliary ion with $n=1$. A single CNOT gate in this setup only requires one MS gate and one intra-ion gate. A Hadamard gate can be implemented with three intra-ion gates (gate parameters provided in Appendix \ref{sec:app_inter_ion}).

Remarkably, two successive CNOT gates with the control qubits inside the same ion can be implemented with just \textit{one} MS gate, as opposed to \textit{two} with single qubit per ion. This leads to an overall reduction in the number of MS gates required, when compared to a larger ion chain with the same $N$. 

An optimized counting -- excluding irrelevant Hadamard gates, and decomposing combined gate operations -- then implies that for a hidden string $s$, the number of $R_{\alpha \beta}$ gates in the ion chain with $n=2$ is $4\sum s_i$ (where $\sum s_i$ counts the number of $1$s in the string $s$). That's the same as the total number of single qubit gates in the ion chain with larger number of ions. 

As for the number of inter-ion gates, because of the ability to combine two CNOT gates sharing a virtual qubit on the same ion, the number of MS gates is less than or equal to  $\min \big(\sum s_i,L\big)$, whereas for the traditional setup, it is $\sum s_i$. 

Therefore, combining the number of intra- and inter-ion gates needed on an average, the implementation for an $N$ bit string requires fewer inter-ion gates when using $N/2$ ions with two virtual qubits, as opposed to $N$ ions with single qubits. That implies lower error rates, as well as faster implementation.

\begin{figure}[t]
    \centering
    \includegraphics{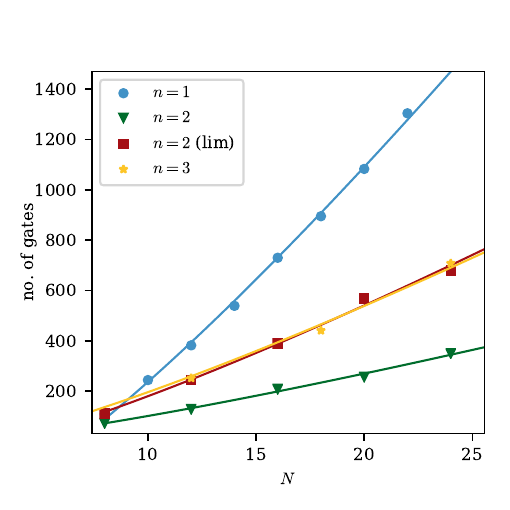}
    \caption{Number of gates required for an ideal noiseless random quantum circuit with $N$ qubits such that its linear cross entropy fidelity  $\mathcal{F}_{\mathrm{XEB}}=2$. The random circuits  have a brick-work architecture. Each ``brick'' unitary is made up using one MS gate and two intra-ion gates applied on neighbouring ions : 
    an MS gate is applied on two neighboring ions, followed by $R_{\alpha \beta}$ gates on the same ions, with randomly chosen parameters. This is done both for the traditional setup with $n=1,L=N$, and using two and three virtual qubits on each ion $n=2,N=2L$ and $n=3,N=3L$. For $n=2,3$ we assume an all-to-all connectivity between internal states of neighboring ions. For $n=2$, we also consider a gate set with the minimal connectivity required for universality (Figure \ref{fig:intro_fig}). 
    For all cases we find a scaling consistent with $O(N\log(N))$ theoretical predictions (fits). More importantly, the overall number of gates are smaller when using $n=2,3$ virtual qubits, compared to $n=1$. }
    \label{fig:XEB_main}
\end{figure}

\subsection{Random quantum circuits}\label{sec:XEB}
How well do we expect the advantage of using a few virtual qubits per ion to hold for more general circuits? In an attempt to understand that, we consider random quantum circuits formed by applying intra- and inter-ion gates with randomly chosen parameters. 

Random quantum circuits have been instrumental in understanding universal aspects of non-equilibrium many-body quantum dynamics \cite{fisher2023random}. Sampling from the distribution of measurements for sufficiently random circuits has been a leading proposal for establishing quantum advantage in near term quantum devices \cite{boixo2018characterizing,google2019supremacy,morvan2023phase}.  

Focusing on the random circuit sampling, we aim to understand how quickly the probability distribution $p(x)$ of measured output bit-strings $x$ for \textit{noiseless} random quantum circuits converges to that for a Haar random unitary, for which the distribution is known to be the Porter-Thomas distribution. Specifically, we look at its modified second moment, the linear cross entropy fidelity $ \mathcal{F}_{\mathrm{XEB}}=2^N\braket{p(x_i)}_{x_i}-1 $, where the averaging is performed over the measured bit-strings $x_i$. 

For a noiseless random circuit, starting with a computational basis state, $\mathcal{F}_{\mathrm{XEB}}=2^N-1$, whereas for the Porter-Thomas distribution, $\mathcal{F}_{\mathrm{XEB}}=1$. As the number of random gates applied in the circuit increases, the possible output bit-strings spread out -- or anti-concentrate -- in the Hilbert space. Prior works have argued that both short-range and long-range random circuits anti-concentrate in $\log$ depth, implying the number of gates scales as $N\log(N)$. This property is also obeyed by $2$-designs of the Haar and Clifford ensemble \cite{dalzell2022random,harrow2009random,harrow2023approximate}.  

Here, we numerically examine the precise number of gates required for  $\mathcal{F}_{\mathrm{XEB}}$ to converge. The number of required gates is an important metric since in the presence of noise, $F_{\mathrm{XEB}}$ tends to decrease exponentially with the number of gates, and the gate error. Therefore, a lower gate count will lead to a higher $\mathcal{F}_{\mathrm{XEB}}$ value for the same number of qubits, which is an important criterion to verify quantum advantage \cite{google2019supremacy,morvan2023phase}.

Our random circuits are constructed using a brick-work architecture. A ``brick" unitary between each pair of neighbouring ions is constructed as follows. An MS gate is applied between the two ions with randomly chosen pairs of states, with a random parameter $J \in [0,2\pi]$. It is followed by two intra-ion $R_{\alpha \beta}$ gates, applied on the same two ions, with randomly chosen $\alpha,\beta,\theta,$ and $\phi$ ($\alpha \neq \beta)$. 

A layer of such ``brick" gates act first on the pairs of ions $(1,2),(3,4),\dots,(L-1, L)$, and then on ion pairs $(2,3),(4,5),\dots,(L, 1)$. New layers are successively added similarly using new random parameters, and finally all qubits are measured in the $Z$ basis.

We look at such circuits with each ion hosting the same number of virtual qubits $n$, and $n\in \{1,2,3\}$.

In Figure \ref{fig:XEB_main}, we plot the number of gates required such that $\mathcal{F}_{\mathrm{XEB}}=2$. We see an $N\log(N)$ scaling for each $n$, consistent with theoretical predictions \cite{dalzell2022random,harrow2009random,harrow2023approximate}.

Even though they scale the same way, the total number of gates required is significantly lower when $n=2$ and $n=3$ virtual qubits are used.

This result is not restricted to the assumed architecture and connectivity of native gates. For example, even if the connectivity is restricted among both MS and $R_{\alpha \beta}$ gates for $n=2$, the number of native gates required for $\mathcal{F}_{\mathrm{XEB}}=2$ is smaller than for $n=1$, as shown in Figure \ref{fig:XEB_main}. For the restricted connectivity curve, we assume a single $MS_{\{0,1\},\{0,1\}}$ gate, and $R_{01},R_{12},R_{23}$ to be allowed, which is still a universal gate set.

With a less severe restriction on connectivity, by allowing  all possible $R_{\alpha \beta}$ but fixing the allowed MS gate to be $MS_{\{0,1\},\{0,1\}}$, we find the number of gates to be similar to when all possible MS gates are assumed (Appendix \ref{sec:app_ng2}). 

Similar conclusions hold when looking at a higher moment of the distribution, or using a long range random circuit, which are presented in Appendix \ref{sec:app_ng2}.

As the number of virtual qubits increases, the required number of gates for $\mathcal{F}_{\mathrm{XEB}}=2$  increases as well. For $n \geq 4$, more number of gates may be needed than for $n=1$. In particular, if the number of ions $L$ is kept fixed and $n$ increases, the number of gates is expected to increase exponentially in $n$. 

Thus, $n=2,3$ virtual qubits in an ion provide a unique middle-ground where random circuits approach Haar designs faster than $n=1$ random circuits of the same type. With access to a larger number of qubits for the same number of ions, this would potentially allow for comparison of quantum supremacy experiments in a trapped ion quantum computer \cite{moses2023race}.

\begin{center}
\begin{figure}[h]
    \centering
    \includegraphics[width=1.00\linewidth,trim = 0cm 0.5cm 0cm -0cm]{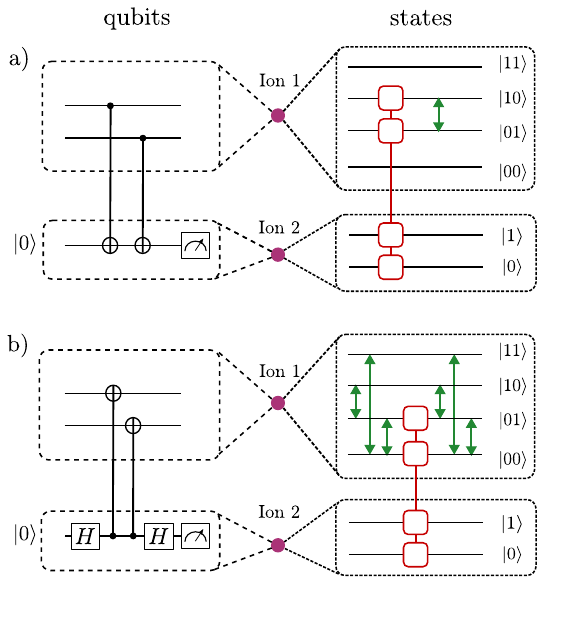}
    \caption{One way to implement a) $ZZ$ and b) $XX$  stabilizer measurements for a bit-flip and a phase-flip repetition code, respectively. Ion 1 serves as a data-ion, with two virtual qubits. And Ion 2 -- with a single qubit -- is used to measure the stabilizer. For both a) and b)  measurements, only \textit{one} inter-ion MS gate is needed! If we used two ions -- with single qubits each -- as data ions, \textit{two} inter-ion gates would be needed. This halving of the number of inter-ion gates then also applies for higher weight stabilizers like $XXXX$ and $ZZZZ$ used in surface codes. }
    \label{fig:stab_meas}
\end{figure}
\end{center}

\subsection{Error correction}\label{sec:bit-flip}

Can we utilize virtual qubits for error correction? Ideally, in an $[N,k,d]$ error correcting code, one might want to increase the logical number of qubits $k$ \textit{and} the code distance $d$. In order to highlight the implications of using virtual qubits clearly, we fix $k$ and the family of codes. Then, one can increase the total number of data-qubits $N=nL$ and the distance $d$, by increasing $n$. Since $nL$ data-qubits will be embedded in $L$ ions, we'll call such ions ``data-ions''.

A larger distance by itself is not a good measure of the performance of the code, and instead the logical error rate must be compared between the two implementations. And that will depend on the natural errors in the setup and the decoding algorithm used.

 We discuss the bit-flip repetition code for simplicity, which protects a single logical qubit ($k=1$) against bit-flip errors on upto $d=\frac{nL-1}{2}$ data-qubits. A higher distance can thus be achieved using the same number of ions by increasing $n$.   

 What does that mean for the logical error rates? Here, it mainly depends on the nature of errors in implementing the stabilizer measurements. 

We assume stabilizer measurements are performed using additional ion(s) with a single-qubit per ion, and refer to them as stabilizer ions. Further, we restrict our discussion to $n=2$ virtual qubits per data-ion. 

Figure \ref{fig:stab_meas} shows the decomposition for $ZZ$ stabilizer measurements -- with full gate parameters provided in the Appendix. We also show the decomposition for an $XX$ stabilizer measurement, which helps us generalize our results for the phase-flip repetition code and the surface code. We find that both the stabilizers can be measured by using just \textit{one} inter-ion gate, as compared to \textit{two} gates required when $n=1$  -- closely related to the discussion in Section \ref{sec:bv}.

Even though fewer gates are required for stabilizer measurements on an average, the multi-qubit nature of errors can potentially contribute to larger logical error rates. For example, a single CNOT gate -- in Figure \ref{fig:stab_meas} --  acts between three qubits : two data-qubits in one data-ion, and one qubit on the auxiliary ion. As a result, an error in that gate is likely to be a three-qubit error. 

How does the competition between fewer gates and multi-qubit errors translate into the performance of the bit-flip repetition code for $n=2$? To answer that, we consider a circuit implementing a logical memory. In such a circuit, no external errors are introduced, and the errors due to the stabilizers themselves determine the logical error rate. The stabilizers are measured for multiple rounds for an encoded state, which is decoded at the end. 

To model the error in stabilizer measurements, we use a three-qubit depolarizing channel, applied after each inter-ion gate between a data-ion ($n=2$) and a stabilizer ion ($n=1$).  

In order to compare with the single-qubit per ion case, we also consider a data-ion with $n=1$, and apply a two-qubit depolarizing error channel after every CNOT gate. 

The error rates for the depolarizing channels are estimated from the number of intra- and inter-ion gates applied. The error models are described in more detail in Appendix \ref{sec:app_err_model}.


Once we have decided the error model, which $L$ and $d$ should be compared between $n=1$ and $n=2$? A natural approach is to keep $L$ fixed. For a given total number of  qubits $N=2d+1$, the repetition code uses $d+1$ data-qubits and $d$ auxiliary qubits for stabilizer measurements. In fact, a single auxiliary ion is sufficient for performing \textit{serial} stabilizer measurements \cite{moses2023race}.

Therefore, the number of ions required using $n=1$ is $L_1=d_1+2$. On the other hand, when using two virtual data-qubits per data-ion, the number of ions is $L_2=\frac{d_2+1}{2} + 1$. $L_1=L_2$ then implies $d_2 = 2d_1+1$.

\begin{center}
\begin{figure}[h]
    \centering
    \includegraphics[width=1.01\linewidth,trim = 0cm -0.0cm 0cm -0cm]{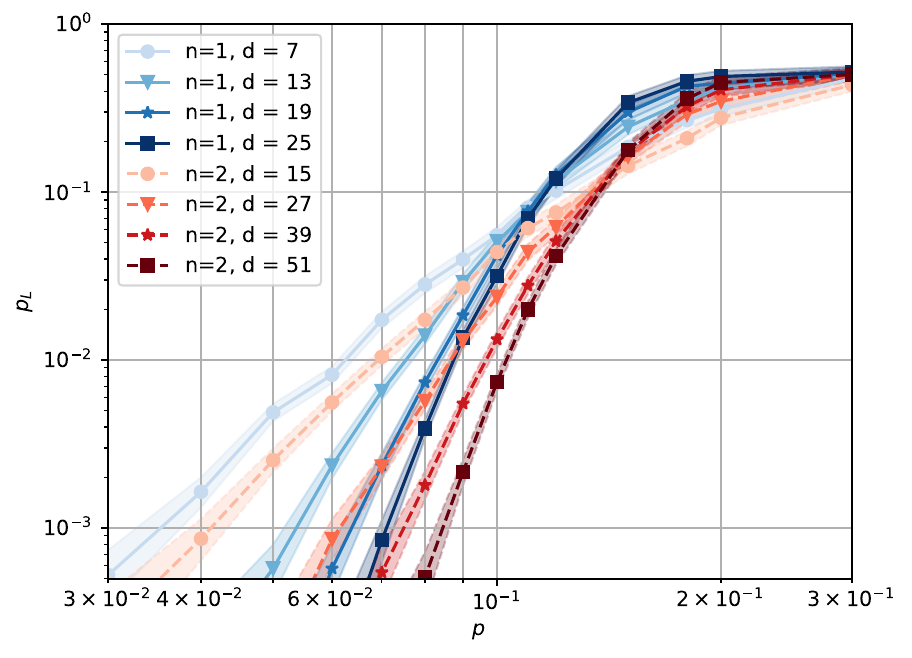}
    \caption{Logical ($p_L$) and physical $(p)$ error rates for a bit-flip repetition code, assuming a depolarizing error channel applied after every inter-ion MS gate, and perfect measurements. The physical error rate corresponds to the infidelity of the two-qubit depolarizing error channel. The blue curves correspond to a single data-qubit per ion, while the red curves correspond to $n=2$ data-qubits per data-ion. The distances are chosen so that the number of ions is the same for each shade.  We assume a single ion is used for stabilizer measurements, such that the total number of ions is $L=d+2$. When using virtual qubits, stabilizer measurements require fewer inter-ion gates when $n=2$ (Figure \ref{fig:stab_meas}, \ref{fig:BV}), but that enhancement competes with the fact the application of an inter-ion gate leads to a three-qubit error as opposed to the two-qubit error model with $n=1$. Under our assumptions, we see that the logical error rate can  be an order of magnitude smaller when using $n=2$ virtual qubits.   }
    \label{fig:threshold_repcode}
\end{figure}
\end{center}

We plot the physical and logical error rates for different $L$ in Figure \ref{fig:threshold_repcode}, using the stabilizer circuit simulator Stim \cite{Gidney2021stimfaststabilizer}, and a minimum weight perfect matching decoder through PyMatching \cite{higgott2022pymatching}. For each $L$, we perform $d_1$ rounds of stabilizer measurements. We also assume that the measurements are error-free. 

We observe that there is a regime -- for a given physical error rate -- where the logical error rate is lower by almost an order of magnitude when using virtual qubits to increase the code distance!

As a by-product of using fewer gates, the time needed to perform a round of stabilizer measurement will also be smaller using $n=2$.

We expect our results to generalize to the phase-flip repetition code and also surface codes, and thus our approach can potentially enhance the performance of codes that can correct arbitrary errors, by optimizing the number of native gates required, without using any additional ions.


\section{Considerations for experimental realization}\label{sec:experiment}

The discussion so far is applicable to any trapped ion species that offers a sufficient number of long-lived non-degenerate states with the required connectivity. In the following section, we outline experimental criteria for manipulating multiple qubits in a single ion. We discuss criteria for selecting the set of states and transitions used to store these qubits (the \textit{computational manifold}) out of a large number of available atomic states. Thereafter we compare different ions, and highlight the advantages of using the metastable level in \Ba ions. 

\subsection{Experimental manipulation protocols}\label{sec:experiment_gates_readout}
To process information with $n>1$ qubits per ion, we must experimentally implement a universal gate set using intra-ion two-state rotations and inter-ion gates, described in Sections \ref{sec:native_intra} and \ref{sec:inter_ion}. Additionally, state-preparation and measurement operations are also necessary. These can be realized similarly to the case of encoding qubits or multi-dimensional qudits \cite{low_2020_practical,low2023control,hrmo2023native,ringbauer2022universal}.

The mechanism employed for driving rotations between two atomic states depends on the transition frequency between them. Trapped-ion qubits are encoded using two long-lived states, which are typically chosen from the ground level (\slevel), the metastable level (\dlevel\ or \flevel) or both. We similarly select the states in the computational manifold from these levels. Thus, two states in the manifold may be in the same level (either Zeeman states, or hyperfine states if the nucleus has non-zero spin) or across the two levels. Transitions frequencies between two Zeeman or hyperfine states in the same level are in the range  $\sim\!10\,\mathrm{MHz} - \sim\!10\,\mathrm{GHz}$. They can be driven with a resonant microwave field \cite{harty2014high} or via a two-photon Raman transition \cite{ballance2016high}. States that are in different levels are separated by an optical transition $\sim\!100\,\mathrm{THz}$ which can be driven directly using a narrow-band laser \cite{manovitz_small_computer, saner_standing_waves}.  

However, the radiation used to drive these rotations also generates AC Zeeman or AC Stark shifts on off-resonant transitions, which induces a phase shift between the states they connect. Driving any transition in the computational manifold will thus induce unwanted shifts in the remaining computational states, which must be accounted for. Furthermore, it has been shown that these shifts can induce decoherence if the amplitude or polarisation of the radiation field fluctuates \cite{vizvary2023eliminating}. While these shifts do not pose fundamental limits, they may need to be experimentally mitigated.

Inter-ion gates require a native two-ion entangling gate using at least one pair of states in the encoding. Following Section \ref{sec:inter_ion}, we choose this to be via the Mølmer–Sørensen (MS) interaction~\cite{molmer_s_1999,molmer_s_2000} for which fidelities well in excess of $99\%$ have been demonstrated for optical \cite{akerman2015universal}, microwave \cite{srinivas2021high} and Raman gates  \cite{ballance2016high}. (Other gate mechanisms \cite{milburn1999simulating, sawyer2021wavelength} with suitable modification to the constructions in Section \ref{sec:inter_ion}). These techniques can be directly applied to the case of encoding more than one qubit per ion. 

In addition, we must be able to execute state-preparation and measurement operations. State-preparation involves transferring all the population to one of the states in the computational manifold, which is the same requirement as when encoding one qubit per ion. Thus, standard optical pumping techniques can be used to achieve fidelities $>99\%$, with the best fidelities of any platform achieved by additional use of coherent operations \cite{an2022high} and heralding \cite{sotirova2024high}. 

Measurement operations are implemented by observing fluorescence while driving a closed cycling transition between the ground level and a short-lived excited level, analogously to how a single qubit can be read out. The operation begins with a \textit{hiding} sequence, which involves transferring any computational states in the ground level (except one) to the metastable level. A single state $\ket{\alpha}$ (the first state to be measured) is left in the ground level, or is transferred from the metastable level if the computational manifold is entirely encoded there. We note that this transfer pulse \textit{does not} need to preserve phase, since no coherent operations occur after it. The lasers that drive the cycling transition are then turned on. If fluorescence is observed (a \textit{bright} outcome), the state is collapsed into $\ket{\alpha}$, so the measurement outcome is $\mathcal{M}(\ket{\alpha})$, where $\mathcal{M}$ is the encoding that maps atomic states to $n$-qubit computational states. If fluorescence is not observed (a \textit{dark} outcome), the population has been projected to the remaining states in the metastable level under the projector $\mathbf{I} - \ket{\alpha}\bra{\alpha}$, since the metastable level does not couple to the fluorescence lasers. This operation is repeated $2^n - 1$ times with the remaining states in the computational manifold, except for one. The measurement outcome is the first state for which fluorescence is observed, or the final state that was not mapped down if no fluorescence is observed. Thus, in addition to any pulses required for the initial hiding sequence, this operation requires $2^n-1$ fluorescence detections and transfer pulses between the metastable and ground levels. This number is halved on average if the operation is terminated after the first observation of fluorescence. Once the ion is read out, the ion can be reinitialised by state preparation.

Reading out an arbitrary subset of ions in the chain can be done by hiding the state of the remaining ions in the metastable level during the readout operation. In contrast, reading out an arbitrary subset of \textit{qubits} is challenging. This is because reading out the $i^{\textrm{th}}$ ion collapses the state of all $n_i$ virtual qubits it encodes, even if we only wish to measure one of them. However, one could think of using SWAP gates (which are enabled by the universal gate set) to move qubits between different ions, such that the subset of qubits to be read out is stored in separate ions to the subset that is to be maintained. While this is not possible for arbitrary subsets of the qubit register, a single auxiliary ion that encodes $n_{\mathrm{aux}} \leq \max{n_i}$ virtual qubits enables this for an arbitrarily large register.

Once a subset of the ions has been read out, the hiding operation must be undone in the remaining ones. Because coherent operations will follow, these hiding operations must be coherent.

\begin{center}
\begin{table*}[t]
\begin{tabular}{| p{1.5cm} | p{1.5cm}| p{2.5cm} | p{3.7cm} | p{3cm}| p{3cm} |}
  \hline
  \hline
  \textit{Ion species} & \textit{Nuclear spin} & \textit{Number of states in} $\mathrm{S}_{1/2}$ & \textit{Metastable level (lifetime)} & \textit{Number of states in the metastable level} & \textit{Maximum number of virtual qubits} \\
  \hline
  \hline
  $^{40}\mathrm{Ca}^+$ & 0 & 2 & $\mathrm{D}_{5/2}$ ($1.2\;\mathrm{s}$ \cite{barton2000Ca40metalifetime}) & 6 & 2 \\ 
  \hline
  $^{43}\mathrm{Ca}^+$ & 7/2 & 16 & $\mathrm{D}_{5/2}$ ($1.2\;\mathrm{s}$) & 48 & 5 \\ 
  \hline
  $^{88}\mathrm{Sr}^+$ & 0 & 2 & $\mathrm{D}_{5/2}$ ($0.391\;\mathrm{s}$ \cite{Letchumanan2005Sr88metalifetime}) & 6 & 2 \\ 
  \hline
  $^{87}\mathrm{Sr}^+$ & 9/2 & 20 & $\mathrm{D}_{5/2}$ ($0.391\;\mathrm{s}$) & 60 & 6 \\ 
  \hline
  $^{138}\mathrm{Ba}^+$ & 0 & 2 & $\mathrm{D}_{5/2}$ ($30.1\;\mathrm{s}$ \cite{Zhang2020Ba138metalifetime}) & 6 & 2 \\ 
  \hline
  $^{137}\mathrm{Ba}^+$ & 3/2 & 8 & $\mathrm{D}_{5/2}$ ($30.1\;\mathrm{s}$) & 24 & 4 \\ 
  \hline
  $^{133}\mathrm{Ba}^+$ & 1/2 & 4 & $\mathrm{D}_{5/2}$ ($30.1\;\mathrm{s}$) & 12 & 3 \\ 
  \hline
  $^{171}\mathrm{Yb}^+$ & 1/2 & 4 & $\mathrm{F}_{7/2}$ ($1.6\;\mathrm{y}$ \cite{Lange2021Yb171metalifetime}) & 16 & 4 \\ 
  \hline
  $^{173}\mathrm{Yb}^+$ & 5/2 & 12 & $\mathrm{F}_{7/2}$ ($> \mathrm{days}$\cite{Dzuba2016Yb173metalifetime}) & 48 & 5 \\ 
  \hline
  \hline
\end{tabular}
\caption{A list of the most commonly used species in trapped-ion quantum computers, together with the number of states in the ground level and in another low-lying long-lived (metastable) level, together with the lifetime of the metastable level. The maximum number of virtual qubits is calculated from the total number of states in the ground and metastable levels minus 1 as at least one state in the ground level needs to be left for readout.}
\label{table:ion-species}
\end{table*}
\end{center}

\subsection{Guidelines for choosing the computational manifold}\label{sec:guidelines_comp_manifold}

To encode $n$ virtual qubits in an ion, we must select $2^n$ suitable states from across the ground and metastable levels to form the computational manifold. For any atomic species used in trapped-ion quantum computing experiments, these atomic levels contain enough states to allow for encoding multiple qubits. This is particularly true if the species has non-zero nuclear spin that results in hyperfine structure. An overview of the properties of the most commonly used isotopes is given in Table \ref{table:ion-species}. Thus, there are many possible combinations of states that can be chosen as the computational manifold. A minimal condition on the manifold is that it is possible to choose enough allowed transitions between the states to form a connected graph, as per Section \ref{sec:native_intra}. Transitions are typically considered allowed or not according to atomic selection rules, but other considerations, such as excessive off-resonant excitation of nearby transitions, may influence whether the transition can be practically driven, and thus included in the computational manifold (see Section \ref{sec:sup_constructing_encoding} for further discussion).

A good choice of computational manifold is one that minimises the experimental complexity, the error rate in intra- and inter-ion gates, and the error rate due to coupling to noisy environments. While some sources of error are independent of the encoding (e.g. inter-ion gates are sensitive to the heating rate and coherence of the motional modes of the ion chain), others depend on the atomic properties of the states in the computational manifold. The most relevant of these are:

\begin{enumerate}
    \item Driving two-state rotations may off-resonantly excite unwanted transitions (both within and outside of the computational manifold). Transitions between states in the manifold thus need to be well-resolved from each other and from transitions to spectator states. Off-resonant excitations can be further mitigated using well-established techniques like composite pulses \cite{Cummins2003Tackling} and pulse shaping \cite{bauer1984Gaussian}.
    \item Coupling to fluctuating external magnetic fields causes states in the computational manifold to acquire unknown differential phases, corrupting the quantum state stored within an ion. For $n=1$, it is possible to choose a pair of states whose frequency splitting is magnetically insensitive to first-order at specific quantisation magnetic fields, a so-called clock qubit \cite{harty2014high}. These states then exhibit significantly reduced dephasing errors. In general, magnetically insensitive manifolds with $n\geq2$ are not accessible in typical trapped ion experiments. The dephasing error can be minimized by choosing states with low relative magnetic field sensitivity. Furthermore, the execution time of algorithms determines the time period over which noisy external fields can couple to the computational manifold. This time can be reduced by maximising the matrix elements of the transitions in the encoding, thus speeding up two-state rotations. However, this may increase off-resonant excitation between states within the computational manifold. Finally, experimental measures such as magnetic field shielding and permanent magnets \cite{ruster2016b_field} can be used to reduce the field fluctuations.
    \item Two-state rotations may be imperfect due to miscalibration and noise in the control fields. Thus, the total error per intra- and inter-ion gate depends on the required number of two-state rotations per gate. This is heavily dependent on the connectivity of the manifold.
    \item Performing a measurement involves carrying out many transfer pulses on the ions being read out. Additional transfer pulses may also be required to protect computational states in the ground level of the ions that are not to be read out. These pulses take up computation time and may be imperfect, leading to readout  errors or errors in the computational state if hiding pulses are required. Thus, the computational manifold should be chosen to minimise the number of pulses required.
\end{enumerate}

While the relative impact of each error source will ultimately depend on the experimental implementation, the expected error can be used as a benchmark between computational manifolds, as discussed in Section \ref{sec:choosing_the_manifold}. 

\subsection{Choosing the ion}
As outlined in Section \ref{sec:XEB}, the benefits of encoding $n$ virtual qubits per ion are greatest for $n=2,3$. Thus, we focus on choosing suitable ion species for encoding computational manifolds of 4 or 8 states. 

Table \ref{table:ion-species} shows the available states for the most commonly used isotopes in trapped ion quantum computing, together with the maximum number of virtual qubits one can encode in the respective ion. Isotopes with no nuclear spin have no hyperfine structure, so they do not have enough states to encode three virtual qubits. Thus, we restrict our discussion to isotopes with non-zero nuclear spin. This comes with many additional advantages. In the absence of hyperfine structure, transitions in the metastable level are degenerate, so driving them directly without excessive off-resonant excitation is challenging. While it is possible to choose a computational manifold of two virtual qubits that forms a connected graph using only optical transitions between the two levels, it cannot be fully connected. Furthermore, hyperfine structure increases the number of available atomic states. On first sight, this has the potential to increase the amount of off-resonant scattering to spectator states, as well as the effect of AC Stark and Zeeman shifts. However, this is not the case for every state in a hyperfine level, since some states have few allowed transitions (e.g. states with $|m_F|=F$). In addition, many states with attractive properties also arise, such as clock transitions. The computational manifold can be carefully chosen to include states which maximise desired properties and avoid deleterious ones. This is discussed in detail in Section \ref{sec:choosing_the_manifold}. As well as this, natural extensions of the protocols in \cite{sotirova2024high, kang2023quantum} can be used to detect and reject shots where leakage outside the computational manifold has occurred, at the cost of a reduction in data rate. 

Another desirable trait is a metastable level with a long lifetime. This is beneficial even if the entire computational manifold is encoded in the ground level, since the manifold must be transferred to the metastable level for measurement operations. State decay from the metastable level increases readout errors and introduces logical errors in spectator ions that are not read out. These effects are particularly important when encoding $n>1$ qubits per ion, since readout operations are significantly longer. 

Furthermore, a long metastable level lifetime allows for the entire computational manifold to be encoded in this level for the entirety of a computation. This is in analogy to so-called \textit{omg} protocols for single qubits \cite{allcock2021omg, yang2022realizing, vizvary2023eliminating}. In doing so, computational states are decoupled from the cooling, state-preparation and readout lasers. Operations such as in-sequence cooling can thus be easily implemented, as well as executing state-preparation and readout on a subset of the ions in the chain \cite{allcock2021omg}. In addition, the only mapping pulses required for readout operations are those that sequentially transfer states from the metastable level to the ground level. However, if part of the manifold is in the ground level, hiding pulses are necessary. These pulses must additionally be coherent when hiding the state in an ion that is not to be read out. These benefits compound when encoding $n>1$ virtual qubits, since transferring an entire computational manifold from the ground to the metastable level and back requires $2\times2^n$ pulses.

Motivated by this, the remainder of the text focuses on encoding the computational manifold entirely within the metastable level. Thus, we are interested in species with long-lived metastable levels and hyperfine structure. Of the ions considered in Table \ref{table:ion-species}, the lanthanide ytterbium has the longest lifetime of the metastable \flevel\ level. However, this ion comes with technical challenges. The metastable state manifold one has to access is \flevel. This requires either a narrow-linewidth high-power laser to directly drive an octupole transition, or a pair of lasers drive the transition via the intermediate \dlevel\ level \cite{roberts2000observation}. Additionally, the main cycling transition used for cooling and readout is in the UV regime. 

Amongst the alkali ions, barium has the longest metastable level lifetime at $30.1\;\mathrm{s}$ \cite{Zhang2020Ba138metalifetime}. Additionally, all transition wavelengths are in the visible or near-infrared part of the spectrum, for which high-quality optics and fibre components are readily available. Transitions within the ground and metastable levels can be driven by microwaves or via a two-photon Raman process. Uniquely for barium, Raman transitions in either level can be driven efficiently at the same wavelength (typically $532\;\mathrm{nm}$ \cite{dietrich2009barium,christensen2020high}). While all logical operations will be carried out in the metastable level, this property can be used for in-sequence resolved-sideband cooling in the ground level of spectator ions.

There are several barium isotopes that have been used in trapped ion experiments \cite{christensen2020high,an2022high}. To work with hyperfine structure we require an odd isotope. Of the naturally abundant barium isotopes, \Ba\ and $^{135}\mathrm{Ba}^+$ have equivalent properties, however \Ba\ is 10 times more abundant \cite{lide2004crc}, therefore $^{135}\mathrm{Ba}^+$ is excluded from further consideration. Finally, the synthetic $^{133}\mathrm{Ba}^+$ has nuclear spin of $I=1/2$ which significantly simplifies the level structure as well as the execution of state preparation and readout operations. However, it is weakly radioactive, which limits the amount of $^{133}\mathrm{Ba}^+$ in the source, making loading difficult \cite{white2022isotope}.

\subsection{Choosing the computational manifold in \Ba}\label{sec:choosing_the_manifold}

\begin{figure*}[t]
    \centering
    \includegraphics[width=\linewidth,trim = 0cm 0cm 0cm 0cm]{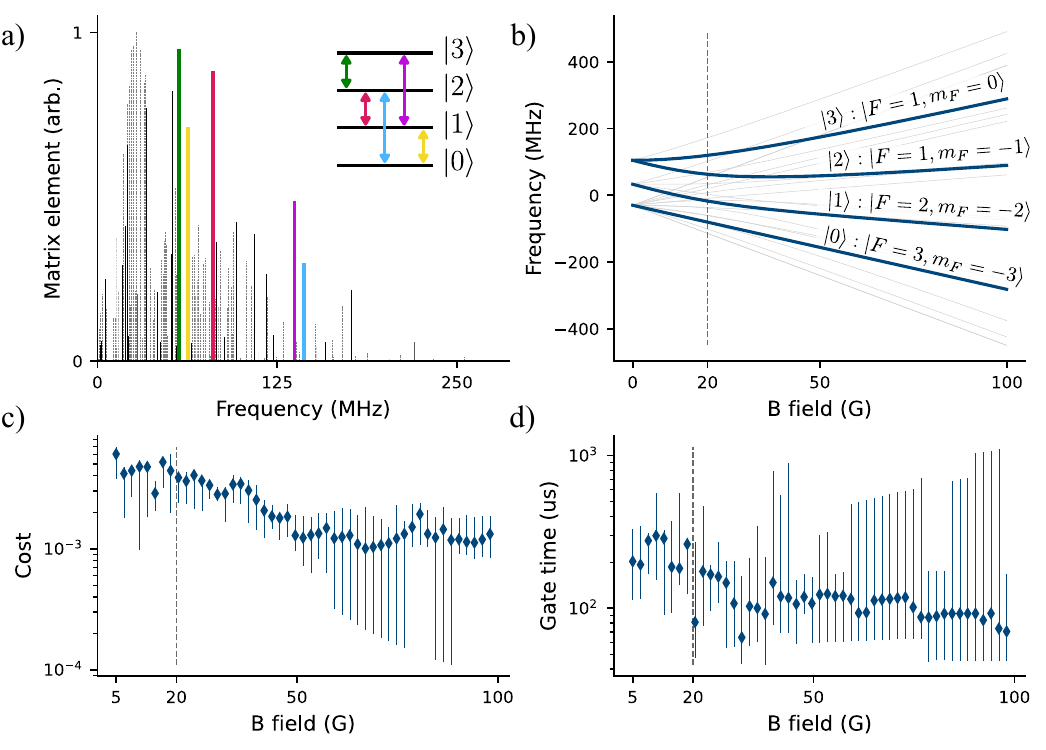}
    \caption{a) One possible computational manifold for $n=2$ qubits using states from the \dlevel\ level in \Ba, chosen via the numerical method outlined in the text. The height of the bars in the main panel shows the matrix element of the transition between two states. Thick colored lines, solid black lines and dashed grey lines denote transitions between two states in the manifold, between a state in the manifold and a spectator, and between two states not connected to the manifold, respectively. The inset shows the connectivity between computational states (see Subfigure b for the corresponding atomic states). The transitions in the manifold have large matrix elements and are separated from spectator transitions by at least 2 MHz. A quantisation magnetic field of 20 G was used to generate this manifold, which is marked by grey dashed lines in b), c) and d). b) Energy of the states in the \dlevel\ level as a function of quantisation magnetic field. In solid blue are the states in the computational manifold, labelled by the $F, m_F$ quantum numbers of the state they adiabatically connect to at zero field. c) The median of the ten lowest cost values as a function of the quantisation field. Error bars show the maximum and minimum error metrics. c) Median of the approximate mean intra-ion gate time for the computational manifolds used to generate b). (See Appendix \ref{sec:sup_memory} for details on how the mean gate time is estimated). Error bars show the maximum and minimum mean gate time. The quantisation field of $20$ G was chosen since it is low enough to be accessible in most ion trapping setups without significant modification, while yielding favourable cost function values and gate times. However, using large quantisation fields $>\!60\, \mathrm{G}$ may result in better performance, as suggested by the reduced error metrics and gate times.}
    \label{fig:Ba_level}
\end{figure*}

Because of the large number of states in the \dlevel\ level in \Ba (24 states) there are $\binom{24}{4} \approx 1\times10^4$ and $\binom{24}{8} \approx 7\times10^5$ ways of choosing the computational manifold for $n=2,3$ virtual qubits, respectively. While the number of connected manifolds is significantly smaller ($\sim1000$ for two virtual qubits), these numbers are too large for a manual evaluation of each computational manifold. As such, we use numerical methods to assess the expected performance of different manifold choices via a heuristic cost function. This allows us to reduce the number of candidate choices to the ten manifolds with the lowest cost. The particular manifold used in an experiment can then be chosen manually from this reduced set.

The cost function estimates various errors that differ between computational manifolds. Of the error sources outlined in \ref{sec:guidelines_comp_manifold}, it takes into account \textit{memory} errors $\varepsilon_M$ due to dephasing from noisy external magnetic fields and \textit{crosstalk} errors $\varepsilon_C$ due to off-resonant coupling in intra-ion gates. These error estimates are functions of experimental parameters, including the magnitude of the quantisation magnetic field, the mechanism used for driving two-state rotations and noise in the quantisation field. Parameters used in the calculations, which are typical for ion trapping experiments, as well as expressions for the individual error estimates are given in Appendix \ref{sec:sup_cost_function}. Errors due to imperfect rotations are not included, since they can arise from multiple different sources (such as pulse area and frequency miscalibrations, and laser noise). We instead pay particular attention to the manifold connectivity and the gate time in the manual post-selection. Additionally, since all the states in the metastable level can be connected to a state in the ground level, errors in the mapping pulses are the same for all manifold choices and are thus neglected. Errors in inter-ion gates are also not included in the cost, since they primarily depend on the motional dynamics of the ion chain. These are also independent of the computational manifold.

We separate crosstalk errors into \textit{internal} crosstalk $\varepsilon_{\textrm{Int}}$ and \textit{spectator} crosstalk $\varepsilon_{\textrm{Spect}}$, which are due to off-resonant excitation of spectator transitions in the manifold and transitions to spectator states, respectively. The latter can be detected and eliminated using natural extensions of the protocols in \cite{kang2023quantum, sotirova2024high}. In contrast, correcting errors due to internal crosstalk requires error correction schemes with a large overhead in the number of ions required per logical qubit. Implementing these may be intractable in small, near-term devices. As such, we add a weight $\kappa \in [0,1]$ to mitigate the cost of spectator crosstalk. The full heuristic cost is given by

\begin{equation}\label{eq:cost_function}
    C= \varepsilon_M + \varepsilon_{\mathrm{Int}} + \kappa \varepsilon_{\mathrm{Spect}}
\end{equation}
  
We then individually select between the ten candidate encodings. Carrying out this procedure for typical experimental conditions yields the computational manifold for two virtual qubits shown in Figure \ref{fig:Ba_level}. While this analysis was done for the metastable level in \Ba with gates driven by Raman transitions, this cost function is applicable to any ion and gate driving mechanism. 

The heuristic cost function can also be used to approximately optimise experimental parameters such as the quantisation field, polarisation of the driving fields and maximum tolerable magnetic field noise. This can be done by comparing the ten lowest cost values obtained at different values of the parameter. We use this to explore the impact of the quantisation field on the performance.

The quantisation field determines the frequency splitting between the atomic eigenstates (see Figure \ref{fig:Ba_level} b)). When a quantisation field $B_Q$ that is large compared to the hyperfine splitting $\hbar\Delta$ between $F$ levels is applied, the eigenstates become superpositions of the zero-field eigenstates \cite{low2023control}. Thus, the magnetic field sensitivity of the eigenstates and the transition matrix elements are modified. This is particularly relevant in the \dlevel\ level in \Ba since the frequency splitting between the $F=3$ and $F=4$ levels at zero quantisation field is only $\sim0.5$ MHz \cite{low2023control}. The states within these levels begin to mix at a quantisation field $ B_Q \approx \hbar\Delta / \mu_B \approx 0.35$ G. The $F=1,2$ levels begin to mix at around $35$ G.

In Figure \ref{fig:Ba_level} c) we plot the median of the ten smallest cost values as a function of the quantisation field magnitude. Increasing the quantisation field leads to reduced cost, even as the quantisation field is increased above 60 G. We attribute this behaviour to a reduction in the crosstalk, which arises from an increased spectral separation beteween transitons and modifications to the selection rules. Figure \ref{fig:Ba_level} d) shows the median of the approximate time required to execute an intra-ion gate for the encodings in c). A reduction in this value with magnetic field is also observed. These two metrics suggest that the use of large quantisation magnetic fields may be advantageous for trapped ion processors encoding more than one qubit per ion, despite the increase in experimental complexity. We expect this advantage to also apply to similar qudit-based processors.


\section{Conclusion}
We have argued that using multiple levels within ions as virtual qubits can be particularly helpful for near term trapped-ion quantum computers, highlighted specific applications with two virtual qubits in each ion, including a circuit decomposition for the Bernstein-Vazirani algorithm, many-body entangling operations, and generating intra-ion entanglement without using vibrational modes. Implementing virtual qubits is particularly profitable in trapped ion quantum computers since the Molmer-Sorensen entangling gate performs better in smaller ion chains. Importantly, a universal and expressive gate set can be formed with minimal modifications to traditional trapped-ion gates: intra-ion gates that perform Rabi oscillations between a pair of states, and inter-ion gates which entangle a pair of states in one ion with another, possibly different, pair in another ion. Since all such gates are multi-qubit in nature, careful considerations must be made when generalizing benchmarking and error correction from traditional setups. For the special case of defining two data-qubits inside a single ion, we argued that stabilizer measurements can be performed using fewer inter-ion gates, and, for the bit-flip repetition code, leads to a lower logical error rate compared to encoding one single logical qubit in the same number of ions.

We propose the long-lived $D_{5/2}$ levels of \Ba as a platform to encode such virtual qubits. \Ba has the advantages of good connectivity, requiring fewer pulses for measurements, low magnetic field fluctuations, mid-circuit measurements in different ions, and detection of leakages to ground states. We also have developed techniques to carefully pick suitable states to form the computational manifold out of a large pool of options.

\section{Acknowledgements}
The authors would like to thank C.W. von Keyserlingk, Raghavendra Srinivas, and Peter Drmota for helpful discussions. S.S. acknowledges resources from Princeton Research Computing. S.L.S. was supported by a Leverhulme
Trust International Professorship, Grant Number LIP-202-014. For the purpose of Open Access, the authors have applied a CC BY public copyright license to any Author Accepted Manuscript version arising from this submission.

\bibliography{main.bib}

\begin{thebibliography}{94}%
\makeatletter
\providecommand \@ifxundefined [1]{%
 \@ifx{#1\undefined}
}%
\providecommand \@ifnum [1]{%
 \ifnum #1\expandafter \@firstoftwo
 \else \expandafter \@secondoftwo
 \fi
}%
\providecommand \@ifx [1]{%
 \ifx #1\expandafter \@firstoftwo
 \else \expandafter \@secondoftwo
 \fi
}%
\providecommand \natexlab [1]{#1}%
\providecommand \enquote  [1]{``#1''}%
\providecommand \bibnamefont  [1]{#1}%
\providecommand \bibfnamefont [1]{#1}%
\providecommand \citenamefont [1]{#1}%
\providecommand \href@noop [0]{\@secondoftwo}%
\providecommand \href [0]{\begingroup \@sanitize@url \@href}%
\providecommand \@href[1]{\@@startlink{#1}\@@href}%
\providecommand \@@href[1]{\endgroup#1\@@endlink}%
\providecommand \@sanitize@url [0]{\catcode `\\12\catcode `\$12\catcode
  `\&12\catcode `\#12\catcode `\^12\catcode `\_12\catcode `\%12\relax}%
\providecommand \@@startlink[1]{}%
\providecommand \@@endlink[0]{}%
\providecommand \url  [0]{\begingroup\@sanitize@url \@url }%
\providecommand \@url [1]{\endgroup\@href {#1}{\urlprefix }}%
\providecommand \urlprefix  [0]{URL }%
\providecommand \Eprint [0]{\href }%
\providecommand \doibase [0]{https://doi.org/}%
\providecommand \selectlanguage [0]{\@gobble}%
\providecommand \bibinfo  [0]{\@secondoftwo}%
\providecommand \bibfield  [0]{\@secondoftwo}%
\providecommand \translation [1]{[#1]}%
\providecommand \BibitemOpen [0]{}%
\providecommand \bibitemStop [0]{}%
\providecommand \bibitemNoStop [0]{.\EOS\space}%
\providecommand \EOS [0]{\spacefactor3000\relax}%
\providecommand \BibitemShut  [1]{\csname bibitem#1\endcsname}%
\let\auto@bib@innerbib\@empty
\bibitem [{\citenamefont {Ballance}(2017)}]{ballance2017high}%
  \BibitemOpen
  \bibfield  {author} {\bibinfo {author} {\bibfnamefont {C.~J.}\ \bibnamefont
  {Ballance}},\ }\href
  {https://link.springer.com/book/10.1007/978-3-319-68216-7} {\emph {\bibinfo
  {title} {High-fidelity quantum logic in Ca+}}}\ (\bibinfo  {publisher}
  {Springer},\ \bibinfo {year} {2017})\BibitemShut {NoStop}%
\bibitem [{\citenamefont {Stephenson}\ \emph {et~al.}(2020)\citenamefont
  {Stephenson}, \citenamefont {Nadlinger}, \citenamefont {Nichol},
  \citenamefont {An}, \citenamefont {Drmota}, \citenamefont {Ballance},
  \citenamefont {Thirumalai}, \citenamefont {Goodwin}, \citenamefont {Lucas},\
  and\ \citenamefont {Ballance}}]{stephenson2020high}%
  \BibitemOpen
  \bibfield  {author} {\bibinfo {author} {\bibfnamefont {L.~J.}\ \bibnamefont
  {Stephenson}}, \bibinfo {author} {\bibfnamefont {D.~P.}\ \bibnamefont
  {Nadlinger}}, \bibinfo {author} {\bibfnamefont {B.~C.}\ \bibnamefont
  {Nichol}}, \bibinfo {author} {\bibfnamefont {S.}~\bibnamefont {An}}, \bibinfo
  {author} {\bibfnamefont {P.}~\bibnamefont {Drmota}}, \bibinfo {author}
  {\bibfnamefont {T.~G.}\ \bibnamefont {Ballance}}, \bibinfo {author}
  {\bibfnamefont {K.}~\bibnamefont {Thirumalai}}, \bibinfo {author}
  {\bibfnamefont {J.~F.}\ \bibnamefont {Goodwin}}, \bibinfo {author}
  {\bibfnamefont {D.~M.}\ \bibnamefont {Lucas}},\ and\ \bibinfo {author}
  {\bibfnamefont {C.~J.}\ \bibnamefont {Ballance}},\ }\bibfield  {title}
  {\bibinfo {title} {High-rate, high-fidelity entanglement of qubits across an
  elementary quantum network},\ }\href
  {https://doi.org/10.1103/PhysRevLett.124.110501} {\bibfield  {journal}
  {\bibinfo  {journal} {Phys. Rev. Lett.}\ }\textbf {\bibinfo {volume} {124}},\
  \bibinfo {pages} {110501} (\bibinfo {year} {2020})}\BibitemShut {NoStop}%
\bibitem [{\citenamefont {Bruzewicz}\ \emph {et~al.}(2019)\citenamefont
  {Bruzewicz}, \citenamefont {Chiaverini}, \citenamefont {McConnell},\ and\
  \citenamefont {Sage}}]{bruzewicz2019trapped}%
  \BibitemOpen
  \bibfield  {author} {\bibinfo {author} {\bibfnamefont {C.~D.}\ \bibnamefont
  {Bruzewicz}}, \bibinfo {author} {\bibfnamefont {J.}~\bibnamefont
  {Chiaverini}}, \bibinfo {author} {\bibfnamefont {R.}~\bibnamefont
  {McConnell}},\ and\ \bibinfo {author} {\bibfnamefont {J.~M.}\ \bibnamefont
  {Sage}},\ }\bibfield  {title} {\bibinfo {title} {Trapped-ion quantum
  computing: Progress and challenges},\ }\href
  {https://doi.org/https://doi.org/10.1063/1.5088164} {\bibfield  {journal}
  {\bibinfo  {journal} {Applied Physics Reviews}\ }\textbf {\bibinfo {volume}
  {6}},\ \bibinfo {pages} {021314} (\bibinfo {year} {2019})}\BibitemShut
  {NoStop}%
\bibitem [{\citenamefont {Wang}\ \emph {et~al.}(2021)\citenamefont {Wang},
  \citenamefont {Luan}, \citenamefont {Qiao}, \citenamefont {Um}, \citenamefont
  {Zhang}, \citenamefont {Wang}, \citenamefont {Yuan}, \citenamefont {Gu},
  \citenamefont {Zhang},\ and\ \citenamefont {Kim}}]{wang2021single}%
  \BibitemOpen
  \bibfield  {author} {\bibinfo {author} {\bibfnamefont {P.}~\bibnamefont
  {Wang}}, \bibinfo {author} {\bibfnamefont {C.-Y.}\ \bibnamefont {Luan}},
  \bibinfo {author} {\bibfnamefont {M.}~\bibnamefont {Qiao}}, \bibinfo {author}
  {\bibfnamefont {M.}~\bibnamefont {Um}}, \bibinfo {author} {\bibfnamefont
  {J.}~\bibnamefont {Zhang}}, \bibinfo {author} {\bibfnamefont
  {Y.}~\bibnamefont {Wang}}, \bibinfo {author} {\bibfnamefont {X.}~\bibnamefont
  {Yuan}}, \bibinfo {author} {\bibfnamefont {M.}~\bibnamefont {Gu}}, \bibinfo
  {author} {\bibfnamefont {J.}~\bibnamefont {Zhang}},\ and\ \bibinfo {author}
  {\bibfnamefont {K.}~\bibnamefont {Kim}},\ }\bibfield  {title} {\bibinfo
  {title} {Single ion qubit with estimated coherence time exceeding one hour},\
  }\href {https://doi.org/https://doi.org/10.1038/s41467-020-20330-w}
  {\bibfield  {journal} {\bibinfo  {journal} {Nature communications}\ }\textbf
  {\bibinfo {volume} {12}},\ \bibinfo {pages} {1} (\bibinfo {year}
  {2021})}\BibitemShut {NoStop}%
\bibitem [{\citenamefont {Cai}\ \emph {et~al.}(2023)\citenamefont {Cai},
  \citenamefont {Luan}, \citenamefont {Ou}, \citenamefont {Tu}, \citenamefont
  {Yin}, \citenamefont {Zhang},\ and\ \citenamefont {Kim}}]{cai2023entangling}%
  \BibitemOpen
  \bibfield  {author} {\bibinfo {author} {\bibfnamefont {Z.}~\bibnamefont
  {Cai}}, \bibinfo {author} {\bibfnamefont {C.-Y.}\ \bibnamefont {Luan}},
  \bibinfo {author} {\bibfnamefont {L.}~\bibnamefont {Ou}}, \bibinfo {author}
  {\bibfnamefont {H.}~\bibnamefont {Tu}}, \bibinfo {author} {\bibfnamefont
  {Z.}~\bibnamefont {Yin}}, \bibinfo {author} {\bibfnamefont {J.-N.}\
  \bibnamefont {Zhang}},\ and\ \bibinfo {author} {\bibfnamefont
  {K.}~\bibnamefont {Kim}},\ }\bibfield  {title} {\bibinfo {title} {Entangling
  gates for trapped-ion quantum computation and quantum simulation},\ }\href
  {https://doi.org/10.1007/s40042-023-00772-3} {\bibfield  {journal} {\bibinfo
  {journal} {Journal of the Korean Physical Society}\ }\textbf {\bibinfo
  {volume} {82}},\ \bibinfo {pages} {882} (\bibinfo {year} {2023})}\BibitemShut
  {NoStop}%
\bibitem [{\citenamefont {Pino}\ \emph {et~al.}(2021)\citenamefont {Pino},
  \citenamefont {Dreiling}, \citenamefont {Figgatt}, \citenamefont {Gaebler},
  \citenamefont {Moses}, \citenamefont {Allman}, \citenamefont {Baldwin},
  \citenamefont {Foss-Feig}, \citenamefont {Hayes}, \citenamefont {Mayer} \emph
  {et~al.}}]{pino2021demonstration}%
  \BibitemOpen
  \bibfield  {author} {\bibinfo {author} {\bibfnamefont {J.~M.}\ \bibnamefont
  {Pino}}, \bibinfo {author} {\bibfnamefont {J.~M.}\ \bibnamefont {Dreiling}},
  \bibinfo {author} {\bibfnamefont {C.}~\bibnamefont {Figgatt}}, \bibinfo
  {author} {\bibfnamefont {J.~P.}\ \bibnamefont {Gaebler}}, \bibinfo {author}
  {\bibfnamefont {S.~A.}\ \bibnamefont {Moses}}, \bibinfo {author}
  {\bibfnamefont {M.}~\bibnamefont {Allman}}, \bibinfo {author} {\bibfnamefont
  {C.}~\bibnamefont {Baldwin}}, \bibinfo {author} {\bibfnamefont
  {M.}~\bibnamefont {Foss-Feig}}, \bibinfo {author} {\bibfnamefont
  {D.}~\bibnamefont {Hayes}}, \bibinfo {author} {\bibfnamefont
  {K.}~\bibnamefont {Mayer}}, \emph {et~al.},\ }\bibfield  {title} {\bibinfo
  {title} {Demonstration of the trapped-ion quantum ccd computer
  architecture},\ }\href
  {https://doi.org/https://doi.org/10.1038/s41586-021-03318-4} {\bibfield
  {journal} {\bibinfo  {journal} {Nature}\ }\textbf {\bibinfo {volume} {592}},\
  \bibinfo {pages} {209} (\bibinfo {year} {2021})}\BibitemShut {NoStop}%
\bibitem [{\citenamefont {Moses}\ \emph {et~al.}(2023)\citenamefont {Moses},
  \citenamefont {Baldwin}, \citenamefont {Allman}, \citenamefont {Ancona},
  \citenamefont {Ascarrunz}, \citenamefont {Barnes}, \citenamefont
  {Bartolotta}, \citenamefont {Bjork}, \citenamefont {Blanchard}, \citenamefont
  {Bohn} \emph {et~al.}}]{moses2023race}%
  \BibitemOpen
  \bibfield  {author} {\bibinfo {author} {\bibfnamefont {S.}~\bibnamefont
  {Moses}}, \bibinfo {author} {\bibfnamefont {C.}~\bibnamefont {Baldwin}},
  \bibinfo {author} {\bibfnamefont {M.}~\bibnamefont {Allman}}, \bibinfo
  {author} {\bibfnamefont {R.}~\bibnamefont {Ancona}}, \bibinfo {author}
  {\bibfnamefont {L.}~\bibnamefont {Ascarrunz}}, \bibinfo {author}
  {\bibfnamefont {C.}~\bibnamefont {Barnes}}, \bibinfo {author} {\bibfnamefont
  {J.}~\bibnamefont {Bartolotta}}, \bibinfo {author} {\bibfnamefont
  {B.}~\bibnamefont {Bjork}}, \bibinfo {author} {\bibfnamefont
  {P.}~\bibnamefont {Blanchard}}, \bibinfo {author} {\bibfnamefont
  {M.}~\bibnamefont {Bohn}}, \emph {et~al.},\ }\bibfield  {title} {\bibinfo
  {title} {A race track trapped-ion quantum processor},\ }\href
  {https://arxiv.org/abs/2305.03828} {\bibfield  {journal} {\bibinfo  {journal}
  {arXiv preprint arXiv:2305.03828}\ } (\bibinfo {year} {2023})}\BibitemShut
  {NoStop}%
\bibitem [{\citenamefont {Malinowski}\ \emph {et~al.}(2023)\citenamefont
  {Malinowski}, \citenamefont {Allcock},\ and\ \citenamefont
  {Ballance}}]{malinowski2023wire}%
  \BibitemOpen
  \bibfield  {author} {\bibinfo {author} {\bibfnamefont {M.}~\bibnamefont
  {Malinowski}}, \bibinfo {author} {\bibfnamefont {D.}~\bibnamefont
  {Allcock}},\ and\ \bibinfo {author} {\bibfnamefont {C.}~\bibnamefont
  {Ballance}},\ }\bibfield  {title} {\bibinfo {title} {How to wire a 1000-qubit
  trapped ion quantum computer},\ }\href {https://arxiv.org/abs/2305.12773}
  {\bibfield  {journal} {\bibinfo  {journal} {arXiv preprint arXiv:2305.12773}\
  } (\bibinfo {year} {2023})}\BibitemShut {NoStop}%
\bibitem [{\citenamefont {Elder}\ \emph {et~al.}(2020)\citenamefont {Elder},
  \citenamefont {Wang}, \citenamefont {Reinhold}, \citenamefont {Hann},
  \citenamefont {Chou}, \citenamefont {Lester}, \citenamefont {Rosenblum},
  \citenamefont {Frunzio}, \citenamefont {Jiang},\ and\ \citenamefont
  {Schoelkopf}}]{elder_mult_sc_2020}%
  \BibitemOpen
  \bibfield  {author} {\bibinfo {author} {\bibfnamefont {S.~S.}\ \bibnamefont
  {Elder}}, \bibinfo {author} {\bibfnamefont {C.~S.}\ \bibnamefont {Wang}},
  \bibinfo {author} {\bibfnamefont {P.}~\bibnamefont {Reinhold}}, \bibinfo
  {author} {\bibfnamefont {C.~T.}\ \bibnamefont {Hann}}, \bibinfo {author}
  {\bibfnamefont {K.~S.}\ \bibnamefont {Chou}}, \bibinfo {author}
  {\bibfnamefont {B.~J.}\ \bibnamefont {Lester}}, \bibinfo {author}
  {\bibfnamefont {S.}~\bibnamefont {Rosenblum}}, \bibinfo {author}
  {\bibfnamefont {L.}~\bibnamefont {Frunzio}}, \bibinfo {author} {\bibfnamefont
  {L.}~\bibnamefont {Jiang}},\ and\ \bibinfo {author} {\bibfnamefont {R.~J.}\
  \bibnamefont {Schoelkopf}},\ }\bibfield  {title} {\bibinfo {title}
  {High-fidelity measurement of qubits encoded in multilevel superconducting
  circuits},\ }\href {https://doi.org/10.1103/PhysRevX.10.011001} {\bibfield
  {journal} {\bibinfo  {journal} {Phys. Rev. X}\ }\textbf {\bibinfo {volume}
  {10}},\ \bibinfo {pages} {011001} (\bibinfo {year} {2020})}\BibitemShut
  {NoStop}%
\bibitem [{\citenamefont {Low}\ \emph {et~al.}(2020{\natexlab{a}})\citenamefont
  {Low}, \citenamefont {White}, \citenamefont {Cox}, \citenamefont {Day},\ and\
  \citenamefont {Senko}}]{low_practical_qudit_2020}%
  \BibitemOpen
  \bibfield  {author} {\bibinfo {author} {\bibfnamefont {P.~J.}\ \bibnamefont
  {Low}}, \bibinfo {author} {\bibfnamefont {B.~M.}\ \bibnamefont {White}},
  \bibinfo {author} {\bibfnamefont {A.~A.}\ \bibnamefont {Cox}}, \bibinfo
  {author} {\bibfnamefont {M.~L.}\ \bibnamefont {Day}},\ and\ \bibinfo {author}
  {\bibfnamefont {C.}~\bibnamefont {Senko}},\ }\bibfield  {title} {\bibinfo
  {title} {Practical trapped-ion protocols for universal qudit-based quantum
  computing},\ }\href {https://doi.org/10.1103/PhysRevResearch.2.033128}
  {\bibfield  {journal} {\bibinfo  {journal} {Phys. Rev. Research}\ }\textbf
  {\bibinfo {volume} {2}},\ \bibinfo {pages} {033128} (\bibinfo {year}
  {2020}{\natexlab{a}})}\BibitemShut {NoStop}%
\bibitem [{\citenamefont {Rambach}\ \emph {et~al.}(2021)\citenamefont
  {Rambach}, \citenamefont {Qaryan}, \citenamefont {Kewming}, \citenamefont
  {Ferrie}, \citenamefont {White},\ and\ \citenamefont
  {Romero}}]{rambach_qudit_tomo_2021}%
  \BibitemOpen
  \bibfield  {author} {\bibinfo {author} {\bibfnamefont {M.}~\bibnamefont
  {Rambach}}, \bibinfo {author} {\bibfnamefont {M.}~\bibnamefont {Qaryan}},
  \bibinfo {author} {\bibfnamefont {M.}~\bibnamefont {Kewming}}, \bibinfo
  {author} {\bibfnamefont {C.}~\bibnamefont {Ferrie}}, \bibinfo {author}
  {\bibfnamefont {A.~G.}\ \bibnamefont {White}},\ and\ \bibinfo {author}
  {\bibfnamefont {J.}~\bibnamefont {Romero}},\ }\bibfield  {title} {\bibinfo
  {title} {Robust and efficient high-dimensional quantum state tomography},\
  }\href {https://doi.org/10.1103/PhysRevLett.126.100402} {\bibfield  {journal}
  {\bibinfo  {journal} {Phys. Rev. Lett.}\ }\textbf {\bibinfo {volume} {126}},\
  \bibinfo {pages} {100402} (\bibinfo {year} {2021})}\BibitemShut {NoStop}%
\bibitem [{\citenamefont {Chi}\ \emph {et~al.}(2022)\citenamefont {Chi},
  \citenamefont {Huang}, \citenamefont {Zhang}, \citenamefont {Mao},
  \citenamefont {Zhou}, \citenamefont {Chen}, \citenamefont {Zhai},
  \citenamefont {Bao}, \citenamefont {Dai}, \citenamefont {Yuan} \emph
  {et~al.}}]{chi2022programmable}%
  \BibitemOpen
  \bibfield  {author} {\bibinfo {author} {\bibfnamefont {Y.}~\bibnamefont
  {Chi}}, \bibinfo {author} {\bibfnamefont {J.}~\bibnamefont {Huang}}, \bibinfo
  {author} {\bibfnamefont {Z.}~\bibnamefont {Zhang}}, \bibinfo {author}
  {\bibfnamefont {J.}~\bibnamefont {Mao}}, \bibinfo {author} {\bibfnamefont
  {Z.}~\bibnamefont {Zhou}}, \bibinfo {author} {\bibfnamefont {X.}~\bibnamefont
  {Chen}}, \bibinfo {author} {\bibfnamefont {C.}~\bibnamefont {Zhai}}, \bibinfo
  {author} {\bibfnamefont {J.}~\bibnamefont {Bao}}, \bibinfo {author}
  {\bibfnamefont {T.}~\bibnamefont {Dai}}, \bibinfo {author} {\bibfnamefont
  {H.}~\bibnamefont {Yuan}}, \emph {et~al.},\ }\bibfield  {title} {\bibinfo
  {title} {A programmable qudit-based quantum processor},\ }\href
  {https://doi.org/10.1038/s41467-022-28767-x} {\bibfield  {journal} {\bibinfo
  {journal} {Nature communications}\ }\textbf {\bibinfo {volume} {13}},\
  \bibinfo {pages} {1} (\bibinfo {year} {2022})}\BibitemShut {NoStop}%
\bibitem [{\citenamefont {Ringbauer}\ \emph {et~al.}(2022)\citenamefont
  {Ringbauer}, \citenamefont {Meth}, \citenamefont {Postler}, \citenamefont
  {Stricker}, \citenamefont {Blatt}, \citenamefont {Schindler},\ and\
  \citenamefont {Monz}}]{ringbauer2022universal}%
  \BibitemOpen
  \bibfield  {author} {\bibinfo {author} {\bibfnamefont {M.}~\bibnamefont
  {Ringbauer}}, \bibinfo {author} {\bibfnamefont {M.}~\bibnamefont {Meth}},
  \bibinfo {author} {\bibfnamefont {L.}~\bibnamefont {Postler}}, \bibinfo
  {author} {\bibfnamefont {R.}~\bibnamefont {Stricker}}, \bibinfo {author}
  {\bibfnamefont {R.}~\bibnamefont {Blatt}}, \bibinfo {author} {\bibfnamefont
  {P.}~\bibnamefont {Schindler}},\ and\ \bibinfo {author} {\bibfnamefont
  {T.}~\bibnamefont {Monz}},\ }\bibfield  {title} {\bibinfo {title} {A
  universal qudit quantum processor with trapped ions},\ }\href
  {https://doi.org/https://doi.org/10.1038/s41567-022-01658-0} {\bibfield
  {journal} {\bibinfo  {journal} {Nature Physics}\ ,\ \bibinfo {pages} {1}}
  (\bibinfo {year} {2022})}\BibitemShut {NoStop}%
\bibitem [{\citenamefont {Gao}\ \emph {et~al.}(2022)\citenamefont {Gao},
  \citenamefont {Appel}, \citenamefont {Friis}, \citenamefont {Ringbauer},\
  and\ \citenamefont {Huber}}]{gao2022role}%
  \BibitemOpen
  \bibfield  {author} {\bibinfo {author} {\bibfnamefont {X.}~\bibnamefont
  {Gao}}, \bibinfo {author} {\bibfnamefont {P.}~\bibnamefont {Appel}}, \bibinfo
  {author} {\bibfnamefont {N.}~\bibnamefont {Friis}}, \bibinfo {author}
  {\bibfnamefont {M.}~\bibnamefont {Ringbauer}},\ and\ \bibinfo {author}
  {\bibfnamefont {M.}~\bibnamefont {Huber}},\ }\bibfield  {title} {\bibinfo
  {title} {On the role of entanglement in qudit-based circuit compression},\
  }\href {https://arxiv.org/abs/2209.14584} {\bibfield  {journal} {\bibinfo
  {journal} {arXiv preprint arXiv:2209.14584}\ } (\bibinfo {year}
  {2022})}\BibitemShut {NoStop}%
\bibitem [{\citenamefont {Nikolaeva}\ \emph {et~al.}(2021)\citenamefont
  {Nikolaeva}, \citenamefont {Kiktenko},\ and\ \citenamefont
  {Fedorov}}]{nikolaeva2021efficient}%
  \BibitemOpen
  \bibfield  {author} {\bibinfo {author} {\bibfnamefont {A.~S.}\ \bibnamefont
  {Nikolaeva}}, \bibinfo {author} {\bibfnamefont {E.~O.}\ \bibnamefont
  {Kiktenko}},\ and\ \bibinfo {author} {\bibfnamefont {A.~K.}\ \bibnamefont
  {Fedorov}},\ }\bibfield  {title} {\bibinfo {title} {Efficient realization of
  quantum algorithms with qudits},\ }\href {https://arxiv.org/abs/2111.04384}
  {\bibfield  {journal} {\bibinfo  {journal} {arXiv preprint arXiv:2111.04384}\
  } (\bibinfo {year} {2021})}\BibitemShut {NoStop}%
\bibitem [{\citenamefont {Low}\ \emph {et~al.}(2023)\citenamefont {Low},
  \citenamefont {White},\ and\ \citenamefont {Senko}}]{low2023control}%
  \BibitemOpen
  \bibfield  {author} {\bibinfo {author} {\bibfnamefont {P.~J.}\ \bibnamefont
  {Low}}, \bibinfo {author} {\bibfnamefont {B.}~\bibnamefont {White}},\ and\
  \bibinfo {author} {\bibfnamefont {C.}~\bibnamefont {Senko}},\ }\bibfield
  {title} {\bibinfo {title} {Control and readout of a 13-level trapped ion
  qudit},\ }\href {https://arxiv.org/abs/2306.03340} {\bibfield  {journal}
  {\bibinfo  {journal} {arXiv preprint arXiv:2306.03340}\ } (\bibinfo {year}
  {2023})}\BibitemShut {NoStop}%
\bibitem [{\citenamefont {de~Fuentes}\ \emph {et~al.}(2023)\citenamefont
  {de~Fuentes}, \citenamefont {Botzem}, \citenamefont {Vaartjes}, \citenamefont
  {Asaad}, \citenamefont {Mourik}, \citenamefont {Hudson}, \citenamefont
  {Itoh}, \citenamefont {Johnson}, \citenamefont {Jakob}, \citenamefont
  {McCallum} \emph {et~al.}}]{de2023navigating}%
  \BibitemOpen
  \bibfield  {author} {\bibinfo {author} {\bibfnamefont {I.~F.}\ \bibnamefont
  {de~Fuentes}}, \bibinfo {author} {\bibfnamefont {T.}~\bibnamefont {Botzem}},
  \bibinfo {author} {\bibfnamefont {A.}~\bibnamefont {Vaartjes}}, \bibinfo
  {author} {\bibfnamefont {S.}~\bibnamefont {Asaad}}, \bibinfo {author}
  {\bibfnamefont {V.}~\bibnamefont {Mourik}}, \bibinfo {author} {\bibfnamefont
  {F.~E.}\ \bibnamefont {Hudson}}, \bibinfo {author} {\bibfnamefont {K.~M.}\
  \bibnamefont {Itoh}}, \bibinfo {author} {\bibfnamefont {B.~C.}\ \bibnamefont
  {Johnson}}, \bibinfo {author} {\bibfnamefont {A.~M.}\ \bibnamefont {Jakob}},
  \bibinfo {author} {\bibfnamefont {J.~C.}\ \bibnamefont {McCallum}}, \emph
  {et~al.},\ }\bibfield  {title} {\bibinfo {title} {Navigating the
  16-dimensional hilbert space of a high-spin donor qudit with electric and
  magnetic fields},\ }\href {https://arxiv.org/abs/2306.07453} {\bibfield
  {journal} {\bibinfo  {journal} {arXiv preprint arXiv:2306.07453}\ } (\bibinfo
  {year} {2023})}\BibitemShut {NoStop}%
\bibitem [{\citenamefont {Hrmo}\ \emph {et~al.}(2023)\citenamefont {Hrmo},
  \citenamefont {Wilhelm}, \citenamefont {Gerster}, \citenamefont {van Mourik},
  \citenamefont {Huber}, \citenamefont {Blatt}, \citenamefont {Schindler},
  \citenamefont {Monz},\ and\ \citenamefont {Ringbauer}}]{hrmo2023native}%
  \BibitemOpen
  \bibfield  {author} {\bibinfo {author} {\bibfnamefont {P.}~\bibnamefont
  {Hrmo}}, \bibinfo {author} {\bibfnamefont {B.}~\bibnamefont {Wilhelm}},
  \bibinfo {author} {\bibfnamefont {L.}~\bibnamefont {Gerster}}, \bibinfo
  {author} {\bibfnamefont {M.~W.}\ \bibnamefont {van Mourik}}, \bibinfo
  {author} {\bibfnamefont {M.}~\bibnamefont {Huber}}, \bibinfo {author}
  {\bibfnamefont {R.}~\bibnamefont {Blatt}}, \bibinfo {author} {\bibfnamefont
  {P.}~\bibnamefont {Schindler}}, \bibinfo {author} {\bibfnamefont
  {T.}~\bibnamefont {Monz}},\ and\ \bibinfo {author} {\bibfnamefont
  {M.}~\bibnamefont {Ringbauer}},\ }\bibfield  {title} {\bibinfo {title}
  {Native qudit entanglement in a trapped ion quantum processor},\ }\href
  {https://doi.org/10.1038/s41467-023-37375-2} {\bibfield  {journal} {\bibinfo
  {journal} {Nature Communications}\ }\textbf {\bibinfo {volume} {14}},\
  \bibinfo {pages} {2242} (\bibinfo {year} {2023})}\BibitemShut {NoStop}%
\bibitem [{\citenamefont {Kiktenko}\ \emph
  {et~al.}(2015{\natexlab{a}})\citenamefont {Kiktenko}, \citenamefont
  {Fedorov}, \citenamefont {Man'ko},\ and\ \citenamefont
  {Man'ko}}]{multilevel_sc_2015}%
  \BibitemOpen
  \bibfield  {author} {\bibinfo {author} {\bibfnamefont {E.~O.}\ \bibnamefont
  {Kiktenko}}, \bibinfo {author} {\bibfnamefont {A.~K.}\ \bibnamefont
  {Fedorov}}, \bibinfo {author} {\bibfnamefont {O.~V.}\ \bibnamefont
  {Man'ko}},\ and\ \bibinfo {author} {\bibfnamefont {V.~I.}\ \bibnamefont
  {Man'ko}},\ }\bibfield  {title} {\bibinfo {title} {Multilevel superconducting
  circuits as two-qubit systems: Operations, state preparation, and entropic
  inequalities},\ }\href {https://doi.org/10.1103/PhysRevA.91.042312}
  {\bibfield  {journal} {\bibinfo  {journal} {Phys. Rev. A}\ }\textbf {\bibinfo
  {volume} {91}},\ \bibinfo {pages} {042312} (\bibinfo {year}
  {2015}{\natexlab{a}})}\BibitemShut {NoStop}%
\bibitem [{\citenamefont {Svetitsky}\ \emph {et~al.}(2014)\citenamefont
  {Svetitsky}, \citenamefont {Suchowski}, \citenamefont {Resh}, \citenamefont
  {Shalibo}, \citenamefont {Martinis},\ and\ \citenamefont
  {Katz}}]{svetitsky2014hidden}%
  \BibitemOpen
  \bibfield  {author} {\bibinfo {author} {\bibfnamefont {E.}~\bibnamefont
  {Svetitsky}}, \bibinfo {author} {\bibfnamefont {H.}~\bibnamefont
  {Suchowski}}, \bibinfo {author} {\bibfnamefont {R.}~\bibnamefont {Resh}},
  \bibinfo {author} {\bibfnamefont {Y.}~\bibnamefont {Shalibo}}, \bibinfo
  {author} {\bibfnamefont {J.~M.}\ \bibnamefont {Martinis}},\ and\ \bibinfo
  {author} {\bibfnamefont {N.}~\bibnamefont {Katz}},\ }\bibfield  {title}
  {\bibinfo {title} {Hidden two-qubit dynamics of a four-level josephson
  circuit},\ }\href {https://doi.org/https://doi.org/10.1038/ncomms6617}
  {\bibfield  {journal} {\bibinfo  {journal} {Nature communications}\ }\textbf
  {\bibinfo {volume} {5}},\ \bibinfo {pages} {1} (\bibinfo {year}
  {2014})}\BibitemShut {NoStop}%
\bibitem [{\citenamefont {Dong}\ \emph {et~al.}(2022)\citenamefont {Dong},
  \citenamefont {Liu}, \citenamefont {Wang}, \citenamefont {Li}, \citenamefont
  {Yu}, \citenamefont {Zheng}, \citenamefont {Li}, \citenamefont {Lan},
  \citenamefont {Tan},\ and\ \citenamefont {Yu}}]{dong2022simulation}%
  \BibitemOpen
  \bibfield  {author} {\bibinfo {author} {\bibfnamefont {Y.}~\bibnamefont
  {Dong}}, \bibinfo {author} {\bibfnamefont {Q.}~\bibnamefont {Liu}}, \bibinfo
  {author} {\bibfnamefont {J.}~\bibnamefont {Wang}}, \bibinfo {author}
  {\bibfnamefont {Q.}~\bibnamefont {Li}}, \bibinfo {author} {\bibfnamefont
  {X.}~\bibnamefont {Yu}}, \bibinfo {author} {\bibfnamefont {W.}~\bibnamefont
  {Zheng}}, \bibinfo {author} {\bibfnamefont {Y.}~\bibnamefont {Li}}, \bibinfo
  {author} {\bibfnamefont {D.}~\bibnamefont {Lan}}, \bibinfo {author}
  {\bibfnamefont {X.}~\bibnamefont {Tan}},\ and\ \bibinfo {author}
  {\bibfnamefont {Y.}~\bibnamefont {Yu}},\ }\bibfield  {title} {\bibinfo
  {title} {Simulation of two-qubit gates with a superconducting qudit},\ }\href
  {https://doi.org/https://doi.org/10.1002/pssb.202100500} {\bibfield
  {journal} {\bibinfo  {journal} {physica status solidi (b)}\ }\textbf
  {\bibinfo {volume} {259}},\ \bibinfo {pages} {2100500} (\bibinfo {year}
  {2022})}\BibitemShut {NoStop}%
\bibitem [{\citenamefont {Rambow}\ and\ \citenamefont
  {Tian}(2021)}]{rambow2021reduction}%
  \BibitemOpen
  \bibfield  {author} {\bibinfo {author} {\bibfnamefont {P.}~\bibnamefont
  {Rambow}}\ and\ \bibinfo {author} {\bibfnamefont {M.}~\bibnamefont {Tian}},\
  }\bibfield  {title} {\bibinfo {title} {Reduction of circuit depth by mapping
  qubit-based quantum gates to a qudit basis},\ }\href
  {https://arxiv.org/abs/2109.09902} {\bibfield  {journal} {\bibinfo  {journal}
  {arXiv preprint arXiv:2109.09902}\ } (\bibinfo {year} {2021})}\BibitemShut
  {NoStop}%
\bibitem [{\citenamefont {Jankovi{\'c}}\ \emph {et~al.}(2023)\citenamefont
  {Jankovi{\'c}}, \citenamefont {Hartmann}, \citenamefont {Ruben},\ and\
  \citenamefont {Hervieux}}]{jankovic2023noisy}%
  \BibitemOpen
  \bibfield  {author} {\bibinfo {author} {\bibfnamefont {D.}~\bibnamefont
  {Jankovi{\'c}}}, \bibinfo {author} {\bibfnamefont {J.-G.}\ \bibnamefont
  {Hartmann}}, \bibinfo {author} {\bibfnamefont {M.}~\bibnamefont {Ruben}},\
  and\ \bibinfo {author} {\bibfnamefont {P.-A.}\ \bibnamefont {Hervieux}},\
  }\bibfield  {title} {\bibinfo {title} {Noisy qudit vs multiple qubits:
  Conditions on gate efficiency},\ }\href {https://arxiv.org/abs/2302.04543}
  {\bibfield  {journal} {\bibinfo  {journal} {arXiv preprint arXiv:2302.04543}\
  } (\bibinfo {year} {2023})}\BibitemShut {NoStop}%
\bibitem [{\citenamefont {Cao}\ \emph {et~al.}(2023)\citenamefont {Cao},
  \citenamefont {Bakr}, \citenamefont {Campanaro}, \citenamefont {Fasciati},
  \citenamefont {Wills}, \citenamefont {Lall}, \citenamefont {Shteynas},
  \citenamefont {Chidambaram}, \citenamefont {Rungger},\ and\ \citenamefont
  {Leek}}]{cao2023emulating}%
  \BibitemOpen
  \bibfield  {author} {\bibinfo {author} {\bibfnamefont {S.}~\bibnamefont
  {Cao}}, \bibinfo {author} {\bibfnamefont {M.}~\bibnamefont {Bakr}}, \bibinfo
  {author} {\bibfnamefont {G.}~\bibnamefont {Campanaro}}, \bibinfo {author}
  {\bibfnamefont {S.~D.}\ \bibnamefont {Fasciati}}, \bibinfo {author}
  {\bibfnamefont {J.}~\bibnamefont {Wills}}, \bibinfo {author} {\bibfnamefont
  {D.}~\bibnamefont {Lall}}, \bibinfo {author} {\bibfnamefont {B.}~\bibnamefont
  {Shteynas}}, \bibinfo {author} {\bibfnamefont {V.}~\bibnamefont
  {Chidambaram}}, \bibinfo {author} {\bibfnamefont {I.}~\bibnamefont
  {Rungger}},\ and\ \bibinfo {author} {\bibfnamefont {P.}~\bibnamefont
  {Leek}},\ }\bibfield  {title} {\bibinfo {title} {Emulating two qubits with a
  four-level transmon qudit for variational quantum algorithms},\ }\href
  {https://arxiv.org/abs/2303.04796} {\bibfield  {journal} {\bibinfo  {journal}
  {arXiv preprint arXiv:2303.04796}\ } (\bibinfo {year} {2023})}\BibitemShut
  {NoStop}%
\bibitem [{\citenamefont {Litteken}\ \emph {et~al.}(2023)\citenamefont
  {Litteken}, \citenamefont {Seifert}, \citenamefont {Chadwick}, \citenamefont
  {Nottingham}, \citenamefont {Roy}, \citenamefont {Li}, \citenamefont
  {Schuster}, \citenamefont {Chong},\ and\ \citenamefont
  {Baker}}]{litteken2023dancing}%
  \BibitemOpen
  \bibfield  {author} {\bibinfo {author} {\bibfnamefont {A.}~\bibnamefont
  {Litteken}}, \bibinfo {author} {\bibfnamefont {L.~M.}\ \bibnamefont
  {Seifert}}, \bibinfo {author} {\bibfnamefont {J.~D.}\ \bibnamefont
  {Chadwick}}, \bibinfo {author} {\bibfnamefont {N.}~\bibnamefont
  {Nottingham}}, \bibinfo {author} {\bibfnamefont {T.}~\bibnamefont {Roy}},
  \bibinfo {author} {\bibfnamefont {Z.}~\bibnamefont {Li}}, \bibinfo {author}
  {\bibfnamefont {D.}~\bibnamefont {Schuster}}, \bibinfo {author}
  {\bibfnamefont {F.~T.}\ \bibnamefont {Chong}},\ and\ \bibinfo {author}
  {\bibfnamefont {J.~M.}\ \bibnamefont {Baker}},\ }\bibfield  {title} {\bibinfo
  {title} {Dancing the quantum waltz: Compiling three-qubit gates on four level
  architectures},\ }\href {https://arxiv.org/abs/2302.04543} {\bibfield
  {journal} {\bibinfo  {journal} {arXiv preprint arXiv:2302.04543}\ } (\bibinfo
  {year} {2023})}\BibitemShut {NoStop}%
\bibitem [{\citenamefont {Kessel}\ and\ \citenamefont
  {Ermakov}(1999)}]{kessel1999multiqubit}%
  \BibitemOpen
  \bibfield  {author} {\bibinfo {author} {\bibfnamefont {A.~R.}\ \bibnamefont
  {Kessel}}\ and\ \bibinfo {author} {\bibfnamefont {V.~L.}\ \bibnamefont
  {Ermakov}},\ }\bibfield  {title} {\bibinfo {title} {Multiqubit spin},\ }\href
  {https://doi.org/https://doi.org/10.1134/1.568130} {\bibfield  {journal}
  {\bibinfo  {journal} {Journal of Experimental and Theoretical Physics
  Letters}\ }\textbf {\bibinfo {volume} {70}},\ \bibinfo {pages} {61} (\bibinfo
  {year} {1999})}\BibitemShut {NoStop}%
\bibitem [{\citenamefont {G{\"u}n}\ \emph {et~al.}(2013)\citenamefont
  {G{\"u}n}, \citenamefont {{\c{C}}akmak},\ and\ \citenamefont
  {Gen{\c{c}}ten}}]{gun2013construction}%
  \BibitemOpen
  \bibfield  {author} {\bibinfo {author} {\bibfnamefont {A.}~\bibnamefont
  {G{\"u}n}}, \bibinfo {author} {\bibfnamefont {S.}~\bibnamefont
  {{\c{C}}akmak}},\ and\ \bibinfo {author} {\bibfnamefont {A.}~\bibnamefont
  {Gen{\c{c}}ten}},\ }\bibfield  {title} {\bibinfo {title} {Construction of
  four-qubit quantum entanglement for si (s= 3/2, i= 3/2) spin system},\ }\href
  {https://doi.org/https://doi.org/10.1007/s11128-012-0367-x} {\bibfield
  {journal} {\bibinfo  {journal} {Quantum information processing}\ }\textbf
  {\bibinfo {volume} {12}},\ \bibinfo {pages} {205} (\bibinfo {year}
  {2013})}\BibitemShut {NoStop}%
\bibitem [{\citenamefont {Campbell}\ and\ \citenamefont
  {Hudson}(2022)}]{campbell2022polyqubit}%
  \BibitemOpen
  \bibfield  {author} {\bibinfo {author} {\bibfnamefont {W.~C.}\ \bibnamefont
  {Campbell}}\ and\ \bibinfo {author} {\bibfnamefont {E.~R.}\ \bibnamefont
  {Hudson}},\ }\bibfield  {title} {\bibinfo {title} {Polyqubit quantum
  processing},\ }\href {https://arxiv.org/abs/2210.15484} {\bibfield  {journal}
  {\bibinfo  {journal} {arXiv preprint arXiv:2210.15484}\ } (\bibinfo {year}
  {2022})}\BibitemShut {NoStop}%
\bibitem [{\citenamefont {Nikolaeva}\ \emph {et~al.}(2023)\citenamefont
  {Nikolaeva}, \citenamefont {Kiktenko},\ and\ \citenamefont
  {Fedorov}}]{nikolaeva2023compiling}%
  \BibitemOpen
  \bibfield  {author} {\bibinfo {author} {\bibfnamefont {A.~S.}\ \bibnamefont
  {Nikolaeva}}, \bibinfo {author} {\bibfnamefont {E.~O.}\ \bibnamefont
  {Kiktenko}},\ and\ \bibinfo {author} {\bibfnamefont {A.~K.}\ \bibnamefont
  {Fedorov}},\ }\bibfield  {title} {\bibinfo {title} {Compiling quantum
  circuits with qubits embedded in trapped-ion qudits},\ }\href
  {https://arxiv.org/abs/2302.02966} {\bibfield  {journal} {\bibinfo  {journal}
  {arXiv preprint arXiv:2302.02966}\ } (\bibinfo {year} {2023})}\BibitemShut
  {NoStop}%
\bibitem [{\citenamefont {Cirac}\ and\ \citenamefont
  {Zoller}(1995)}]{cirac1995quantum}%
  \BibitemOpen
  \bibfield  {author} {\bibinfo {author} {\bibfnamefont {J.~I.}\ \bibnamefont
  {Cirac}}\ and\ \bibinfo {author} {\bibfnamefont {P.}~\bibnamefont {Zoller}},\
  }\bibfield  {title} {\bibinfo {title} {Quantum computations with cold trapped
  ions},\ }\href {https://doi.org/10.1103/PhysRevLett.74.4091} {\bibfield
  {journal} {\bibinfo  {journal} {Phys. Rev. Lett.}\ }\textbf {\bibinfo
  {volume} {74}},\ \bibinfo {pages} {4091} (\bibinfo {year}
  {1995})}\BibitemShut {NoStop}%
\bibitem [{\citenamefont {Islam}(2012)}]{islam2012quantum}%
  \BibitemOpen
  \bibfield  {author} {\bibinfo {author} {\bibfnamefont {K.~R.}\ \bibnamefont
  {Islam}},\ }\href@noop {} {\emph {\bibinfo {title} {Quantum simulation of
  interacting spin models with trapped ions}}}\ (\bibinfo  {publisher}
  {University of Maryland, College Park},\ \bibinfo {year} {2012})\BibitemShut
  {NoStop}%
\bibitem [{\citenamefont {Islam}\ \emph {et~al.}(2011)\citenamefont {Islam},
  \citenamefont {Edwards}, \citenamefont {Kim}, \citenamefont {Korenblit},
  \citenamefont {Noh}, \citenamefont {Carmichael}, \citenamefont {Lin},
  \citenamefont {Duan}, \citenamefont {Joseph~Wang}, \citenamefont {Freericks}
  \emph {et~al.}}]{islam2011onset}%
  \BibitemOpen
  \bibfield  {author} {\bibinfo {author} {\bibfnamefont {R.}~\bibnamefont
  {Islam}}, \bibinfo {author} {\bibfnamefont {E.}~\bibnamefont {Edwards}},
  \bibinfo {author} {\bibfnamefont {K.}~\bibnamefont {Kim}}, \bibinfo {author}
  {\bibfnamefont {S.}~\bibnamefont {Korenblit}}, \bibinfo {author}
  {\bibfnamefont {C.}~\bibnamefont {Noh}}, \bibinfo {author} {\bibfnamefont
  {H.}~\bibnamefont {Carmichael}}, \bibinfo {author} {\bibfnamefont {G.-D.}\
  \bibnamefont {Lin}}, \bibinfo {author} {\bibfnamefont {L.-M.}\ \bibnamefont
  {Duan}}, \bibinfo {author} {\bibfnamefont {C.-C.}\ \bibnamefont
  {Joseph~Wang}}, \bibinfo {author} {\bibfnamefont {J.}~\bibnamefont
  {Freericks}}, \emph {et~al.},\ }\bibfield  {title} {\bibinfo {title} {Onset
  of a quantum phase transition with a trapped ion quantum simulator},\ }\href
  {https://doi.org/10.1038/ncomms1374} {\bibfield  {journal} {\bibinfo
  {journal} {Nature communications}\ }\textbf {\bibinfo {volume} {2}},\
  \bibinfo {pages} {1} (\bibinfo {year} {2011})}\BibitemShut {NoStop}%
\bibitem [{Note1()}]{Note1}%
  \BibitemOpen
  \bibinfo {note} {Note that since the two qubits inside an ion cannot be
  separated, applications such as quantum teleportation are not considered
  here.}\BibitemShut {Stop}%
\bibitem [{Note2()}]{Note2}%
  \BibitemOpen
  \bibinfo {note} {The native intra-ion $R_{\alpha \beta }$ can only be used to
  construct gates belonging to the group SU($d$)}\BibitemShut {NoStop}%
\bibitem [{\citenamefont {Blume-Kohout}\ \emph {et~al.}(2002)\citenamefont
  {Blume-Kohout}, \citenamefont {Caves},\ and\ \citenamefont
  {Deutsch}}]{blume2002climbing}%
  \BibitemOpen
  \bibfield  {author} {\bibinfo {author} {\bibfnamefont {R.}~\bibnamefont
  {Blume-Kohout}}, \bibinfo {author} {\bibfnamefont {C.~M.}\ \bibnamefont
  {Caves}},\ and\ \bibinfo {author} {\bibfnamefont {I.~H.}\ \bibnamefont
  {Deutsch}},\ }\bibfield  {title} {\bibinfo {title} {Climbing mount scalable:
  Physical resource requirements for a scalable quantum computer},\ }\href
  {https://doi.org/10.1023/A:1021471621587} {\bibfield  {journal} {\bibinfo
  {journal} {Foundations of Physics}\ }\textbf {\bibinfo {volume} {32}},\
  \bibinfo {pages} {1641} (\bibinfo {year} {2002})}\BibitemShut {NoStop}%
\bibitem [{\citenamefont {Kiktenko}\ \emph
  {et~al.}(2015{\natexlab{b}})\citenamefont {Kiktenko}, \citenamefont
  {Fedorov}, \citenamefont {Strakhov},\ and\ \citenamefont
  {Man'Ko}}]{kiktenko2015single}%
  \BibitemOpen
  \bibfield  {author} {\bibinfo {author} {\bibfnamefont {E.}~\bibnamefont
  {Kiktenko}}, \bibinfo {author} {\bibfnamefont {A.}~\bibnamefont {Fedorov}},
  \bibinfo {author} {\bibfnamefont {A.}~\bibnamefont {Strakhov}},\ and\
  \bibinfo {author} {\bibfnamefont {V.}~\bibnamefont {Man'Ko}},\ }\bibfield
  {title} {\bibinfo {title} {Single qudit realization of the deutsch algorithm
  using superconducting many-level quantum circuits},\ }\href
  {https://doi.org/10.1016/j.physleta.2015.03.023} {\bibfield  {journal}
  {\bibinfo  {journal} {Physics Letters A}\ }\textbf {\bibinfo {volume}
  {379}},\ \bibinfo {pages} {1409} (\bibinfo {year}
  {2015}{\natexlab{b}})}\BibitemShut {NoStop}%
\bibitem [{\citenamefont {Wright}\ \emph {et~al.}(2019)\citenamefont {Wright},
  \citenamefont {Beck}, \citenamefont {Debnath}, \citenamefont {Amini},
  \citenamefont {Nam}, \citenamefont {Grzesiak}, \citenamefont {Chen},
  \citenamefont {Pisenti}, \citenamefont {Chmielewski}, \citenamefont {Collins}
  \emph {et~al.}}]{wright2019benchmarking}%
  \BibitemOpen
  \bibfield  {author} {\bibinfo {author} {\bibfnamefont {K.}~\bibnamefont
  {Wright}}, \bibinfo {author} {\bibfnamefont {K.~M.}\ \bibnamefont {Beck}},
  \bibinfo {author} {\bibfnamefont {S.}~\bibnamefont {Debnath}}, \bibinfo
  {author} {\bibfnamefont {J.}~\bibnamefont {Amini}}, \bibinfo {author}
  {\bibfnamefont {Y.}~\bibnamefont {Nam}}, \bibinfo {author} {\bibfnamefont
  {N.}~\bibnamefont {Grzesiak}}, \bibinfo {author} {\bibfnamefont {J.-S.}\
  \bibnamefont {Chen}}, \bibinfo {author} {\bibfnamefont {N.}~\bibnamefont
  {Pisenti}}, \bibinfo {author} {\bibfnamefont {M.}~\bibnamefont
  {Chmielewski}}, \bibinfo {author} {\bibfnamefont {C.}~\bibnamefont
  {Collins}}, \emph {et~al.},\ }\bibfield  {title} {\bibinfo {title}
  {Benchmarking an 11-qubit quantum computer},\ }\href
  {https://doi.org/https://doi.org/10.1038/s41467-019-13534-2} {\bibfield
  {journal} {\bibinfo  {journal} {Nature communications}\ }\textbf {\bibinfo
  {volume} {10}},\ \bibinfo {pages} {1} (\bibinfo {year} {2019})}\BibitemShut
  {NoStop}%
\bibitem [{\citenamefont {Harty}\ \emph {et~al.}(2014)\citenamefont {Harty},
  \citenamefont {Allcock}, \citenamefont {Ballance}, \citenamefont {Guidoni},
  \citenamefont {Janacek}, \citenamefont {Linke}, \citenamefont {Stacey},\ and\
  \citenamefont {Lucas}}]{harty2014high}%
  \BibitemOpen
  \bibfield  {author} {\bibinfo {author} {\bibfnamefont {T.~P.}\ \bibnamefont
  {Harty}}, \bibinfo {author} {\bibfnamefont {D.~T.~C.}\ \bibnamefont
  {Allcock}}, \bibinfo {author} {\bibfnamefont {C.~J.}\ \bibnamefont
  {Ballance}}, \bibinfo {author} {\bibfnamefont {L.}~\bibnamefont {Guidoni}},
  \bibinfo {author} {\bibfnamefont {H.~A.}\ \bibnamefont {Janacek}}, \bibinfo
  {author} {\bibfnamefont {N.~M.}\ \bibnamefont {Linke}}, \bibinfo {author}
  {\bibfnamefont {D.~N.}\ \bibnamefont {Stacey}},\ and\ \bibinfo {author}
  {\bibfnamefont {D.~M.}\ \bibnamefont {Lucas}},\ }\bibfield  {title} {\bibinfo
  {title} {High-fidelity preparation, gates, memory, and readout of a
  trapped-ion quantum bit},\ }\href
  {https://doi.org/10.1103/PhysRevLett.113.220501} {\bibfield  {journal}
  {\bibinfo  {journal} {Phys. Rev. Lett.}\ }\textbf {\bibinfo {volume} {113}},\
  \bibinfo {pages} {220501} (\bibinfo {year} {2014})}\BibitemShut {NoStop}%
\bibitem [{\citenamefont {M\o{}lmer}\ and\ \citenamefont
  {S\o{}rensen}(1999)}]{molmer_s_1999}%
  \BibitemOpen
  \bibfield  {author} {\bibinfo {author} {\bibfnamefont {K.}~\bibnamefont
  {M\o{}lmer}}\ and\ \bibinfo {author} {\bibfnamefont {A.}~\bibnamefont
  {S\o{}rensen}},\ }\bibfield  {title} {\bibinfo {title} {Multiparticle
  entanglement of hot trapped ions},\ }\href
  {https://doi.org/10.1103/PhysRevLett.82.1835} {\bibfield  {journal} {\bibinfo
   {journal} {Phys. Rev. Lett.}\ }\textbf {\bibinfo {volume} {82}},\ \bibinfo
  {pages} {1835} (\bibinfo {year} {1999})}\BibitemShut {NoStop}%
\bibitem [{\citenamefont {S\o{}rensen}\ and\ \citenamefont
  {M\o{}lmer}(2000)}]{molmer_s_2000}%
  \BibitemOpen
  \bibfield  {author} {\bibinfo {author} {\bibfnamefont {A.}~\bibnamefont
  {S\o{}rensen}}\ and\ \bibinfo {author} {\bibfnamefont {K.}~\bibnamefont
  {M\o{}lmer}},\ }\bibfield  {title} {\bibinfo {title} {Entanglement and
  quantum computation with ions in thermal motion},\ }\href
  {https://doi.org/10.1103/PhysRevA.62.022311} {\bibfield  {journal} {\bibinfo
  {journal} {Phys. Rev. A}\ }\textbf {\bibinfo {volume} {62}},\ \bibinfo
  {pages} {022311} (\bibinfo {year} {2000})}\BibitemShut {NoStop}%
\bibitem [{\citenamefont {Brylinski}\ and\ \citenamefont
  {Brylinski}(2002)}]{brylinski2002universal}%
  \BibitemOpen
  \bibfield  {author} {\bibinfo {author} {\bibfnamefont {J.-L.}\ \bibnamefont
  {Brylinski}}\ and\ \bibinfo {author} {\bibfnamefont {R.}~\bibnamefont
  {Brylinski}},\ }\bibfield  {title} {\bibinfo {title} {Universal quantum
  gates},\ }in\ \href@noop {} {\emph {\bibinfo {booktitle} {Mathematics of
  quantum computation}}}\ (\bibinfo  {publisher} {Chapman and Hall/CRC},\
  \bibinfo {year} {2002})\ pp.\ \bibinfo {pages} {117--134}\BibitemShut
  {NoStop}%
\bibitem [{\citenamefont {von Keyserlingk}\ and\ \citenamefont
  {Sondhi}(2016{\natexlab{a}})}]{CvK2016phase1}%
  \BibitemOpen
  \bibfield  {author} {\bibinfo {author} {\bibfnamefont {C.~W.}\ \bibnamefont
  {von Keyserlingk}}\ and\ \bibinfo {author} {\bibfnamefont {S.~L.}\
  \bibnamefont {Sondhi}},\ }\bibfield  {title} {\bibinfo {title} {Phase
  structure of one-dimensional interacting floquet systems. i. abelian
  symmetry-protected topological phases},\ }\href
  {https://doi.org/10.1103/PhysRevB.93.245145} {\bibfield  {journal} {\bibinfo
  {journal} {Phys. Rev. B}\ }\textbf {\bibinfo {volume} {93}},\ \bibinfo
  {pages} {245145} (\bibinfo {year} {2016}{\natexlab{a}})}\BibitemShut
  {NoStop}%
\bibitem [{\citenamefont {von Keyserlingk}\ and\ \citenamefont
  {Sondhi}(2016{\natexlab{b}})}]{CvK2016phase2}%
  \BibitemOpen
  \bibfield  {author} {\bibinfo {author} {\bibfnamefont {C.~W.}\ \bibnamefont
  {von Keyserlingk}}\ and\ \bibinfo {author} {\bibfnamefont {S.~L.}\
  \bibnamefont {Sondhi}},\ }\bibfield  {title} {\bibinfo {title} {Phase
  structure of one-dimensional interacting floquet systems. ii. symmetry-broken
  phases},\ }\href {https://doi.org/10.1103/PhysRevB.93.245146} {\bibfield
  {journal} {\bibinfo  {journal} {Phys. Rev. B}\ }\textbf {\bibinfo {volume}
  {93}},\ \bibinfo {pages} {245146} (\bibinfo {year}
  {2016}{\natexlab{b}})}\BibitemShut {NoStop}%
\bibitem [{\citenamefont {Bahri}\ \emph {et~al.}(2015)\citenamefont {Bahri},
  \citenamefont {Vosk}, \citenamefont {Altman},\ and\ \citenamefont
  {Vishwanath}}]{bahri2015localization}%
  \BibitemOpen
  \bibfield  {author} {\bibinfo {author} {\bibfnamefont {Y.}~\bibnamefont
  {Bahri}}, \bibinfo {author} {\bibfnamefont {R.}~\bibnamefont {Vosk}},
  \bibinfo {author} {\bibfnamefont {E.}~\bibnamefont {Altman}},\ and\ \bibinfo
  {author} {\bibfnamefont {A.}~\bibnamefont {Vishwanath}},\ }\bibfield  {title}
  {\bibinfo {title} {Localization and topology protected quantum coherence at
  the edge of hot matter},\ }\href
  {https://doi.org/https://doi.org/10.1038/ncomms8341} {\bibfield  {journal}
  {\bibinfo  {journal} {Nature communications}\ }\textbf {\bibinfo {volume}
  {6}},\ \bibinfo {pages} {1} (\bibinfo {year} {2015})}\BibitemShut {NoStop}%
\bibitem [{\citenamefont {Debnath}\ \emph {et~al.}(2016)\citenamefont
  {Debnath}, \citenamefont {Linke}, \citenamefont {Figgatt}, \citenamefont
  {Landsman}, \citenamefont {Wright},\ and\ \citenamefont
  {Monroe}}]{debnath2016demonstration}%
  \BibitemOpen
  \bibfield  {author} {\bibinfo {author} {\bibfnamefont {S.}~\bibnamefont
  {Debnath}}, \bibinfo {author} {\bibfnamefont {N.~M.}\ \bibnamefont {Linke}},
  \bibinfo {author} {\bibfnamefont {C.}~\bibnamefont {Figgatt}}, \bibinfo
  {author} {\bibfnamefont {K.~A.}\ \bibnamefont {Landsman}}, \bibinfo {author}
  {\bibfnamefont {K.}~\bibnamefont {Wright}},\ and\ \bibinfo {author}
  {\bibfnamefont {C.}~\bibnamefont {Monroe}},\ }\bibfield  {title} {\bibinfo
  {title} {Demonstration of a small programmable quantum computer with atomic
  qubits},\ }\href {https://doi.org/10.1038/nature18648} {\bibfield  {journal}
  {\bibinfo  {journal} {Nature}\ }\textbf {\bibinfo {volume} {536}},\ \bibinfo
  {pages} {63} (\bibinfo {year} {2016})}\BibitemShut {NoStop}%
\bibitem [{\citenamefont {Low}\ \emph {et~al.}(2020{\natexlab{b}})\citenamefont
  {Low}, \citenamefont {White}, \citenamefont {Cox}, \citenamefont {Day},\ and\
  \citenamefont {Senko}}]{low_2020_practical}%
  \BibitemOpen
  \bibfield  {author} {\bibinfo {author} {\bibfnamefont {P.~J.}\ \bibnamefont
  {Low}}, \bibinfo {author} {\bibfnamefont {B.~M.}\ \bibnamefont {White}},
  \bibinfo {author} {\bibfnamefont {A.~A.}\ \bibnamefont {Cox}}, \bibinfo
  {author} {\bibfnamefont {M.~L.}\ \bibnamefont {Day}},\ and\ \bibinfo {author}
  {\bibfnamefont {C.}~\bibnamefont {Senko}},\ }\bibfield  {title} {\bibinfo
  {title} {Practical trapped-ion protocols for universal qudit-based quantum
  computing},\ }\href {https://doi.org/10.1103/PhysRevResearch.2.033128}
  {\bibfield  {journal} {\bibinfo  {journal} {Phys. Rev. Research}\ }\textbf
  {\bibinfo {volume} {2}},\ \bibinfo {pages} {033128} (\bibinfo {year}
  {2020}{\natexlab{b}})}\BibitemShut {NoStop}%
\bibitem [{\citenamefont {Katz}\ \emph {et~al.}(2023)\citenamefont {Katz},
  \citenamefont {Cetina},\ and\ \citenamefont {Monroe}}]{katz2023programmable}%
  \BibitemOpen
  \bibfield  {author} {\bibinfo {author} {\bibfnamefont {O.}~\bibnamefont
  {Katz}}, \bibinfo {author} {\bibfnamefont {M.}~\bibnamefont {Cetina}},\ and\
  \bibinfo {author} {\bibfnamefont {C.}~\bibnamefont {Monroe}},\ }\bibfield
  {title} {\bibinfo {title} {Programmable $n$-body interactions with trapped
  ions},\ }\href {https://doi.org/10.1103/PRXQuantum.4.030311} {\bibfield
  {journal} {\bibinfo  {journal} {PRX Quantum}\ }\textbf {\bibinfo {volume}
  {4}},\ \bibinfo {pages} {030311} (\bibinfo {year} {2023})}\BibitemShut
  {NoStop}%
\bibitem [{\citenamefont {Figgatt}\ \emph {et~al.}(2019)\citenamefont
  {Figgatt}, \citenamefont {Ostrander}, \citenamefont {Linke}, \citenamefont
  {Landsman}, \citenamefont {Zhu}, \citenamefont {Maslov},\ and\ \citenamefont
  {Monroe}}]{figgatt2019parallel}%
  \BibitemOpen
  \bibfield  {author} {\bibinfo {author} {\bibfnamefont {C.}~\bibnamefont
  {Figgatt}}, \bibinfo {author} {\bibfnamefont {A.}~\bibnamefont {Ostrander}},
  \bibinfo {author} {\bibfnamefont {N.~M.}\ \bibnamefont {Linke}}, \bibinfo
  {author} {\bibfnamefont {K.~A.}\ \bibnamefont {Landsman}}, \bibinfo {author}
  {\bibfnamefont {D.}~\bibnamefont {Zhu}}, \bibinfo {author} {\bibfnamefont
  {D.}~\bibnamefont {Maslov}},\ and\ \bibinfo {author} {\bibfnamefont
  {C.}~\bibnamefont {Monroe}},\ }\bibfield  {title} {\bibinfo {title} {Parallel
  entangling operations on a universal ion-trap quantum computer},\ }\href
  {https://doi.org/https://doi.org/10.1038/s41586-019-1427-5} {\bibfield
  {journal} {\bibinfo  {journal} {Nature}\ }\textbf {\bibinfo {volume} {572}},\
  \bibinfo {pages} {368} (\bibinfo {year} {2019})}\BibitemShut {NoStop}%
\bibitem [{\citenamefont {Grzesiak}\ \emph {et~al.}(2020)\citenamefont
  {Grzesiak}, \citenamefont {Bl{\"u}mel}, \citenamefont {Wright}, \citenamefont
  {Beck}, \citenamefont {Pisenti}, \citenamefont {Li}, \citenamefont {Chaplin},
  \citenamefont {Amini}, \citenamefont {Debnath}, \citenamefont {Chen} \emph
  {et~al.}}]{grzesiak2020efficient}%
  \BibitemOpen
  \bibfield  {author} {\bibinfo {author} {\bibfnamefont {N.}~\bibnamefont
  {Grzesiak}}, \bibinfo {author} {\bibfnamefont {R.}~\bibnamefont
  {Bl{\"u}mel}}, \bibinfo {author} {\bibfnamefont {K.}~\bibnamefont {Wright}},
  \bibinfo {author} {\bibfnamefont {K.~M.}\ \bibnamefont {Beck}}, \bibinfo
  {author} {\bibfnamefont {N.~C.}\ \bibnamefont {Pisenti}}, \bibinfo {author}
  {\bibfnamefont {M.}~\bibnamefont {Li}}, \bibinfo {author} {\bibfnamefont
  {V.}~\bibnamefont {Chaplin}}, \bibinfo {author} {\bibfnamefont {J.~M.}\
  \bibnamefont {Amini}}, \bibinfo {author} {\bibfnamefont {S.}~\bibnamefont
  {Debnath}}, \bibinfo {author} {\bibfnamefont {J.-S.}\ \bibnamefont {Chen}},
  \emph {et~al.},\ }\bibfield  {title} {\bibinfo {title} {Efficient arbitrary
  simultaneously entangling gates on a trapped-ion quantum computer},\ }\href
  {https://doi.org/10.1038/s41467-020-16790-9} {\bibfield  {journal} {\bibinfo
  {journal} {Nature communications}\ }\textbf {\bibinfo {volume} {11}},\
  \bibinfo {pages} {2963} (\bibinfo {year} {2020})}\BibitemShut {NoStop}%
\bibitem [{\citenamefont {Pogorelov}\ \emph {et~al.}(2021)\citenamefont
  {Pogorelov}, \citenamefont {Feldker}, \citenamefont {Marciniak},
  \citenamefont {Postler}, \citenamefont {Jacob}, \citenamefont
  {Krieglsteiner}, \citenamefont {Podlesnic}, \citenamefont {Meth},
  \citenamefont {Negnevitsky}, \citenamefont {Stadler}, \citenamefont
  {H\"ofer}, \citenamefont {W\"achter}, \citenamefont {Lakhmanskiy},
  \citenamefont {Blatt}, \citenamefont {Schindler},\ and\ \citenamefont
  {Monz}}]{pogorelov2021compact}%
  \BibitemOpen
  \bibfield  {author} {\bibinfo {author} {\bibfnamefont {I.}~\bibnamefont
  {Pogorelov}}, \bibinfo {author} {\bibfnamefont {T.}~\bibnamefont {Feldker}},
  \bibinfo {author} {\bibfnamefont {C.~D.}\ \bibnamefont {Marciniak}}, \bibinfo
  {author} {\bibfnamefont {L.}~\bibnamefont {Postler}}, \bibinfo {author}
  {\bibfnamefont {G.}~\bibnamefont {Jacob}}, \bibinfo {author} {\bibfnamefont
  {O.}~\bibnamefont {Krieglsteiner}}, \bibinfo {author} {\bibfnamefont
  {V.}~\bibnamefont {Podlesnic}}, \bibinfo {author} {\bibfnamefont
  {M.}~\bibnamefont {Meth}}, \bibinfo {author} {\bibfnamefont {V.}~\bibnamefont
  {Negnevitsky}}, \bibinfo {author} {\bibfnamefont {M.}~\bibnamefont
  {Stadler}}, \bibinfo {author} {\bibfnamefont {B.}~\bibnamefont {H\"ofer}},
  \bibinfo {author} {\bibfnamefont {C.}~\bibnamefont {W\"achter}}, \bibinfo
  {author} {\bibfnamefont {K.}~\bibnamefont {Lakhmanskiy}}, \bibinfo {author}
  {\bibfnamefont {R.}~\bibnamefont {Blatt}}, \bibinfo {author} {\bibfnamefont
  {P.}~\bibnamefont {Schindler}},\ and\ \bibinfo {author} {\bibfnamefont
  {T.}~\bibnamefont {Monz}},\ }\bibfield  {title} {\bibinfo {title} {Compact
  ion-trap quantum computing demonstrator},\ }\href
  {https://doi.org/10.1103/PRXQuantum.2.020343} {\bibfield  {journal} {\bibinfo
   {journal} {PRX Quantum}\ }\textbf {\bibinfo {volume} {2}},\ \bibinfo {pages}
  {020343} (\bibinfo {year} {2021})}\BibitemShut {NoStop}%
\bibitem [{\citenamefont {Fisher}\ \emph {et~al.}(2023)\citenamefont {Fisher},
  \citenamefont {Khemani}, \citenamefont {Nahum},\ and\ \citenamefont
  {Vijay}}]{fisher2023random}%
  \BibitemOpen
  \bibfield  {author} {\bibinfo {author} {\bibfnamefont {M.~P.}\ \bibnamefont
  {Fisher}}, \bibinfo {author} {\bibfnamefont {V.}~\bibnamefont {Khemani}},
  \bibinfo {author} {\bibfnamefont {A.}~\bibnamefont {Nahum}},\ and\ \bibinfo
  {author} {\bibfnamefont {S.}~\bibnamefont {Vijay}},\ }\bibfield  {title}
  {\bibinfo {title} {Random quantum circuits},\ }\href
  {https://doi.org/10.1146/annurev-conmatphys-031720-030658} {\bibfield
  {journal} {\bibinfo  {journal} {Annual Review of Condensed Matter Physics}\
  }\textbf {\bibinfo {volume} {14}},\ \bibinfo {pages} {335} (\bibinfo {year}
  {2023})}\BibitemShut {NoStop}%
\bibitem [{\citenamefont {Boixo}\ \emph {et~al.}(2018)\citenamefont {Boixo},
  \citenamefont {Isakov}, \citenamefont {Smelyanskiy}, \citenamefont {Babbush},
  \citenamefont {Ding}, \citenamefont {Jiang}, \citenamefont {Bremner},
  \citenamefont {Martinis},\ and\ \citenamefont
  {Neven}}]{boixo2018characterizing}%
  \BibitemOpen
  \bibfield  {author} {\bibinfo {author} {\bibfnamefont {S.}~\bibnamefont
  {Boixo}}, \bibinfo {author} {\bibfnamefont {S.~V.}\ \bibnamefont {Isakov}},
  \bibinfo {author} {\bibfnamefont {V.~N.}\ \bibnamefont {Smelyanskiy}},
  \bibinfo {author} {\bibfnamefont {R.}~\bibnamefont {Babbush}}, \bibinfo
  {author} {\bibfnamefont {N.}~\bibnamefont {Ding}}, \bibinfo {author}
  {\bibfnamefont {Z.}~\bibnamefont {Jiang}}, \bibinfo {author} {\bibfnamefont
  {M.~J.}\ \bibnamefont {Bremner}}, \bibinfo {author} {\bibfnamefont {J.~M.}\
  \bibnamefont {Martinis}},\ and\ \bibinfo {author} {\bibfnamefont
  {H.}~\bibnamefont {Neven}},\ }\bibfield  {title} {\bibinfo {title}
  {Characterizing quantum supremacy in near-term devices},\ }\href
  {https://doi.org/https://doi.org/10.1038/s41567-018-0124-x} {\bibfield
  {journal} {\bibinfo  {journal} {Nature Physics}\ }\textbf {\bibinfo {volume}
  {14}},\ \bibinfo {pages} {595} (\bibinfo {year} {2018})}\BibitemShut
  {NoStop}%
\bibitem [{\citenamefont {Arute}\ \emph {et~al.}(2019)\citenamefont {Arute},
  \citenamefont {Arya}, \citenamefont {Babbush}, \citenamefont {Bacon},
  \citenamefont {Bardin}, \citenamefont {Barends}, \citenamefont {Biswas},
  \citenamefont {Boixo}, \citenamefont {Brandao}, \citenamefont {Buell} \emph
  {et~al.}}]{google2019supremacy}%
  \BibitemOpen
  \bibfield  {author} {\bibinfo {author} {\bibfnamefont {F.}~\bibnamefont
  {Arute}}, \bibinfo {author} {\bibfnamefont {K.}~\bibnamefont {Arya}},
  \bibinfo {author} {\bibfnamefont {R.}~\bibnamefont {Babbush}}, \bibinfo
  {author} {\bibfnamefont {D.}~\bibnamefont {Bacon}}, \bibinfo {author}
  {\bibfnamefont {J.~C.}\ \bibnamefont {Bardin}}, \bibinfo {author}
  {\bibfnamefont {R.}~\bibnamefont {Barends}}, \bibinfo {author} {\bibfnamefont
  {R.}~\bibnamefont {Biswas}}, \bibinfo {author} {\bibfnamefont
  {S.}~\bibnamefont {Boixo}}, \bibinfo {author} {\bibfnamefont {F.~G.}\
  \bibnamefont {Brandao}}, \bibinfo {author} {\bibfnamefont {D.~A.}\
  \bibnamefont {Buell}}, \emph {et~al.},\ }\bibfield  {title} {\bibinfo {title}
  {Quantum supremacy using a programmable superconducting processor},\ }\href
  {https://doi.org/https://doi.org/10.1038/s41586-019-1666-5} {\bibfield
  {journal} {\bibinfo  {journal} {Nature}\ }\textbf {\bibinfo {volume} {574}},\
  \bibinfo {pages} {505} (\bibinfo {year} {2019})}\BibitemShut {NoStop}%
\bibitem [{\citenamefont {Morvan}\ \emph {et~al.}(2023)\citenamefont {Morvan},
  \citenamefont {Villalonga}, \citenamefont {Mi}, \citenamefont {Mandra},
  \citenamefont {Bengtsson}, \citenamefont {Klimov}, \citenamefont {Chen},
  \citenamefont {Hong}, \citenamefont {Erickson}, \citenamefont {Drozdov} \emph
  {et~al.}}]{morvan2023phase}%
  \BibitemOpen
  \bibfield  {author} {\bibinfo {author} {\bibfnamefont {A.}~\bibnamefont
  {Morvan}}, \bibinfo {author} {\bibfnamefont {B.}~\bibnamefont {Villalonga}},
  \bibinfo {author} {\bibfnamefont {X.}~\bibnamefont {Mi}}, \bibinfo {author}
  {\bibfnamefont {S.}~\bibnamefont {Mandra}}, \bibinfo {author} {\bibfnamefont
  {A.}~\bibnamefont {Bengtsson}}, \bibinfo {author} {\bibfnamefont
  {P.}~\bibnamefont {Klimov}}, \bibinfo {author} {\bibfnamefont
  {Z.}~\bibnamefont {Chen}}, \bibinfo {author} {\bibfnamefont {S.}~\bibnamefont
  {Hong}}, \bibinfo {author} {\bibfnamefont {C.}~\bibnamefont {Erickson}},
  \bibinfo {author} {\bibfnamefont {I.}~\bibnamefont {Drozdov}}, \emph
  {et~al.},\ }\bibfield  {title} {\bibinfo {title} {Phase transition in random
  circuit sampling},\ }\href {https://arxiv.org/abs/2304.11119} {\bibfield
  {journal} {\bibinfo  {journal} {arXiv preprint arXiv:2304.11119}\ } (\bibinfo
  {year} {2023})}\BibitemShut {NoStop}%
\bibitem [{\citenamefont {Dalzell}\ \emph {et~al.}(2022)\citenamefont
  {Dalzell}, \citenamefont {Hunter-Jones},\ and\ \citenamefont
  {Brand\~ao}}]{dalzell2022random}%
  \BibitemOpen
  \bibfield  {author} {\bibinfo {author} {\bibfnamefont {A.~M.}\ \bibnamefont
  {Dalzell}}, \bibinfo {author} {\bibfnamefont {N.}~\bibnamefont
  {Hunter-Jones}},\ and\ \bibinfo {author} {\bibfnamefont {F.~G. S.~L.}\
  \bibnamefont {Brand\~ao}},\ }\bibfield  {title} {\bibinfo {title} {Random
  quantum circuits anticoncentrate in log depth},\ }\href
  {https://doi.org/10.1103/PRXQuantum.3.010333} {\bibfield  {journal} {\bibinfo
   {journal} {PRX Quantum}\ }\textbf {\bibinfo {volume} {3}},\ \bibinfo {pages}
  {010333} (\bibinfo {year} {2022})}\BibitemShut {NoStop}%
\bibitem [{\citenamefont {Harrow}\ and\ \citenamefont
  {Low}(2009)}]{harrow2009random}%
  \BibitemOpen
  \bibfield  {author} {\bibinfo {author} {\bibfnamefont {A.~W.}\ \bibnamefont
  {Harrow}}\ and\ \bibinfo {author} {\bibfnamefont {R.~A.}\ \bibnamefont
  {Low}},\ }\bibfield  {title} {\bibinfo {title} {Random quantum circuits are
  approximate 2-designs},\ }\href
  {https://doi.org/https://doi.org/10.1007/s00220-009-0873-6} {\bibfield
  {journal} {\bibinfo  {journal} {Communications in Mathematical Physics}\
  }\textbf {\bibinfo {volume} {291}},\ \bibinfo {pages} {257} (\bibinfo {year}
  {2009})}\BibitemShut {NoStop}%
\bibitem [{\citenamefont {Harrow}\ and\ \citenamefont
  {Mehraban}(2023)}]{harrow2023approximate}%
  \BibitemOpen
  \bibfield  {author} {\bibinfo {author} {\bibfnamefont {A.~W.}\ \bibnamefont
  {Harrow}}\ and\ \bibinfo {author} {\bibfnamefont {S.}~\bibnamefont
  {Mehraban}},\ }\bibfield  {title} {\bibinfo {title} {Approximate unitary
  t-designs by short random quantum circuits using nearest-neighbor and
  long-range gates},\ }\href
  {https://doi.org/https://doi.org/10.1007/s00220-023-04675-z} {\bibfield
  {journal} {\bibinfo  {journal} {Communications in Mathematical Physics}\ ,\
  \bibinfo {pages} {1}} (\bibinfo {year} {2023})}\BibitemShut {NoStop}%
\bibitem [{\citenamefont {Gidney}(2021)}]{Gidney2021stimfaststabilizer}%
  \BibitemOpen
  \bibfield  {author} {\bibinfo {author} {\bibfnamefont {C.}~\bibnamefont
  {Gidney}},\ }\bibfield  {title} {\bibinfo {title} {Stim: a fast stabilizer
  circuit simulator},\ }\href {https://doi.org/10.22331/q-2021-07-06-497}
  {\bibfield  {journal} {\bibinfo  {journal} {{Quantum}}\ }\textbf {\bibinfo
  {volume} {5}},\ \bibinfo {pages} {497} (\bibinfo {year} {2021})}\BibitemShut
  {NoStop}%
\bibitem [{\citenamefont {Higgott}(2022)}]{higgott2022pymatching}%
  \BibitemOpen
  \bibfield  {author} {\bibinfo {author} {\bibfnamefont {O.}~\bibnamefont
  {Higgott}},\ }\bibfield  {title} {\bibinfo {title} {Pymatching: A python
  package for decoding quantum codes with minimum-weight perfect matching},\
  }\href {https://doi.org/10.1145/3505637} {\bibfield  {journal} {\bibinfo
  {journal} {ACM Transactions on Quantum Computing}\ }\textbf {\bibinfo
  {volume} {3}},\ \bibinfo {pages} {1} (\bibinfo {year} {2022})}\BibitemShut
  {NoStop}%
\bibitem [{\citenamefont {Ballance}\ \emph {et~al.}(2016)\citenamefont
  {Ballance}, \citenamefont {Harty}, \citenamefont {Linke}, \citenamefont
  {Sepiol},\ and\ \citenamefont {Lucas}}]{ballance2016high}%
  \BibitemOpen
  \bibfield  {author} {\bibinfo {author} {\bibfnamefont {C.~J.}\ \bibnamefont
  {Ballance}}, \bibinfo {author} {\bibfnamefont {T.~P.}\ \bibnamefont {Harty}},
  \bibinfo {author} {\bibfnamefont {N.~M.}\ \bibnamefont {Linke}}, \bibinfo
  {author} {\bibfnamefont {M.~A.}\ \bibnamefont {Sepiol}},\ and\ \bibinfo
  {author} {\bibfnamefont {D.~M.}\ \bibnamefont {Lucas}},\ }\bibfield  {title}
  {\bibinfo {title} {High-fidelity quantum logic gates using trapped-ion
  hyperfine qubits},\ }\href {https://doi.org/10.1103/PhysRevLett.117.060504}
  {\bibfield  {journal} {\bibinfo  {journal} {Phys. Rev. Lett.}\ }\textbf
  {\bibinfo {volume} {117}},\ \bibinfo {pages} {060504} (\bibinfo {year}
  {2016})}\BibitemShut {NoStop}%
\bibitem [{\citenamefont {Manovitz}\ \emph {et~al.}(2022)\citenamefont
  {Manovitz}, \citenamefont {Shapira}, \citenamefont {Gazit}, \citenamefont
  {Akerman},\ and\ \citenamefont {Ozeri}}]{manovitz_small_computer}%
  \BibitemOpen
  \bibfield  {author} {\bibinfo {author} {\bibfnamefont {T.}~\bibnamefont
  {Manovitz}}, \bibinfo {author} {\bibfnamefont {Y.}~\bibnamefont {Shapira}},
  \bibinfo {author} {\bibfnamefont {L.}~\bibnamefont {Gazit}}, \bibinfo
  {author} {\bibfnamefont {N.}~\bibnamefont {Akerman}},\ and\ \bibinfo {author}
  {\bibfnamefont {R.}~\bibnamefont {Ozeri}},\ }\bibfield  {title} {\bibinfo
  {title} {Trapped-ion quantum computer with robust entangling gates and
  quantum coherent feedback},\ }\href
  {https://doi.org/10.1103/PRXQuantum.3.010347} {\bibfield  {journal} {\bibinfo
   {journal} {PRX Quantum}\ }\textbf {\bibinfo {volume} {3}},\ \bibinfo {pages}
  {010347} (\bibinfo {year} {2022})}\BibitemShut {NoStop}%
\bibitem [{\citenamefont {Saner}\ \emph {et~al.}(2023)\citenamefont {Saner},
  \citenamefont {B\ifmmode \u{a}\else \u{a}\fi{}z\ifmmode~\u{a}\else
  \u{a}\fi{}van}, \citenamefont {Minder}, \citenamefont {Drmota}, \citenamefont
  {Webb}, \citenamefont {Araneda}, \citenamefont {Srinivas}, \citenamefont
  {Lucas},\ and\ \citenamefont {Ballance}}]{saner_standing_waves}%
  \BibitemOpen
  \bibfield  {author} {\bibinfo {author} {\bibfnamefont {S.}~\bibnamefont
  {Saner}}, \bibinfo {author} {\bibfnamefont {O.}~\bibnamefont {B\ifmmode
  \u{a}\else \u{a}\fi{}z\ifmmode~\u{a}\else \u{a}\fi{}van}}, \bibinfo {author}
  {\bibfnamefont {M.}~\bibnamefont {Minder}}, \bibinfo {author} {\bibfnamefont
  {P.}~\bibnamefont {Drmota}}, \bibinfo {author} {\bibfnamefont {D.~J.}\
  \bibnamefont {Webb}}, \bibinfo {author} {\bibfnamefont {G.}~\bibnamefont
  {Araneda}}, \bibinfo {author} {\bibfnamefont {R.}~\bibnamefont {Srinivas}},
  \bibinfo {author} {\bibfnamefont {D.~M.}\ \bibnamefont {Lucas}},\ and\
  \bibinfo {author} {\bibfnamefont {C.~J.}\ \bibnamefont {Ballance}},\
  }\bibfield  {title} {\bibinfo {title} {Breaking the entangling gate speed
  limit for trapped-ion qubits using a phase-stable standing wave},\ }\href
  {https://doi.org/10.1103/PhysRevLett.131.220601} {\bibfield  {journal}
  {\bibinfo  {journal} {Phys. Rev. Lett.}\ }\textbf {\bibinfo {volume} {131}},\
  \bibinfo {pages} {220601} (\bibinfo {year} {2023})}\BibitemShut {NoStop}%
\bibitem [{\citenamefont {Vizvary}\ \emph {et~al.}(2023)\citenamefont
  {Vizvary}, \citenamefont {Wall}, \citenamefont {Boguslawski}, \citenamefont
  {Bareian}, \citenamefont {Derevianko}, \citenamefont {Campbell},\ and\
  \citenamefont {Hudson}}]{vizvary2023eliminating}%
  \BibitemOpen
  \bibfield  {author} {\bibinfo {author} {\bibfnamefont {S.~R.}\ \bibnamefont
  {Vizvary}}, \bibinfo {author} {\bibfnamefont {Z.~J.}\ \bibnamefont {Wall}},
  \bibinfo {author} {\bibfnamefont {M.~J.}\ \bibnamefont {Boguslawski}},
  \bibinfo {author} {\bibfnamefont {M.}~\bibnamefont {Bareian}}, \bibinfo
  {author} {\bibfnamefont {A.}~\bibnamefont {Derevianko}}, \bibinfo {author}
  {\bibfnamefont {W.~C.}\ \bibnamefont {Campbell}},\ and\ \bibinfo {author}
  {\bibfnamefont {E.~R.}\ \bibnamefont {Hudson}},\ }\href@noop {} {\bibinfo
  {title} {Eliminating qubit type cross-talk in the $\textit{omg}$ protocol}}
  (\bibinfo {year} {2023}),\ \Eprint {https://arxiv.org/abs/2310.10905}
  {arXiv:2310.10905 [quant-ph]} \BibitemShut {NoStop}%
\bibitem [{\citenamefont {Akerman}\ \emph {et~al.}(2015)\citenamefont
  {Akerman}, \citenamefont {Navon}, \citenamefont {Kotler}, \citenamefont
  {Glickman},\ and\ \citenamefont {Ozeri}}]{akerman2015universal}%
  \BibitemOpen
  \bibfield  {author} {\bibinfo {author} {\bibfnamefont {N.}~\bibnamefont
  {Akerman}}, \bibinfo {author} {\bibfnamefont {N.}~\bibnamefont {Navon}},
  \bibinfo {author} {\bibfnamefont {S.}~\bibnamefont {Kotler}}, \bibinfo
  {author} {\bibfnamefont {Y.}~\bibnamefont {Glickman}},\ and\ \bibinfo
  {author} {\bibfnamefont {R.}~\bibnamefont {Ozeri}},\ }\bibfield  {title}
  {\bibinfo {title} {Universal gate-set for trapped-ion qubits using a narrow
  linewidth diode laser},\ }\href
  {https://doi.org/10.1088/1367-2630/17/11/113060} {\bibfield  {journal}
  {\bibinfo  {journal} {New Journal of Physics}\ }\textbf {\bibinfo {volume}
  {17}},\ \bibinfo {pages} {113060} (\bibinfo {year} {2015})}\BibitemShut
  {NoStop}%
\bibitem [{\citenamefont {Srinivas}\ \emph {et~al.}(2021)\citenamefont
  {Srinivas}, \citenamefont {Burd}, \citenamefont {Knaack}, \citenamefont
  {Sutherland}, \citenamefont {Kwiatkowski}, \citenamefont {Glancy},
  \citenamefont {Knill}, \citenamefont {Wineland}, \citenamefont {Leibfried},
  \citenamefont {Wilson}, \citenamefont {Allcock},\ and\ \citenamefont
  {Slichter}}]{srinivas2021high}%
  \BibitemOpen
  \bibfield  {author} {\bibinfo {author} {\bibfnamefont {R.}~\bibnamefont
  {Srinivas}}, \bibinfo {author} {\bibfnamefont {S.}~\bibnamefont {Burd}},
  \bibinfo {author} {\bibfnamefont {H.}~\bibnamefont {Knaack}}, \bibinfo
  {author} {\bibfnamefont {R.}~\bibnamefont {Sutherland}}, \bibinfo {author}
  {\bibfnamefont {A.}~\bibnamefont {Kwiatkowski}}, \bibinfo {author}
  {\bibfnamefont {S.}~\bibnamefont {Glancy}}, \bibinfo {author} {\bibfnamefont
  {E.}~\bibnamefont {Knill}}, \bibinfo {author} {\bibfnamefont
  {D.}~\bibnamefont {Wineland}}, \bibinfo {author} {\bibfnamefont
  {D.}~\bibnamefont {Leibfried}}, \bibinfo {author} {\bibfnamefont
  {A.}~\bibnamefont {Wilson}}, \bibinfo {author} {\bibfnamefont
  {D.}~\bibnamefont {Allcock}},\ and\ \bibinfo {author} {\bibfnamefont
  {D.}~\bibnamefont {Slichter}},\ }\bibfield  {title} {\bibinfo {title}
  {High-fidelity laser-free universal control of trapped ion qubits},\
  }\bibfield  {journal} {\bibinfo  {journal} {Nature}\ }\textbf {\bibinfo
  {volume} {597}},\ \href
  {https://doi.org/https://doi.org/10.1038/s41586-021-03809-4}
  {https://doi.org/10.1038/s41586-021-03809-4} (\bibinfo {year}
  {2021})\BibitemShut {NoStop}%
\bibitem [{\citenamefont {Milburn}(1999)}]{milburn1999simulating}%
  \BibitemOpen
  \bibfield  {author} {\bibinfo {author} {\bibfnamefont {G.~J.}\ \bibnamefont
  {Milburn}},\ }\href@noop {} {\bibinfo {title} {Simulating nonlinear spin
  models in an ion trap}} (\bibinfo {year} {1999}),\ \Eprint
  {https://arxiv.org/abs/quant-ph/9908037} {arXiv:quant-ph/9908037 [quant-ph]}
  \BibitemShut {NoStop}%
\bibitem [{\citenamefont {Sawyer}\ and\ \citenamefont
  {Brown}(2021)}]{sawyer2021wavelength}%
  \BibitemOpen
  \bibfield  {author} {\bibinfo {author} {\bibfnamefont {B.~C.}\ \bibnamefont
  {Sawyer}}\ and\ \bibinfo {author} {\bibfnamefont {K.~R.}\ \bibnamefont
  {Brown}},\ }\bibfield  {title} {\bibinfo {title} {Wavelength-insensitive,
  multispecies entangling gate for group-2 atomic ions},\ }\href
  {https://doi.org/10.1103/PhysRevA.103.022427} {\bibfield  {journal} {\bibinfo
   {journal} {Phys. Rev. A}\ }\textbf {\bibinfo {volume} {103}},\ \bibinfo
  {pages} {022427} (\bibinfo {year} {2021})}\BibitemShut {NoStop}%
\bibitem [{\citenamefont {An}\ \emph {et~al.}(2022)\citenamefont {An},
  \citenamefont {Ransford}, \citenamefont {Schaffer}, \citenamefont {Sletten},
  \citenamefont {Gaebler}, \citenamefont {Hostetter},\ and\ \citenamefont
  {Vittorini}}]{an2022high}%
  \BibitemOpen
  \bibfield  {author} {\bibinfo {author} {\bibfnamefont {F.~A.}\ \bibnamefont
  {An}}, \bibinfo {author} {\bibfnamefont {A.}~\bibnamefont {Ransford}},
  \bibinfo {author} {\bibfnamefont {A.}~\bibnamefont {Schaffer}}, \bibinfo
  {author} {\bibfnamefont {L.~R.}\ \bibnamefont {Sletten}}, \bibinfo {author}
  {\bibfnamefont {J.}~\bibnamefont {Gaebler}}, \bibinfo {author} {\bibfnamefont
  {J.}~\bibnamefont {Hostetter}},\ and\ \bibinfo {author} {\bibfnamefont
  {G.}~\bibnamefont {Vittorini}},\ }\bibfield  {title} {\bibinfo {title} {High
  fidelity state preparation and measurement of ion hyperfine qubits with
  $i>\frac{1}{2}$},\ }\href {https://doi.org/10.1103/PhysRevLett.129.130501}
  {\bibfield  {journal} {\bibinfo  {journal} {Phys. Rev. Lett.}\ }\textbf
  {\bibinfo {volume} {129}},\ \bibinfo {pages} {130501} (\bibinfo {year}
  {2022})}\BibitemShut {NoStop}%
\bibitem [{\citenamefont {Sotirova}\ and\ \citenamefont
  {Leppard}(2024)}]{sotirova2024high}%
  \BibitemOpen
  \bibfield  {author} {\bibinfo {author} {\bibfnamefont {A.}~\bibnamefont
  {Sotirova}}\ and\ \bibinfo {author} {\bibfnamefont {J.}~\bibnamefont
  {Leppard}},\ }\bibfield  {title} {\bibinfo {title} {High-fidelity heralded
  quantum state preparation and measurement using multiple qubit encodings
  complete when published},\ }\href@noop {} {\bibfield  {journal} {\bibinfo
  {journal} {Article in preparation}\ } (\bibinfo {year} {2024})}\BibitemShut
  {NoStop}%
\bibitem [{\citenamefont {Barton}\ \emph {et~al.}(2000)\citenamefont {Barton},
  \citenamefont {Donald}, \citenamefont {Lucas}, \citenamefont {Stevens},
  \citenamefont {Steane},\ and\ \citenamefont
  {Stacey}}]{barton2000Ca40metalifetime}%
  \BibitemOpen
  \bibfield  {author} {\bibinfo {author} {\bibfnamefont {P.~A.}\ \bibnamefont
  {Barton}}, \bibinfo {author} {\bibfnamefont {C.~J.~S.}\ \bibnamefont
  {Donald}}, \bibinfo {author} {\bibfnamefont {D.~M.}\ \bibnamefont {Lucas}},
  \bibinfo {author} {\bibfnamefont {D.~A.}\ \bibnamefont {Stevens}}, \bibinfo
  {author} {\bibfnamefont {A.~M.}\ \bibnamefont {Steane}},\ and\ \bibinfo
  {author} {\bibfnamefont {D.~N.}\ \bibnamefont {Stacey}},\ }\bibfield  {title}
  {\bibinfo {title} {Measurement of the lifetime of the $3d{}^{2}{D}_{5/2}$
  state in ${}^{40}{\mathrm{ca}}^{+}$},\ }\href
  {https://link.aps.org/doi/10.1103/PhysRevA.62.032503} {\bibfield  {journal}
  {\bibinfo  {journal} {Phys. Rev. A}\ }\textbf {\bibinfo {volume} {62}},\
  \bibinfo {pages} {032503} (\bibinfo {year} {2000})}\BibitemShut {NoStop}%
\bibitem [{\citenamefont {Letchumanan}\ \emph {et~al.}(2005)\citenamefont
  {Letchumanan}, \citenamefont {Wilson}, \citenamefont {Gill},\ and\
  \citenamefont {Sinclair}}]{Letchumanan2005Sr88metalifetime}%
  \BibitemOpen
  \bibfield  {author} {\bibinfo {author} {\bibfnamefont {V.}~\bibnamefont
  {Letchumanan}}, \bibinfo {author} {\bibfnamefont {M.~A.}\ \bibnamefont
  {Wilson}}, \bibinfo {author} {\bibfnamefont {P.}~\bibnamefont {Gill}},\ and\
  \bibinfo {author} {\bibfnamefont {A.~G.}\ \bibnamefont {Sinclair}},\
  }\bibfield  {title} {\bibinfo {title} {Lifetime measurement of the metastable
  $4d^{2} {D}_{5/2}$ state in $^{88}\mathrm{Sr}^{+}$ using a single trapped
  ion},\ }\href {https://doi.org/10.1103/PhysRevA.72.012509} {\bibfield
  {journal} {\bibinfo  {journal} {Phys. Rev. A}\ }\textbf {\bibinfo {volume}
  {72}},\ \bibinfo {pages} {012509} (\bibinfo {year} {2005})}\BibitemShut
  {NoStop}%
\bibitem [{\citenamefont {Zhang}\ \emph {et~al.}(2020)\citenamefont {Zhang},
  \citenamefont {Arnold}, \citenamefont {Chanu}, \citenamefont {Kaewuam},
  \citenamefont {Safronova},\ and\ \citenamefont
  {Barrett}}]{Zhang2020Ba138metalifetime}%
  \BibitemOpen
  \bibfield  {author} {\bibinfo {author} {\bibfnamefont {Z.}~\bibnamefont
  {Zhang}}, \bibinfo {author} {\bibfnamefont {K.~J.}\ \bibnamefont {Arnold}},
  \bibinfo {author} {\bibfnamefont {S.~R.}\ \bibnamefont {Chanu}}, \bibinfo
  {author} {\bibfnamefont {R.}~\bibnamefont {Kaewuam}}, \bibinfo {author}
  {\bibfnamefont {M.~S.}\ \bibnamefont {Safronova}},\ and\ \bibinfo {author}
  {\bibfnamefont {M.~D.}\ \bibnamefont {Barrett}},\ }\bibfield  {title}
  {\bibinfo {title} {Branching fractions for ${P}_{3/2}$ decays in
  ${\mathrm{ba}}^{+}$},\ }\href {https://doi.org/10.1103/PhysRevA.101.062515}
  {\bibfield  {journal} {\bibinfo  {journal} {Phys. Rev. A}\ }\textbf {\bibinfo
  {volume} {101}},\ \bibinfo {pages} {062515} (\bibinfo {year}
  {2020})}\BibitemShut {NoStop}%
\bibitem [{\citenamefont {Lange}\ \emph {et~al.}(2021)\citenamefont {Lange},
  \citenamefont {Peshkov}, \citenamefont {Huntemann}, \citenamefont {Tamm},
  \citenamefont {Surzhykov},\ and\ \citenamefont
  {Peik}}]{Lange2021Yb171metalifetime}%
  \BibitemOpen
  \bibfield  {author} {\bibinfo {author} {\bibfnamefont {R.}~\bibnamefont
  {Lange}}, \bibinfo {author} {\bibfnamefont {A.~A.}\ \bibnamefont {Peshkov}},
  \bibinfo {author} {\bibfnamefont {N.}~\bibnamefont {Huntemann}}, \bibinfo
  {author} {\bibfnamefont {C.}~\bibnamefont {Tamm}}, \bibinfo {author}
  {\bibfnamefont {A.}~\bibnamefont {Surzhykov}},\ and\ \bibinfo {author}
  {\bibfnamefont {E.}~\bibnamefont {Peik}},\ }\bibfield  {title} {\bibinfo
  {title} {Lifetime of the $^{2}{F}_{7/2}$ level in ${\mathrm{yb}}^{+}$ for
  spontaneous emission of electric octupole radiation},\ }\href
  {https://doi.org/10.1103/PhysRevLett.127.213001} {\bibfield  {journal}
  {\bibinfo  {journal} {Phys. Rev. Lett.}\ }\textbf {\bibinfo {volume} {127}},\
  \bibinfo {pages} {213001} (\bibinfo {year} {2021})}\BibitemShut {NoStop}%
\bibitem [{\citenamefont {Dzuba}\ and\ \citenamefont
  {Flambaum}(2016)}]{Dzuba2016Yb173metalifetime}%
  \BibitemOpen
  \bibfield  {author} {\bibinfo {author} {\bibfnamefont {V.~A.}\ \bibnamefont
  {Dzuba}}\ and\ \bibinfo {author} {\bibfnamefont {V.~V.}\ \bibnamefont
  {Flambaum}},\ }\bibfield  {title} {\bibinfo {title} {Hyperfine-induced
  electric dipole contributions to the electric octupole and magnetic
  quadrupole atomic clock transitions},\ }\href
  {https://doi.org/10.1103/PhysRevA.93.052517} {\bibfield  {journal} {\bibinfo
  {journal} {Phys. Rev. A}\ }\textbf {\bibinfo {volume} {93}},\ \bibinfo
  {pages} {052517} (\bibinfo {year} {2016})}\BibitemShut {NoStop}%
\bibitem [{\citenamefont {Cummins}\ \emph {et~al.}(2003)\citenamefont
  {Cummins}, \citenamefont {Llewellyn},\ and\ \citenamefont
  {Jones}}]{Cummins2003Tackling}%
  \BibitemOpen
  \bibfield  {author} {\bibinfo {author} {\bibfnamefont {H.~K.}\ \bibnamefont
  {Cummins}}, \bibinfo {author} {\bibfnamefont {G.}~\bibnamefont {Llewellyn}},\
  and\ \bibinfo {author} {\bibfnamefont {J.~A.}\ \bibnamefont {Jones}},\
  }\bibfield  {title} {\bibinfo {title} {Tackling systematic errors in quantum
  logic gates with composite rotations},\ }\href
  {https://doi.org/10.1103/PhysRevA.67.042308} {\bibfield  {journal} {\bibinfo
  {journal} {Phys. Rev. A}\ }\textbf {\bibinfo {volume} {67}},\ \bibinfo
  {pages} {042308} (\bibinfo {year} {2003})}\BibitemShut {NoStop}%
\bibitem [{\citenamefont {Bauer}\ \emph {et~al.}(1984)\citenamefont {Bauer},
  \citenamefont {Freeman}, \citenamefont {Frenkiel}, \citenamefont {Keeler},\
  and\ \citenamefont {Shaka}}]{bauer1984Gaussian}%
  \BibitemOpen
  \bibfield  {author} {\bibinfo {author} {\bibfnamefont {C.}~\bibnamefont
  {Bauer}}, \bibinfo {author} {\bibfnamefont {R.}~\bibnamefont {Freeman}},
  \bibinfo {author} {\bibfnamefont {T.}~\bibnamefont {Frenkiel}}, \bibinfo
  {author} {\bibfnamefont {J.}~\bibnamefont {Keeler}},\ and\ \bibinfo {author}
  {\bibfnamefont {A.}~\bibnamefont {Shaka}},\ }\bibfield  {title} {\bibinfo
  {title} {Gaussian pulses},\ }\href
  {https://doi.org/https://doi.org/10.1016/0022-2364(84)90148-3} {\bibfield
  {journal} {\bibinfo  {journal} {Journal of Magnetic Resonance (1969)}\
  }\textbf {\bibinfo {volume} {58}},\ \bibinfo {pages} {442} (\bibinfo {year}
  {1984})}\BibitemShut {NoStop}%
\bibitem [{\citenamefont {Ruster}\ \emph {et~al.}(2016)\citenamefont {Ruster},
  \citenamefont {Schmegelow}, \citenamefont {Kaufmann}, \citenamefont
  {Warschburger}, \citenamefont {Schmidt-Kaler},\ and\ \citenamefont
  {Poschinger}}]{ruster2016b_field}%
  \BibitemOpen
  \bibfield  {author} {\bibinfo {author} {\bibfnamefont {T.}~\bibnamefont
  {Ruster}}, \bibinfo {author} {\bibfnamefont {C.}~\bibnamefont {Schmegelow}},
  \bibinfo {author} {\bibfnamefont {H.}~\bibnamefont {Kaufmann}}, \bibinfo
  {author} {\bibfnamefont {C.}~\bibnamefont {Warschburger}}, \bibinfo {author}
  {\bibfnamefont {F.}~\bibnamefont {Schmidt-Kaler}},\ and\ \bibinfo {author}
  {\bibfnamefont {U.}~\bibnamefont {Poschinger}},\ }\bibfield  {title}
  {\bibinfo {title} {A long-lived zeeman trapped-ion qubit},\ }\bibfield
  {journal} {\bibinfo  {journal} {Applied Physics B}\ }\textbf {\bibinfo
  {volume} {122}},\ \href {https://doi.org/10.1007/s00340-016-6527-4}
  {10.1007/s00340-016-6527-4} (\bibinfo {year} {2016})\BibitemShut {NoStop}%
\bibitem [{\citenamefont {Kang}\ \emph {et~al.}(2023)\citenamefont {Kang},
  \citenamefont {Campbell},\ and\ \citenamefont {Brown}}]{kang2023quantum}%
  \BibitemOpen
  \bibfield  {author} {\bibinfo {author} {\bibfnamefont {M.}~\bibnamefont
  {Kang}}, \bibinfo {author} {\bibfnamefont {W.~C.}\ \bibnamefont {Campbell}},\
  and\ \bibinfo {author} {\bibfnamefont {K.~R.}\ \bibnamefont {Brown}},\
  }\bibfield  {title} {\bibinfo {title} {Quantum error correction with
  metastable states of trapped ions using erasure conversion},\ }\href
  {https://doi.org/10.1103/PRXQuantum.4.020358} {\bibfield  {journal} {\bibinfo
   {journal} {PRX Quantum}\ }\textbf {\bibinfo {volume} {4}},\ \bibinfo {pages}
  {020358} (\bibinfo {year} {2023})}\BibitemShut {NoStop}%
\bibitem [{\citenamefont {Allcock}\ \emph {et~al.}(2021)\citenamefont
  {Allcock}, \citenamefont {Campbell}, \citenamefont {Chiaverini},
  \citenamefont {Chuang}, \citenamefont {Hudson}, \citenamefont {Moore},
  \citenamefont {Ransford}, \citenamefont {Roman}, \citenamefont {Sage},\ and\
  \citenamefont {Wineland}}]{allcock2021omg}%
  \BibitemOpen
  \bibfield  {author} {\bibinfo {author} {\bibfnamefont {D.}~\bibnamefont
  {Allcock}}, \bibinfo {author} {\bibfnamefont {W.}~\bibnamefont {Campbell}},
  \bibinfo {author} {\bibfnamefont {J.}~\bibnamefont {Chiaverini}}, \bibinfo
  {author} {\bibfnamefont {I.}~\bibnamefont {Chuang}}, \bibinfo {author}
  {\bibfnamefont {E.}~\bibnamefont {Hudson}}, \bibinfo {author} {\bibfnamefont
  {I.}~\bibnamefont {Moore}}, \bibinfo {author} {\bibfnamefont
  {A.}~\bibnamefont {Ransford}}, \bibinfo {author} {\bibfnamefont
  {C.}~\bibnamefont {Roman}}, \bibinfo {author} {\bibfnamefont
  {J.}~\bibnamefont {Sage}},\ and\ \bibinfo {author} {\bibfnamefont
  {D.}~\bibnamefont {Wineland}},\ }\bibfield  {title} {\bibinfo {title} {omg
  blueprint for trapped ion quantum computing with metastable states},\ }\href
  {https://doi.org/https://doi.org/10.1063/5.0069544} {\bibfield  {journal}
  {\bibinfo  {journal} {Applied Physics Letters}\ }\textbf {\bibinfo {volume}
  {119}},\ \bibinfo {pages} {214002} (\bibinfo {year} {2021})}\BibitemShut
  {NoStop}%
\bibitem [{\citenamefont {Yang}\ \emph {et~al.}(2022)\citenamefont {Yang},
  \citenamefont {Ma}, \citenamefont {Wu}, \citenamefont {Wang}, \citenamefont
  {Cao}, \citenamefont {Guo}, \citenamefont {Huang}, \citenamefont {Feng},
  \citenamefont {Zhou},\ and\ \citenamefont {Duan}}]{yang2022realizing}%
  \BibitemOpen
  \bibfield  {author} {\bibinfo {author} {\bibfnamefont {H.-X.}\ \bibnamefont
  {Yang}}, \bibinfo {author} {\bibfnamefont {J.-Y.}\ \bibnamefont {Ma}},
  \bibinfo {author} {\bibfnamefont {Y.-K.}\ \bibnamefont {Wu}}, \bibinfo
  {author} {\bibfnamefont {Y.}~\bibnamefont {Wang}}, \bibinfo {author}
  {\bibfnamefont {M.-M.}\ \bibnamefont {Cao}}, \bibinfo {author} {\bibfnamefont
  {W.-X.}\ \bibnamefont {Guo}}, \bibinfo {author} {\bibfnamefont {Y.-Y.}\
  \bibnamefont {Huang}}, \bibinfo {author} {\bibfnamefont {L.}~\bibnamefont
  {Feng}}, \bibinfo {author} {\bibfnamefont {Z.-C.}\ \bibnamefont {Zhou}},\
  and\ \bibinfo {author} {\bibfnamefont {L.-M.}\ \bibnamefont {Duan}},\
  }\bibfield  {title} {\bibinfo {title} {Realizing coherently convertible
  dual-type qubits with the same ion species},\ }\href
  {https://doi.org/https://doi.org/10.1038/s41567-022-01661-5} {\bibfield
  {journal} {\bibinfo  {journal} {Nature Physics}\ ,\ \bibinfo {pages} {1}}
  (\bibinfo {year} {2022})}\BibitemShut {NoStop}%
\bibitem [{\citenamefont {Roberts}\ \emph {et~al.}(2000)\citenamefont
  {Roberts}, \citenamefont {Taylor}, \citenamefont {Barwood}, \citenamefont
  {Rowley},\ and\ \citenamefont {Gill}}]{roberts2000observation}%
  \BibitemOpen
  \bibfield  {author} {\bibinfo {author} {\bibfnamefont {M.}~\bibnamefont
  {Roberts}}, \bibinfo {author} {\bibfnamefont {P.}~\bibnamefont {Taylor}},
  \bibinfo {author} {\bibfnamefont {G.~P.}\ \bibnamefont {Barwood}}, \bibinfo
  {author} {\bibfnamefont {W.~R.~C.}\ \bibnamefont {Rowley}},\ and\ \bibinfo
  {author} {\bibfnamefont {P.}~\bibnamefont {Gill}},\ }\bibfield  {title}
  {\bibinfo {title} {Observation of the
  ${}^{2}{S}_{1/2}{\ensuremath{-}}^{2}{F}_{7/2}$ electric octupole transition
  in a single ${}^{171}{\mathrm{yb}}^{+}$ ion},\ }\href
  {https://doi.org/10.1103/PhysRevA.62.020501} {\bibfield  {journal} {\bibinfo
  {journal} {Phys. Rev. A}\ }\textbf {\bibinfo {volume} {62}},\ \bibinfo
  {pages} {020501} (\bibinfo {year} {2000})}\BibitemShut {NoStop}%
\bibitem [{\citenamefont {Dietrich}\ \emph {et~al.}(2009)\citenamefont
  {Dietrich}, \citenamefont {Avril}, \citenamefont {Bowler}, \citenamefont
  {Kurz}, \citenamefont {Salacka}, \citenamefont {Shu},\ and\ \citenamefont
  {Blinov}}]{dietrich2009barium}%
  \BibitemOpen
  \bibfield  {author} {\bibinfo {author} {\bibfnamefont {M.~R.}\ \bibnamefont
  {Dietrich}}, \bibinfo {author} {\bibfnamefont {A.}~\bibnamefont {Avril}},
  \bibinfo {author} {\bibfnamefont {R.}~\bibnamefont {Bowler}}, \bibinfo
  {author} {\bibfnamefont {N.}~\bibnamefont {Kurz}}, \bibinfo {author}
  {\bibfnamefont {J.}~\bibnamefont {Salacka}}, \bibinfo {author} {\bibfnamefont
  {G.}~\bibnamefont {Shu}},\ and\ \bibinfo {author} {\bibfnamefont
  {B.}~\bibnamefont {Blinov}},\ }\bibfield  {title} {\bibinfo {title} {Barium
  ions for quantum computation},\ }in\ \href
  {https://doi.org/10.1063/1.3122286} {\emph {\bibinfo {booktitle} {AIP
  Conference Proceedings}}},\ Vol.\ \bibinfo {volume} {1114}\ (\bibinfo
  {organization} {American Institute of Physics},\ \bibinfo {year} {2009})\
  pp.\ \bibinfo {pages} {25--30}\BibitemShut {NoStop}%
\bibitem [{\citenamefont {Christensen}\ \emph {et~al.}(2020)\citenamefont
  {Christensen}, \citenamefont {Hucul}, \citenamefont {Campbell},\ and\
  \citenamefont {Hudson}}]{christensen2020high}%
  \BibitemOpen
  \bibfield  {author} {\bibinfo {author} {\bibfnamefont {J.~E.}\ \bibnamefont
  {Christensen}}, \bibinfo {author} {\bibfnamefont {D.}~\bibnamefont {Hucul}},
  \bibinfo {author} {\bibfnamefont {W.~C.}\ \bibnamefont {Campbell}},\ and\
  \bibinfo {author} {\bibfnamefont {E.~R.}\ \bibnamefont {Hudson}},\ }\bibfield
   {title} {\bibinfo {title} {High-fidelity manipulation of a qubit enabled by
  a manufactured nucleus},\ }\href {https://doi.org/10.1038/s41534-020-0265-5}
  {\bibfield  {journal} {\bibinfo  {journal} {npj Quantum Information}\
  }\textbf {\bibinfo {volume} {6}},\ \bibinfo {pages} {35} (\bibinfo {year}
  {2020})}\BibitemShut {NoStop}%
\bibitem [{\citenamefont {Lide}(2004)}]{lide2004crc}%
  \BibitemOpen
  \bibfield  {author} {\bibinfo {author} {\bibfnamefont {D.~R.}\ \bibnamefont
  {Lide}},\ }\href@noop {} {\emph {\bibinfo {title} {CRC handbook of chemistry
  and physics}}},\ Vol.~\bibinfo {volume} {85}\ (\bibinfo  {publisher} {CRC
  press},\ \bibinfo {year} {2004})\BibitemShut {NoStop}%
\bibitem [{\citenamefont {White}\ \emph {et~al.}(2022)\citenamefont {White},
  \citenamefont {Low}, \citenamefont {de~Sereville}, \citenamefont {Day},
  \citenamefont {Greenberg}, \citenamefont {Rademacher},\ and\ \citenamefont
  {Senko}}]{white2022isotope}%
  \BibitemOpen
  \bibfield  {author} {\bibinfo {author} {\bibfnamefont {B.~M.}\ \bibnamefont
  {White}}, \bibinfo {author} {\bibfnamefont {P.~J.}\ \bibnamefont {Low}},
  \bibinfo {author} {\bibfnamefont {Y.}~\bibnamefont {de~Sereville}}, \bibinfo
  {author} {\bibfnamefont {M.~L.}\ \bibnamefont {Day}}, \bibinfo {author}
  {\bibfnamefont {N.}~\bibnamefont {Greenberg}}, \bibinfo {author}
  {\bibfnamefont {R.}~\bibnamefont {Rademacher}},\ and\ \bibinfo {author}
  {\bibfnamefont {C.}~\bibnamefont {Senko}},\ }\bibfield  {title} {\bibinfo
  {title} {Isotope-selective laser ablation ion-trap loading of
  $^{137}\mathrm{Ba}^{+}$ using a ${\mathrm{bacl}}_{2}$ target},\ }\href
  {https://doi.org/10.1103/PhysRevA.105.033102} {\bibfield  {journal} {\bibinfo
   {journal} {Phys. Rev. A}\ }\textbf {\bibinfo {volume} {105}},\ \bibinfo
  {pages} {033102} (\bibinfo {year} {2022})}\BibitemShut {NoStop}%
\bibitem [{\citenamefont {Martinez}\ \emph {et~al.}(2016)\citenamefont
  {Martinez}, \citenamefont {Monz}, \citenamefont {Nigg}, \citenamefont
  {Schindler},\ and\ \citenamefont {Blatt}}]{martinez2016compiling}%
  \BibitemOpen
  \bibfield  {author} {\bibinfo {author} {\bibfnamefont {E.~A.}\ \bibnamefont
  {Martinez}}, \bibinfo {author} {\bibfnamefont {T.}~\bibnamefont {Monz}},
  \bibinfo {author} {\bibfnamefont {D.}~\bibnamefont {Nigg}}, \bibinfo {author}
  {\bibfnamefont {P.}~\bibnamefont {Schindler}},\ and\ \bibinfo {author}
  {\bibfnamefont {R.}~\bibnamefont {Blatt}},\ }\bibfield  {title} {\bibinfo
  {title} {Compiling quantum algorithms for architectures with multi-qubit
  gates},\ }\href {https://doi.org/10.1088/1367-2630/18/6/063029} {\bibfield
  {journal} {\bibinfo  {journal} {New Journal of Physics}\ }\textbf {\bibinfo
  {volume} {18}},\ \bibinfo {pages} {063029} (\bibinfo {year}
  {2016})}\BibitemShut {NoStop}%
\bibitem [{\citenamefont {Hagberg}\ \emph {et~al.}(2008)\citenamefont
  {Hagberg}, \citenamefont {Schult},\ and\ \citenamefont {Swart}}]{networkx}%
  \BibitemOpen
  \bibfield  {author} {\bibinfo {author} {\bibfnamefont {A.}~\bibnamefont
  {Hagberg}}, \bibinfo {author} {\bibfnamefont {D.}~\bibnamefont {Schult}},\
  and\ \bibinfo {author} {\bibfnamefont {P.}~\bibnamefont {Swart}},\ }\bibfield
   {title} {\bibinfo {title} {Exploring network structure, dynamics, and
  function using networkx},\ }in\ \href@noop {} {\emph {\bibinfo {booktitle}
  {Proceedings of the 7th Python in Science Conference (SciPy2008)}}}\
  (\bibinfo {year} {2008})\ p.\ \bibinfo {pages} {11–15}\BibitemShut
  {NoStop}%
\bibitem [{\citenamefont {Harty}(2017)}]{atomic-physics}%
  \BibitemOpen
  \bibfield  {author} {\bibinfo {author} {\bibfnamefont {T.}~\bibnamefont
  {Harty}},\ }\href {https://github.com/OxfordIonTrapGroup/atomic_physics/}
  {\bibinfo {title} {Atomic physics}} (\bibinfo {year} {2017})\BibitemShut
  {NoStop}%
\bibitem [{\citenamefont {Wineland}\ \emph {et~al.}(2003)\citenamefont
  {Wineland}, \citenamefont {Barrett}, \citenamefont {Britton}, \citenamefont
  {Chiaverini}, \citenamefont {DeMarco}, \citenamefont {Itano}, \citenamefont
  {Jelenković}, \citenamefont {Langer}, \citenamefont {Leibfried},
  \citenamefont {Meyer}, \citenamefont {Rosenband},\ and\ \citenamefont
  {Schätz}}]{wineland2003quantum}%
  \BibitemOpen
  \bibfield  {author} {\bibinfo {author} {\bibfnamefont {D.}~\bibnamefont
  {Wineland}}, \bibinfo {author} {\bibfnamefont {M.}~\bibnamefont {Barrett}},
  \bibinfo {author} {\bibfnamefont {J.}~\bibnamefont {Britton}}, \bibinfo
  {author} {\bibfnamefont {J.}~\bibnamefont {Chiaverini}}, \bibinfo {author}
  {\bibfnamefont {B.}~\bibnamefont {DeMarco}}, \bibinfo {author} {\bibfnamefont
  {W.}~\bibnamefont {Itano}}, \bibinfo {author} {\bibfnamefont
  {B.}~\bibnamefont {Jelenković}}, \bibinfo {author} {\bibfnamefont
  {C.}~\bibnamefont {Langer}}, \bibinfo {author} {\bibfnamefont
  {D.}~\bibnamefont {Leibfried}}, \bibinfo {author} {\bibfnamefont
  {V.}~\bibnamefont {Meyer}}, \bibinfo {author} {\bibfnamefont
  {T.}~\bibnamefont {Rosenband}},\ and\ \bibinfo {author} {\bibfnamefont
  {T.}~\bibnamefont {Schätz}},\ }\bibfield  {title} {\bibinfo {title} {Quantum
  information processing with trapped ions},\ }\bibfield  {journal} {\bibinfo
  {journal} {Phil. Trans. R. Soc. A}\ }\textbf {\bibinfo {volume} {361}},\
  \href {https://doi.org/https://doi.org/10.1098/rsta.2003.1205}
  {https://doi.org/10.1098/rsta.2003.1205} (\bibinfo {year} {2003})\BibitemShut
  {NoStop}%
\bibitem [{\citenamefont {Harty}(2013)}]{harty_thesis}%
  \BibitemOpen
  \bibfield  {author} {\bibinfo {author} {\bibfnamefont {T.}~\bibnamefont
  {Harty}},\ }\emph {\bibinfo {title} {High-Fidelity Microwave-Driven Quantum
  Logic in Intermediate-Field $^{43}Ca^+$}},\ \href
  {https://ora.ox.ac.uk/objects/uuid:55264c2d-bb42-4439-bf49-731b9f66de74}
  {Ph.D. thesis},\ \bibinfo  {school} {University of Oxford} (\bibinfo {year}
  {2013})\BibitemShut {NoStop}%
\bibitem [{\citenamefont {Wu}\ \emph {et~al.}(2022)\citenamefont {Wu},
  \citenamefont {Kolkowitz}, \citenamefont {Puri},\ and\ \citenamefont
  {Thompson}}]{wu2022erasure}%
  \BibitemOpen
  \bibfield  {author} {\bibinfo {author} {\bibfnamefont {Y.}~\bibnamefont
  {Wu}}, \bibinfo {author} {\bibfnamefont {S.}~\bibnamefont {Kolkowitz}},
  \bibinfo {author} {\bibfnamefont {S.}~\bibnamefont {Puri}},\ and\ \bibinfo
  {author} {\bibfnamefont {J.~D.}\ \bibnamefont {Thompson}},\ }\bibfield
  {title} {\bibinfo {title} {Erasure conversion for fault-tolerant quantum
  computing in alkaline earth rydberg atom arrays},\ }\href
  {https://doi.org/https://doi.org/10.1038/s41467-022-32094-6} {\bibfield
  {journal} {\bibinfo  {journal} {Nature communications}\ }\textbf {\bibinfo
  {volume} {13}},\ \bibinfo {pages} {4657} (\bibinfo {year}
  {2022})}\BibitemShut {NoStop}%
\bibitem [{\citenamefont {Monz}(2011)}]{monz_thesis}%
  \BibitemOpen
  \bibfield  {author} {\bibinfo {author} {\bibfnamefont {T.}~\bibnamefont
  {Monz}},\ }\emph {\bibinfo {title} {Quantum information processing beyond ten
  ion-qubits}},\ \href@noop {} {Ph.D. thesis},\ \bibinfo  {school} {University
  of Innsbruck} (\bibinfo {year} {2011})\BibitemShut {NoStop}%
\bibitem [{\citenamefont {Hughes}(2021)}]{hughes_thesis}%
  \BibitemOpen
  \bibfield  {author} {\bibinfo {author} {\bibfnamefont {A.}~\bibnamefont
  {Hughes}},\ }\emph {\bibinfo {title} {Benchmarking Memory and Logic Gates for
  Trapped-Ion Quantum Computing}},\ \href@noop {} {Ph.D. thesis},\ \bibinfo
  {school} {University of Oxford} (\bibinfo {year} {2021})\BibitemShut
  {NoStop}%
\bibitem [{\citenamefont {Hedemann}(2013)}]{hedemann2013hyperspherical}%
  \BibitemOpen
  \bibfield  {author} {\bibinfo {author} {\bibfnamefont {S.~R.}\ \bibnamefont
  {Hedemann}},\ }\href@noop {} {\bibinfo {title} {Hyperspherical
  parameterization of unitary matrices}} (\bibinfo {year} {2013}),\ \Eprint
  {https://arxiv.org/abs/1303.5904} {arXiv:1303.5904 [quant-ph]} \BibitemShut
  {NoStop}%
\end{thebibliography}%

\newpage\hbox{}\thispagestyle{empty}\newpage
\newpage
\onecolumngrid

\appendix

\begin{center}
\textbf{Appendix}     
\end{center}

In the following sections, we provide details about results described in the main text, including explicit decompositions of gates drawn as pulse sequences in Figures \ref{fig:all_to_all}, \ref{fig:BV}, \ref{fig:stab_meas} and \ref{fig:single_qubit_gate}, along with possible decompositions for other standard gates. We describe the error model used for the bit-flip repetition code for $n=2$, along with additional plots. We also discuss the utility of $n>2$ virtual qubits in the context of random quantum circuits and error correction.

In order to make the Appendix more readable, let's revisit the notations used in the main text. The $d$ levels being used for computation inside an ion are labeled in increasing order of their energies by $\alpha = 0,1,...,d-1$, and an encoding map $\mathcal{M}$ assigns each level as a computational basis state in an $n=\log_2 d$ qubit Hilbert space. For example, we define $\mathcal{M}_1$ for $n=2$ so that $\mathcal{M}_1(\ket{0})=\ket{00}$, $\mathcal{M}_1(\ket{1})=\ket{01}$, $\mathcal{M}_1(\ket{2})=\ket{10}$ and $\mathcal{M}_1(\ket{3})=\ket{11}$. The other map we use is $\mathcal{M}_2$ for $n=2$ so that $\mathcal{M}_2(\ket{0})=\ket{00}$, $\mathcal{M}_1(\ket{1})=\ket{11}$, $\mathcal{M}_2(\ket{2})=\ket{01}$ and $\mathcal{M}_1(\ket{3})=\ket{10}$. The ions are labeled by indices $i,j$, and the total number of ions is $L$. The total number of qubits on all ions is $N$, so if all the ions have $n$ virtual qubits, $N=nL$. When describing native gates,  we refer to the states $\alpha$, which are independent of the map used. Intra-ion gates $R_{i,\alpha \beta}(\theta,\phi)$ act on states $\ket{\alpha}$ and $\ket{\beta}$ inside the ith ion, such that $R_{i,\alpha \beta}(\theta,\phi)=\cos(\theta) \ket{\alpha}_i\bra{\alpha}_i+\cos(\theta) \ket{\beta}_i\bra{\beta}_i - ie^{-i\phi}\sin(\theta) \ket{\alpha}_i\bra{\beta}_i- ie^{i\phi}\sin(\theta) \ket{\beta}_i\bra{\alpha}_i$.  Inter-ion MS gate is denoted by $\mathrm{MS}_{\{\alpha_i \beta_i\}\{\alpha_j \beta_j\}}(J)$, acting between ions i and j, and is explicitly given  by  $U=\exp(-iJ_{ij}(\ket{\alpha}_i\otimes \ket{\alpha}_j\bra{\beta}_i\otimes \bra{\beta}_j+\ket{\alpha}_i\otimes \ket{\beta}_j\bra{\beta}_i\otimes \bra{\alpha}_j+\mathrm{h.c.}))$. 

\section{Intra-ion gate decompositions}\label{sec:app_intra-ion}
Table \ref{tab:m1_all_intra} shows possible decompositions of various standard one and two qubit gates implemented using the native intra-ion gates in an ion with two virtual qubits ($n=2$), using the encoding map $\mathcal{M}_1$ defined in Figure \ref{fig:intro_fig}, and at the beginning of the Appendix. Since all $R_{\alpha \beta}$ are allowed, the decompositions under a different map such as $\mathcal{M}_2$ can be easily found by replacing the corresponding gates. For example, from Figure \ref{fig:intro_fig}, we can see that $R_{01}$ under $\mathcal{M}_1$ can be replaced by $R_{03}$ under $\mathcal{M}_2$. After performing all the replacements, the number of native gates should remain the same, and hence the expected fidelity should be similar (ignoring effects such as magnetic noise). Further, the number of gates for a single qubit gate is independent of the particular virtual qubit, a result which does not hold when all to all connectivity is absent. Equivalently, with all to all connectivity, all virtual qubits inside an ion are equivalent in terms of the number of native gates needed to perform a single/multi-qubit intra-ion gate. \\

The decompositions are performed using an approach which applies successive $R_{\alpha \beta}$ (assuming all $\alpha \neq \beta$ allowed) of the form $ R_{1 2}\dots R_{d-3 d-2}\dots R_{1 2}R_{0 1} R_{d-2 d-1}\dots R_{1 2}R_{0 1}$, with the parameters of each gate chosen such that when multiplied with $U^{\dagger}$ ($U$ being the target unitary), the resulting columns of the product become either exactly equal to those corresponding to the identity operation, or equivalent upto a phase. The exact parameters to choose have been outlined in \cite{low_2020_practical}, but just by the form, we can conclude that the number of gates required to make the product with $U^{\dagger}$ a diagonal phase matrix is $n_g=d(d-1)/2$. Further, the diagonal phases can be made uniform such that the resultant decomposition is equivalent to $U$, by applying two pulses with specially chosen parameters for each $\ket{\alpha}$. Suppose we want to eliminate a phase $e^{i\theta}$ on the state the $\ket{d-1}$. We could apply $R_{\alpha d-1}(\pi/2,\theta+\pi)R_{\alpha d-1}(\pi/2,0)$ for any $\alpha \neq d-1$, to cancel out the phase, leaving just $\ket{d-1}$. Even if all the diagonal phases of the full decomposition are different, in principle we only need to make sure all of them are the same, which only requires $2(d-1)$ pulses, making the total number of pulses required for an arbitrary unitary to be $n_g= d(d-1)/2+2(d-1)$. 

Single qubit gates have more structure and often end up requiring fewer gates. A single qubit gate of the form $U=U_{2\times 2} \otimes \mathbf{I}_{\frac{d}{2} \times \frac{d}{2}} $ needs only $d/2$ non-trivial gates to make the product of the form $ R_{1 2}\dots R_{d-3 d-2}\dots R_{1 2}R_{0 1} R_{d-2 d-1}\dots R_{1 2}R_{0 1}$ equivalent to $U$, upto phases on the diagonal, implying a minimum of $d/2$ native gates needed. The phases can similarly be cancelled with another $2(d-1)$ gates, thus an arbitrary single qubit gate can be decomposed using $d/2+2(d-1)$ gates. This argument can be generalized easily to when $U_{2\times 2}$ acts on a different virtual qubit. For $n=2$ $(d=4)$, the minimum number of native gates required to implement a single qubit gate is thus $n_g=2$, which we see for many gates in Table \ref{tab:m1_all_intra}.


\begin{table}[h]
\begin{center}
\begin{tabular}{ |c|c|c| } 
 \hline
 \textit{No.} & \textit{Gate} & \textit{Decomposition} \\ 
 \hline
 \hline
 \rule{0pt}{4ex}    
 1 & $X\otimes \mathbf{I}$ &  $R_{23}(\frac{\pi}{2},-\frac{\pi}{2})$  $R_{01}(\frac{\pi}{2},-\frac{\pi}{2})$  \\[4pt] 
 \hline
 \rule{0pt}{4ex}    
 2 & $\mathbf{I}\otimes X$ & $R_{12}(\frac{\pi}{2},-\frac{\pi}{2})$  $R_{03}(\frac{\pi}{2},-\frac{\pi}{2})$ \\[4pt]
 
 \hline
 \rule{0pt}{4ex}
 3 & $H\otimes \mathbf{I}$ &  $R_{23}(\frac{5\pi}{4},\frac{\pi}{2})$  $R_{12}(\frac{5\pi}{4},\frac{\pi}{2})$  $R_{03}(\pi,0)$ \\[4pt]
 \hline 
 \rule{0pt}{4ex}
 4 & $ \mathbf{I} \otimes H$ &  $R_{12}(\frac{5\pi}{4},\frac{\pi}{2})$  $R_{03}(\frac{5\pi}{4},\frac{\pi}{2})$  $R_{01}(\pi,0)$  \\[4pt]
 \hline
 \rule{0pt}{4ex} 
 5 & $e^{-i\theta X} \otimes \mathbf{I}$ &  $R_{23}(\theta,0)$  $R_{01}(\theta,0)$  \\[4pt] 
 \hline
 \rule{0pt}{4ex} 
 6 & $\mathbf{I} \otimes e^{-i\theta X} $ &  $R_{12}(\theta,0)$  $R_{03}(\theta,0)$  \\[4pt] 
 \hline
 \rule{0pt}{4ex} 
 7 & $e^{-i\theta Y} \otimes \mathbf{I}$ & $R_{23}(\theta,\frac{\pi}{2})$  $R_{01}(\theta,\frac{\pi}{2})$ \\[4pt] 
 \hline
 \rule{0pt}{4ex} 
 8 & $\mathbf{I} \otimes e^{-i\theta Y} $ & $R_{12}(\theta,\frac{\pi}{2})$  $R_{03}(\theta,\frac{\pi}{2})$  \\[4pt] 
 \hline 
 \rule{0pt}{4ex} 
  9 & $e^{i h Z} \otimes \mathbf{I}$ & $R_{01}(\frac{\pi}{2},\frac{\pi}{2}-h)$ $R_{01}(\frac{\pi}{2},-\frac{\pi}{2})$  $R_{23}(\frac{\pi}{2},\frac{\pi}{2}-h)$ $R_{23}(\frac{\pi}{2},-\frac{\pi}{2})$  \\[4pt]
 \hline
 \rule{0pt}{4ex}
 10 & $\mathbf{I} \otimes e^{i h  Z} $ & $R_{01}(\frac{\pi}{2},\frac{\pi}{2}-h)$ $R_{01}(\frac{\pi}{2},-\frac{\pi}{2})$  $R_{23}(-\frac{\pi}{2},\frac{\pi}{2}+h)$ $R_{23}(\frac{\pi}{2},\frac{\pi}{2})$  \\[4pt]
\hline
 \rule{0pt}{4ex}    
 11 & CNOT &  
  $e^{i3\pi/4}$ $R_{23}(\frac{\pi}{2},\frac{5\pi}{4})$ $R_{02}(\frac{\pi}{2},\pi)$ $R_{01}(\frac{\pi}{2},\frac{3\pi}{2})$  $R_{02}(\frac{\pi}{2},\frac{\pi}{4})$ $R_{23}(\frac{\pi}{2},\frac{\pi}{2})$ $R_{02}(\frac{\pi}{2},\frac{\pi}{4})$ \\[4pt]
 \hline
 \rule{0pt}{4ex}    
 11 & CNOT &  
  $e^{i3\pi/4}$ $R_{03}(\frac{3\pi}{2},0)$ $R_{12}(\frac{3\pi}{2},\frac{5\pi}{4})$ $R_{03}(\frac{3\pi}{2},\frac{3\pi}{4})$  $R_{12}(-\frac{\pi}{2},0)$ $R_{23}(\frac{3\pi}{2},\pi)$ \\[4pt]
 \hline
 \rule{0pt}{4ex} 
 12 & $e^{-i J  Z_1Z_2}$ & $R_{02}(\frac{\pi}{2},2J)$ $R_{02}(-\frac{\pi}{2},0)$  $R_{01}(\frac{\pi}{2},\frac{\pi}{2}-J)$ $R_{01}(\frac{\pi}{2},-\frac{\pi}{2})$ $R_{23}(-\frac{\pi}{2},-\frac{\pi}{2}+J)$ $R_{23}(\frac{\pi}{2},\frac{\pi}{2})$ \\[4pt]
 \hline 
 \rule{0pt}{4ex} 
 13 & $e^{-i J  X_1X_2}$ & $R_{02}(J,0)$  $R_{13}(J,0)$ \\[4pt] 
 \hline 
\end{tabular}
\end{center}
\caption{Decomposition of some standard gates in terms of native intra ion gates for a single ion, with the number of virtual qubits $n=2$, using the encoding map $\mathcal{M}_1$(Figure \ref{fig:intro_fig}), and assuming all to all connectivity such that the allowed gates are $R_{01},R_{02},R_{03},R_{12},R_{13},R_{23}$. Some of the gates used in Figures \ref{fig:all_to_all} and \ref{fig:BV} appear here.}
\label{tab:m1_all_intra}
\end{table}

\subsection{The role of limited connectivity for intra-ion gates}
In Sec \ref{sec:lim-conn}, we pointed out that the nature of gate decompositions can be different when connectivity between different states in an ion is limited, which would lead to differing performance of physically equivalent gates acting on different virtual qubits. Here we provide the details of the decompositions in Figure \ref{fig:single_qubit_gate}. Table \ref{tab:decomp_lim} shows possible decompositions in terms of the native gates using encoding maps $\mathcal{M}_1$ and $\mathcal{M}_2$, where only the gates shown in Figure $\ref{fig:intro_fig}$ are allowed ($R_{01},R_{02},R_{23}$). We see that for a single qubit gate, the number of native gates can be different based on the encoding map, and also on the \textit{specific qubit} the gate acts on, breaking the equivalence when all states are connected by $R_{\alpha \beta}$.

How does the number of native gates scale with the number of qubits/levels ($n/d$) when connectivity is limited? To address this question, consider a minimal connectivity universal gate set $R_{0 1},R_{1 2},\dots ,R_{d-2 d-1}$ (assuming arbitrary possible $\theta,\phi$ for each gate). But the form of the decomposition written in the previous section $ R_{1 2}\dots R_{d-3 d-2}\dots R_{1 2}R_{0 1} R_{d-2 d-1}\dots R_{1 2}R_{0 1}$, for an arbitrary target unitary $U$, uses precisely these gates, implying that the maximum number of native gates needed is be the same. When the native gate set is different, the decomposition form is still equivalent to traversing the connectivity graph, and thus the number of maximum gates is still the same. What about the minimum number for single qubit gates? Again, if the $d/2$ non-trivial gates are part of the gate set, $d/2$ gates will suffice. So infact, while the number of native gates could be different depending on the connectivity and the map, the \textit{minimum} and \textit{maximum} number of gates for a given $d$ will always be the same, as long as the native gate set is universal.

\begin{table}[h]
    \centering
    \begin{tabular}{|c|c|c|c|}
    \hline
    Gate & Map\hspace{5pt}     & Decomposition (limited) \hspace{5pt} & Decomposition (all to all)   \\
    \hline
     \rule{0pt}{4ex} 
     $e^{-i\theta X} \otimes \mathbf{I}$ & $\mathcal{M}_1$ &  $R_{01}(-\frac{\pi}{2},-\frac{\pi}{2})$ $R_{02}(-\frac{\pi}{2},\pi)$ $R_{23}(\theta,0)$ $R_{01}(\theta,0)$ $R_{02}(\frac{\pi}{2},\frac{\pi}{2})$ $R_{01}(\frac{\pi}{2},\frac{\pi}{2})$ & $R_{12}(\theta,0)$   $R_{03}(\theta,0)$ \\[4pt]
     \hline

     \rule{0pt}{4ex} 
    $\mathbf{I} \otimes e^{-i\theta X} $  & $\mathcal{M}_1$ &  $R_{23}(\theta,0)$   $R_{01}(\theta,0)$ & $R_{23}(\theta,0)$   $R_{01}(\theta,0)$ \\[4pt]
     \hline 
     \rule{0pt}{4ex}      
     $e^{-i\theta X} \otimes \mathbf{I}$ & $\mathcal{M}_2$ &  $R_{02}(-\frac{\pi}{2},-\frac{\pi}{2})$ $R_{01}(\theta,0)$ $R_{23}(\theta,\pi)$ $R_{02}(\frac{\pi}{2},-\frac{\pi}{2})$ & $R_{23}(\theta,0)$   $R_{01}(\theta,0)$ \\[4pt]
     \hline
     \rule{0pt}{4ex}      
     $\mathbf{I} \otimes e^{-i\theta X} $ & $\mathcal{M}_2$ &  $R_{23}(\frac{\pi}{2},\frac{\pi}{2})$  $R_{02}(-\frac{\pi}{2},-\frac{\pi}{2})$ $R_{01}(\theta,0)$ $R_{02}(\frac{\pi}{2},-\frac{\pi}{2})$ $R_{23}(\frac{\pi}{2},-\frac{\pi}{2})$ $R_{02}(\theta,0)$ & $R_{13}(\theta,0)$   $R_{02}(\theta,0)$ \\[4pt]
     \hline
        \end{tabular}
    \caption{Decomposition of standard gates under limited connectivity (as sketched in Figure \ref{fig:intro_fig}), and shown as a pulse sequence in Figure \ref{fig:single_qubit_gate}. When all to all connectivity is assumed (so that all $R_{\alpha \beta}$ are allowed), the number of native gates decreases (minimum being 2), and the number doesn't depend on \textit{which} virtual qubit the gate is applied to, as also visible in Table \ref{tab:m1_all_intra}.  }
    \label{tab:decomp_lim}
\end{table}

\newpage

\section{Inter ion gates}\label{sec:app_inter_ion}
The native inter-ion gate we consider is simply the MS gate used in traditional trapped ion setups, with the possibility of coupling different pairs of levels on each ion. Since $\mathrm{MS}_{\{\alpha_i \beta_i\}\{\alpha_j \beta_j\}}(J_{ij})=\exp(-iJ_{ij}(\ket{\alpha}_i\otimes \ket{\alpha}_j\bra{\beta}_i\otimes \bra{\beta}_j+\ket{\alpha}_i\otimes \ket{\beta}_j\bra{\beta}_i\otimes \bra{\alpha}_j+\mathrm{h.c.}))$, when viewed as a term $H_{ij}$ in a interacting Hamiltonian $H=\sum H_{ij}$ such that $U=\exp(-J_{ij}H_{ij})$, because of the projective nature of the individual terms, we can easily see that $[H_{ij},H_{kl}]=0$, $\forall i\neq j,k \neq l$ . This is relevant for stabilizer Hamiltonians discussed in Sec  \ref{sec:inter_ion}.

Next we consider two ions and the decomposition of some inter-ion gates in terms of the native gates, such as those drawn in Figures \ref{fig:BV} and \ref{fig:stab_meas}. We start out with the simplest case when only one of the ions has two qubits, while the other has a single qubit, and then consider the case when both the ions have two virtual qubits. The decomposition is obtained by writing down an ansatz $U_{\mathrm{an}}$, repeated with indepdentent parameters in layers \cite{martinez2016compiling}, which numerically minimizes $(\frac{1}{2^N}|\mathrm{Tr}[U^{\dagger}U_{\mathrm{an}}]|-1)^2$, in order to find a unitary which is equivalent to the target unitary $U$, upto a global phase.  We assume all $R_{\alpha \beta}$ gates are allowed as intra-ion gates on both ions, along with all $\mathrm{MS}_{\{\alpha_i \beta_i\}\{\alpha_j \beta_j\}}$ gates. 

\subsection{Inter ion gates between an ion with $n=2$ and a ion with $n=1$}

We consider three qubits, such that the first two qubits belong to ion 1, and the third qubit belongs to ion 2. Table \ref{tab:nv2_nv1_all} shows the decompositions for some of the gates used in Figures \ref{fig:BV} and \ref{fig:stab_meas}, along with additional inter-ion gates. We see that there are some gates which only require a single MS gate, as compared to two MS gates that might be needed if implemented using three ions. The decompositions are obtained numerically using ansatz such as $
    U_{\mathrm{an}}(\{\alpha\},\{\beta\},\{\theta\},\{\phi\},J)= \mathrm{MS}_{\{\alpha \beta\}\{01\}}(J)
    \times R_{1,\alpha_1\beta_1}(\theta_{\alpha_1\beta_1},\phi_{\alpha_1\beta_1}) \times R_{2,0,1}(\theta_{2}\phi_{2})  $ and $
    U_{\mathrm{an}}(\{\theta\},\{\phi\},J)= \mathrm{MS}_{\{01\}\{01\}}(J)
    \times R_{1,01}(\theta_{01},\phi_{01}) \times R_{1,03}(\theta_{03},\phi_{03}) \times R_{1,12}(\theta_{12},\phi_{12}) \times R_{2,01}(\theta_{2},\phi_{2})  $ which are repeated as layers with independent parameters until varying the parameters leads to the target unitary $U$, with the cost function being zero within a tolerance $\epsilon$.

\begin{table}[h]
\begin{center}
\begin{tabular}{ |c|c|c| } 
 \hline
 \textit{No.} & \textit{Gate}  & \textit{Decomposition} \\ 
 \hline
 \hline
 \rule{0pt}{4ex}    
 1 & ${}_{1,1}\mathrm{CNOT}_{2,1}$  & 
  $R_{1,23}(\frac{\pi}{2},0)$  $\mathrm{MS}_{\{23\}\{01\}}(-\frac{\pi}{2})$ \\[4pt]
 \hline
\rule{0pt}{4ex}    
 2 & ${}_{1,2}\mathrm{CNOT}_{2,1}$  & $R_{1,13}(\frac{\pi}{2},0)$  $\mathrm{MS}_{\{13\}\{01\}}(-\frac{\pi}{2})$  \\[4pt]
 \hline
\rule{0pt}{4ex}    
 3 & (${}_{1,2}\mathrm{CNOT}_{2,1} $) (${}_{1,1}\mathrm{CNOT}_{2,1}$) &  $R_{1,12}(\frac{\pi}{2},0)$  $\mathrm{MS}_{\{12\}\{01\}}(-\frac{\pi}{2})$  \\[4pt]
 \hline

 \rule{0pt}{4ex}    
 4 & ($\mathbf{I}_{4x4} \otimes H$) $ {}_{2,1}\mathrm{CNOT}_{1,1}$ ($\mathbf{I}_{4x4} \otimes H$)  &  $R_{1,01}(1.5\pi,\pi)$ $R_{1,03}(1.75\pi,1.0 \pi)$ $R_{1,12}(1.25\pi,0)$ $R_{2,01}(0.5\pi,\pi)$  $\times$ \\[4pt] 
  &   & $\mathrm{MS}_{\{01\}\{01\}}(0.5\pi)$ $R_{1,13}(0.75\pi,1.5\pi)$  $R_{1,23}(\pi,0)$ $R_{1,02}(0.25\pi,1.5\pi)$\\[4pt] 
 \hline 
 \rule{0pt}{4ex}    
 4 & ($\mathbf{I}_{4x4} \otimes H$) $ {}_{2,1}\mathrm{CNOT}_{1,2}$ ($\mathbf{I}_{4x4} \otimes H$)  &  $R_{1,34}(0.25\pi,0.5\pi)$ $R_{1,13}(0.5\pi,0)$ $R_{1,03}(0.25\pi,0)$ $R_{2,01}(1.5\pi,1.5\pi)$  $\times$ \\[4pt] 
  &   & $\mathrm{MS}_{\{13\}\{01\}}(0.5\pi)$ $R_{1,01}(0.75\pi,0.5\pi)$  $R_{1,23}(1.25\pi,1.5\pi)$ $R_{2,01}(1.5\pi,0.5\pi)$\\[4pt] 
 \hline  
\rule{0pt}{4ex}    
 5 & ($\mathbf{I}_{4x4} \otimes H$) (${}_{2,1}\mathrm{CNOT}_{1,1}$) (${}_{2,1}\mathrm{CNOT}_{1,2} $) ($\mathbf{I}_{4x4} \otimes H$) &  $R_{1,01}(\pi,0)$ $R_{1,03}(0.25\pi,0.5 \pi)$ $R_{1,12}(0.75\pi,1.5\pi)$ $\mathrm{MS}_{\{01\}\{01\}}(0.5\pi)$ $\times$ \\[4pt] 
  &   & $R_{1,01}(0.5\pi,0)$  $R_{1,03}(0.25\pi,0.5 \pi)$ $R_{1,12}(1.25\pi,0.5\pi)$\\[4pt] 
 \hline 
 
 \rule{0pt}{4ex} 
 6 & $e^{-i J  X_{1,1}X_{2,1}}$  & $\mathrm{MS}_{\{02\}\{01\}}(J)$ $\mathrm{MS}_{\{13\}\{01\}}(J)$\\[4pt] 
 \hline 
 \rule{0pt}{4ex} 
 7 & $e^{-i J  X_{1,2}X_{2,1}}$  & $\mathrm{MS}_{\{01\}\{01\}}(J)$ $\mathrm{MS}_{\{23\}\{01\}}(J)$ \\[4pt] 
 \hline
 \rule{0pt}{4ex} 
 8 & $e^{-i \frac{\pi}{4}  (X_{1,1}X_{2,1}+X_{1,2}X_{2,1})}$  & $R_{1,01}(\pi,0)$ $R_{1,03}(1.25\pi,1.5 \pi)$ $R_{1,12}(0.25\pi,1.5\pi)$ $\mathrm{MS}_{\{01\}\{01\}}(0.5\pi)$ $\times$ \\[4pt] 
  &   &  $R_{1,12}(1.75\pi,0.5\pi)$ $R_{1,03}(0.75\pi,0.5 \pi)$\\[4pt] 
 \hline 
 \rule{0pt}{4ex} 
 9 & $e^{-i J  X_{1,1}X_{1,2}X_{2,1}}$  & $R_{1,23}(\frac{3\pi}{2},\frac{\pi}{2})$ $\mathrm{MS}_{\{12\}\{01\}}(J)$ $R_{1,23}(\frac{\pi}{2},\frac{\pi}{2})$ $\mathrm{MS}_{\{02\}\{01\}}(J)$ \\[4pt] 
 \hline  
\end{tabular}
\end{center}
\caption{ Decomposition of some two and three qubit inter-ion gates (from Figures \ref{fig:BV} and \ref{fig:stab_meas}) between two ions, one with $n=2$, and the other ion with $n=1$. The first two qubits belong to ion 1, and encoding map $\mathcal{M}_1$ is used for all gates. ${}_{1,1}\mathrm{CNOT}_{2,1}$ refers to a $\mathrm{CNOT}$ gate with the control qubit being the 1st virtual qubit on ion 1, and the controlled X gate applied on the 1st qubit on ion 2.  }
\label{tab:nv2_nv1_all}
\end{table}

\subsection{Inter ion gates between two ions with $n=2$}\label{sec:compiler}

In the main text, we saw that for generic random circuits with all ions having $n=2$ qubits, the number of gates required for convergence of linear cross entropy fidelity is smaller than when all ions have a single qubit, if the total number of qubits is the same (sec. \ref{sec:XEB}). However, we see that  the same advantage does not carry over for come of the standard gates discussed in the paper, at least using our approach to decomposition . In Table \ref{tab:nv2_all}, we see that some standard gates between two qubits on different ions can take $2-4$ inter-ion gates, or $12-16$ total gates, whereas they may only require one inter-ion gate along with a few intra-ion gates when $n=1$ on all ions, or even when one ion has $n=2$ and the other has $n=1$ (Table \ref{tab:nv2_nv1_all}). The layer ansatz we use here has forms such as $
    U_{\mathrm{an}}(\{\theta\},\{\phi\},J)= \mathrm{MS}_{\{01\}\{01\}}(J)
    \times R_{1,01}(\theta_{01},\phi_{01}) \times R_{1,03}(\theta_{03},\phi_{03}) \times R_{1,12}(\theta_{12},\phi_{12}) \times \mathrm{MS}_{\{23\}\{23\}}(J) \times R_{2,01}(\theta_{01},\phi_{01}) \times R_{2,03}(\theta_{03},\phi_{03}) \times R_{2,12}(\theta_{12},\phi_{12})  $
such that an MS gate is followed by some intra-ion gates. Fixing the states on which the gates act eases the optimization over the continuous set of parameters $\{\theta\},\{\phi\},J$. A number of random initial conditions are attempted, and if the tolerance of the cost function is not achieved, an additional layer is added with independent parameters but having the same structure, and the new circuit is optimized.

\begin{table}[h]
\begin{center}
\begin{tabular}{ |c|c|c|c| } 
 \hline
 \textit{No.} & \textit{Gate}  & \textit{Decomposition} \\ 
 \hline
 \hline
 \rule{0pt}{4ex}    
 1 & ${}_{1,2}\mathrm{CNOT}_{2,1}$  & 
  $R_{1,01}(\pi,0)$ $  R_{2,01}(\pi,0) $ $R_{2,01}(\frac{3\pi}{2},\frac{\pi}{2}) $ $ R_{2,12}(\frac{2\pi}{3},0) $ $\times$\\[4pt]
 & & $\mathrm{MS}_{\{02\}\{03\}}(\frac{3\pi}{2})$ $R_{1,12}(\frac{3\pi}{2},\frac{\pi}{6})$ $R_{1,01}(\frac{3\pi}{2},\frac{\pi}{2})$ $R_{2,23}(\pi,0)$ $R_{2,03}(\frac{3\pi}{2},0) $ $\times$\\[4pt]
 & &  $R_{1,12}(\frac{\pi}{2},\frac{\pi}{6})$ $R_{1,23}(\pi,0)$ $R_{2,12}(\frac{5\pi}{6},\pi) $ $\times$\\[4pt]
 & & $\mathrm{MS}_{\{02\}\{12\}}(\frac{\pi}{2})$ $R_{2,01}(\frac{3\pi}{2},\frac{3\pi}{2})$ $R_{1,01}(\frac{3\pi}{2},\frac{\pi}{2}) $ $R_{2,23}(\pi,0) $ $R_{2,03}(\frac{3\pi}{2},0) $ \\[4pt]
 \hline
 \rule{0pt}{4ex}    
 2 & ${}_{1,2}\mathrm{CNOT}_{2,2}$  & 
  $R_{1,01}(\pi,0)$ $  R_{1,03}(0.5\pi,0) $ $R_{1,12}(\pi,0) $  $\times$\\[4pt]
 & & $R_{2,01}(\pi,0)$ $  R_{2,03}(1.5\pi,0) $ $R_{2,12}(1.5\pi,0) $  $\times$\\[4pt]  
 & & $\mathrm{MS}_{\{01\}\{01\}}(\frac{3\pi}{2})$ $R_{1,01}(\frac{3\pi}{2},\pi)$  $R_{2,03}(-0.5\pi,0.5\pi)$ $R_{2,12}(0.5\pi,0.5\pi) $ $\times$\\[4pt]

 & & $\mathrm{MS}_{\{01\}\{01\}}(\frac{\pi}{2})$ $R_{1,01}(\pi,0)$ $  R_{1,03}(0.5\pi,0) $ $R_{1,12}(\pi,0) $  $R_{2,12}(\pi,0) $ \\[4pt]
 \hline 
\rule{0pt}{4ex}
 3 & $e^{-i\frac{\pi}{4}X_{1,2}X_{2,2}}$  &    $\mathrm{MS}_{\{01\}\{01\}}( 1.25 \pi)$   $ R_{1, 0   3 }( 1.5 \pi, 1.0 \pi)$  $ R_{1, 1   2 }( 1.5 \pi, 0.0 \pi)$    $\times$\\[4pt]

&  & 
$\mathrm{MS}_{\{23\}\{23\}}( 0.75 \pi)$ $\times$  \\[4pt]

& & $ R_{2, 0   1 }( 1.0 \pi, 0 \pi)$   $ R_{2, 1   2 }( 1.0 \pi, 1.0 \pi)$  $\times$\\[4pt]

 & & $\mathrm{MS}_{\{01\}\{01\}}( 0.25 \pi)$ $ R_{1, 0    }( 1.0 \pi, 0.0 \pi)$ $ R_{1, 0   3 }( 0.5 \pi, 1.0 \pi)$  $ R_{1, 1   2 }( 0.5 \pi, 0.0 \pi)$    $\times$\\[4pt]
&  & 
    $ R_{2, 0   1 }( 1.0 \pi, 0.0 \pi)$   $ R_{2, 1   2 }( 1.0 \pi, 0 \pi)$  $\times$\\[4pt]
&  & 
$\mathrm{MS}_{\{23\}\{23\}}( 0.25 \pi)$   \\[4pt]
\hline
\rule{0pt}{4ex}
4 & $e^{-i\frac{\pi}{4}X_{1,1}X_{1,2}X_{2,1}}$ &    $\mathrm{MS}_{\{01\}\{01\}}( 1.0 \pi)$  $ R_{1, 0   1 }( 1.0 \pi, 0 \pi)$ $ R_{1, 0   3 }( 0.75 \pi, 1.0 \pi)$  $ R_{1, 1   2 }( 1.25 \pi, 0.0 \pi)$    $\times$\\[4pt]

&  & 
$\mathrm{MS}_{\{23\}\{23\}}( 1.0 \pi)$ $\times$  \\[4pt]

& & $ R_{2, 0   1 }( 0.5 \pi, 0 \pi)$  $ R_{2, 0   3 }( 0.75 \pi, 1.0 \pi)$ $ R_{2, 1   2 }( 1.75 \pi, 1.0 \pi)$  $\times$\\[4pt]

 & & $\mathrm{MS}_{\{23\}\{23\}}( 1.0 \pi)$  $ R_{1, 0   3 }( 1.25 \pi, 1.0 \pi)$  $ R_{1, 1   2 }( 1.25 \pi, 1.0 \pi)$    $\times$\\[4pt]
&  & 
    $ R_{2, 0   1 }( 1.5 \pi, 0.0 \pi)$   $ R_{2, 1   2 }( 1.0 \pi, 0 \pi)$  $\times$\\[4pt]
&  & 
$\mathrm{MS}_{\{23\}\{23\}}( 1.0 \pi)$   \\[4pt]
\hline
\rule{0pt}{4ex}
5 & $e^{-i\frac{\pi}{4}X_{1,1}X_{1,2}X_{2,1}X_{2,2}}$ &  $\mathrm{MS}_{\{01\}\{01\}}( 1.0 \pi)$  $ R_{2, 0   1 }( 1.0 \pi, 0 \pi)$  $ R_{2, 0   3 }( 0.75 \pi, 1.5 \pi)$ $ R_{2, 1   2 }( 1.75 \pi, 1.0 \pi)$  $\times$\\[4pt]

 & & $\mathrm{MS}_{\{01\}\{01\}}( 1.0 \pi)$  $ R_{1, 0   3 }( 0.25 \pi, 1.5 \pi)$  $ R_{1, 1   2 }( 1.75 \pi, 0.5 \pi)$    $\times$\\[4pt]
&  & 
 $\mathrm{MS}_{\{23\}\{23\}}( 1.0 \pi)$    $ R_{2, 0   1 }( 1.0 \pi, 0.0 \pi)$  $ R_{2, 0   3 }( -0.25 \pi, 0.0 \pi)$  $ R_{2, 1   2 }( 1.25 \pi, 1.0 \pi)$  $\times$\\[4pt]
&  & 
$\mathrm{MS}_{\{23\}\{23\}}( 1.0 \pi)$   \\[4pt]
\hline
\rule{0pt}{4ex}
6 & $e^{-i(J_{1}X_{1,1}X_{1,2}+J_2X_{2,1}X_{2,2})}$ &    $ R_{1, 0   2 }( J_1, 0)$ $ R_{1, 1   3 }( J_1, 0)$ $ R_{2, 0   2 }( J_2, 0)$ $ R_{2, 1   3 }( J_2, 0)$ \\[4pt]
\hline
\rule{0pt}{4ex}
7 & $e^{-i\frac{\pi}{4}(X_{1,2}X_{2,2}+X_{1,1}X_{2,1})}$ &    $ R_{1, 0   1 }( 1.0 \pi, 0.0 \pi)$   $ R_{1, 0   3 }( 1.75 \pi, 1.5 \pi)$  $ R_{1, 1   2 }( 0.75 \pi, 1.5 \pi)$    $\times$\\[4pt]

& & $ R_{2, 0   3 }( 0.75 \pi, 0.5 \pi)$   $ R_{2, 1   2 }( 1.25 \pi, 1.5 \pi)$  $\times$\\[4pt]

&  & 
$\mathrm{MS}_{\{23\}\{23\}}( 0.5 \pi)$  $\times$\ \\[4pt]

 & & $\mathrm{MS}_{\{01\}\{01\}}( 0.5 \pi)$ $ R_{1, 0   3 }( 0.75 \pi, 1.5 \pi)$  $ R_{1, 1   2 }( 0.5 2\pi, 0.5 \pi)$    $\times$\\[4pt]
&  & 
    $ R_{2, 0   1 }( 1.0 \pi, 0.0 \pi)$  $ R_{2, 0   3 }(-0.25 \pi, 1.5 \pi)$  $ R_{2, 1   2 }( -0.25 \pi, -0.5 \pi)$  \\[4pt]

\hline
\end{tabular}
\end{center}
\caption{ Construction of multi-qubit inter-ion gates for two ions, both with $n=2$ qubits, using encoding map $\mathcal{M}_1$, assuming all $R_{1,\alpha \beta}$, $R_{1,\alpha \beta}$ and $\mathrm{MS}_{\{\alpha_1 \beta_1 \},\{\alpha_2 \beta_2\}}$ are allowed. ${}_{1,2}\mathrm{CNOT}_{2,1}$ refers to a $\mathrm{CNOT}$ gate with the control qubit being the 2nd virtual qubit on ion 1, and the controlled X gate applied on the 1st qubit on ion 2. Since we have assumed an all to all connectivity, the number and form of the gates would remain the same if the same gate acting on a different virtual qubit within the same ion is desired, or if a different encoding map $\mathcal{M}$ is used.}
\label{tab:nv2_all}
\end{table}

\newpage
\section{Random circuit sampling}\label{sec:app_ng2}

Consider the number of gates required for a random circuit constructed out of intra- and inter-ion gates to achieve an $F_{\mathrm{XEB}}=2$, as plotted in Figure \ref{fig:XEB_main}. Recall that $F_{\mathrm{XEB}}=2^N\braket{p(x_i)}_i-1$, where $p(x_i)$ is the probability of the measured bit-string $x_i$ (Z basis measurement of all qubits). In the main text, we considered a short range brick-work type random circuit, which applies the MS gate on ions $1,2$ followed by $3,4,\dots$ and eventually $L-1,L$. The MS gate is applied between the two ions with randomly chosen levels, and a random parameter $J$. This is followed by $R_{\alpha \beta}$ gates on both the ions, with levels and parameters chosen randomly. The same is repeated next for ions $2,3$, then $4,5$ and eventually $L,1$.  In the results shown in Figure \ref{fig:XEB_main}, we considered $N_{\mathrm{c}}=5-50$ random circuits, and each circuit was sampled around $N_{\mathrm{m}}=100-500$ times.

\begin{figure}
    \centering
    \includegraphics[width=0.9\linewidth,trim = 0cm 0.25cm 0cm 0cm]{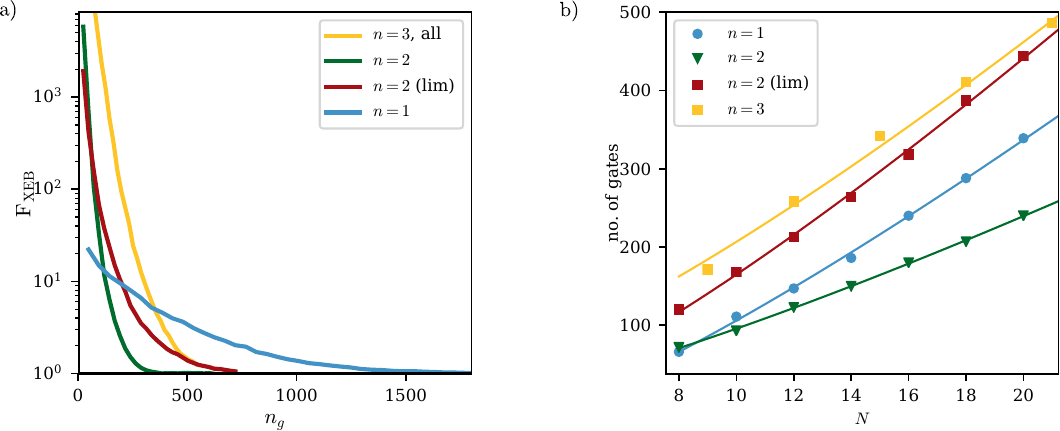}
    \caption{a) Convergence of linear cross entropy fidelity using an ideal noiseless random quantum circuit where each layer corresponds to applying a random MS gate, followed by randomly chosen $R_{\alpha \beta}$ on neighboring ions in a brick-work layer architecture. This is done both for the traditional setup $n=1,L=16$ and using two virtual qubits on each ion $n=2,L=8$, both with a gate set having an all to all connectivity(red), and a limited minimal connectivity required for universality (Figure \ref{fig:intro_fig}), and also for $n=3, L=6$ with an all to all connectivity. With an all to all connectivity $F_{\mathrm{XEB}}$ starts out to be larger than for a traditional setup, but eventually converges faster. b) Number of gates required for $F_{\mathrm{XEB}}=2$ using a random circuit where two ions are picked at random, instead of a brick-work architecture. In this case, with limited connectivity for $n=2$, the convergence is slower than both $n=1$ and $n=2$ with an all to all connectivity. }
    \label{fig:XEB_app1}
\end{figure}

Here, we provide additional results. In Figure \ref{fig:XEB_app1}, the left panel shows the convergence of $F_{\mathrm{XEB}}$ for $n=1$, $n=2,3$ with all possible gates allowed, and with $n=2$ with a minimal connectivity required for a universal gate set. The number of qubits is fixed -- $N=16$. 

In the same figure, the right panel plots the number of gates needed for $F_{\mathrm{XEB}}=2$, for a random circuit in which two random ions are chosen, then an MS gate is applied between those two ions with randomly chosen levels, and a random parameter $J$. This is followed by $R_{\alpha \beta}$ gates on both the ions, with levels and parameters chosen randomly. The gates can be both long and short-range in such a random circuit. This architecture requires more measurement samples ($N_{c} = 256-1000$) than the brick-work architecture. We see that $n=2$ with a maximum connectivity requires fewer gates than $n=1$ to get $F_{\mathrm{XEB}}=2$, however with limited connectivity or when $n=3$, more gates are required than $n=1$.

\begin{figure}
    \centering
    \includegraphics[width=0.9\linewidth,trim = 0cm 0.25cm 0cm 0cm]{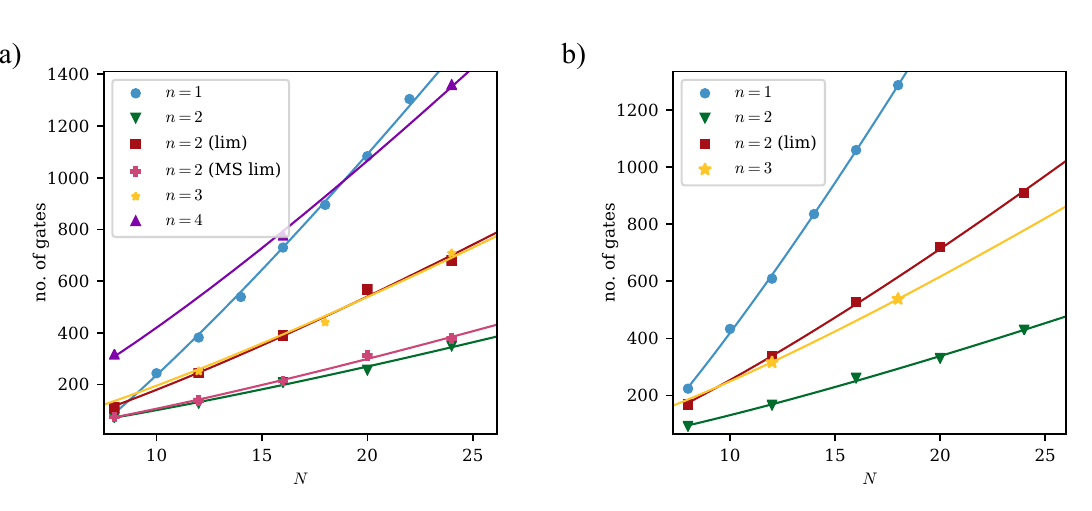}
    \caption{a) Number of gates required for $F_{\mathrm{XEB}}=2$ using a brick-work random circuit -- including data from Figure \ref{fig:XEB_main}. The $n=4$ curve indicates a limit when using no virtual qubits requires fewer gates. The $n=2$ (MS lim) curve shows the gates required when the MS gate is fixed to be $\mathrm{MS}_{\{01\}\{01\}}$ -- as opposed to the states chosen randomly. This has negligible effect compared to randomly chosen states ($n=2$ curve). But that makes it easier to implement practically. b) Number of gates required for $2^{2N}\mathrm{var}(p(x_i))=4$, probing the second moment of the probability distribution instead of linear XEB. The circuit has a brick-work architecture, and assumes full connectivity except for $n=2$ (lim) curve, which assumes a minimal connectivity. As with the XEB on the left, the number of gates required is smaller for $n=2$ and $n=3$, although larger than required for XEB to converge.}
    \label{fig:XEB_app2}
\end{figure}

In the left panel of Figure \ref{fig:XEB_app2}, we provide additional data for Figure \ref{fig:XEB_main}. This was left out in the main text to highlight only the important aspects. But in this panel, we also see a plot for $n=4$, which requires more gates than $n=1$ for around $N<20$.

Additionally, the same panel shows (for $n=2$) the number of gates when the MS gate is fixed to be $\mathrm{MS}_{\{01\}\{01\}}$ -- as opposed to the states chosen randomly. The number of gates doesn't change much in that case, and and is practically helpful.

In the right panel of Figure \ref{fig:XEB_app2}, we argue for the generality of our results by probing a higher moment of the distribution $p(x_i)$. We do this by plotting the number of gates required in a brick-work architecture such that $2^{2N}\mathrm{var}(p(x_i))=4$. For the Porter-Thomas distribution the right hand side is one. We find that the number of gates required is less than $n=1$, just as in Figure \ref{fig:XEB_main}. Note, however, that larger number of gates are required than when looking at $F_{\mathrm{XEB}}$, which is expected since we're looking at a higher moment.

\section{Error model for the repetition code}\label{sec:app_err_model}

\begin{figure}
    \centering
    \includegraphics[width=0.95\linewidth,trim = 0cm 0.8cm 0cm 0cm]{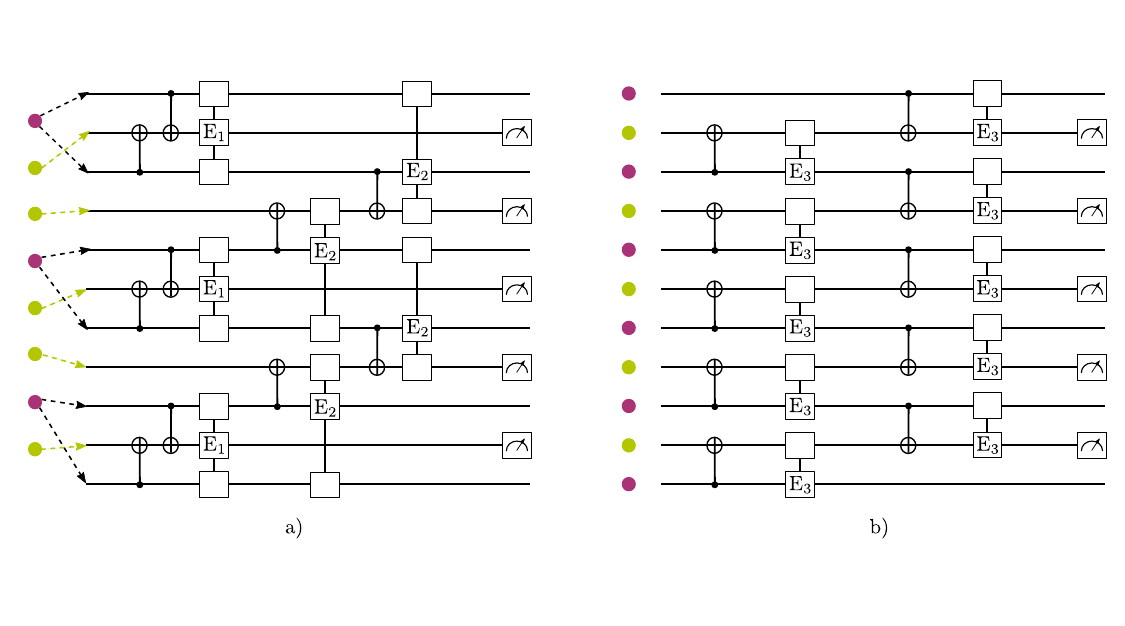}
    \caption{A circuit showing a cycle of error correction using bit-flip repetition code when $n=2$ data qubits are embedded inside a single ion (red ions in a)), and when each ion has a single qubit (red ions in b)). The stabilizer measurements are performed using one or more auxiliary ions (green) which we nevertheless depicted as distinct for convenience. Using the fact that two CNOT gates in the first layer on the left need only a single MS gate (Figure \ref{fig:stab_meas}), we model the error as a three qubit depolarizing error channel $E_1$ after the first layer, and similarly for the subsequent two layers, which require a single MS gate for each CNOT gate since they're acting on different pairs of ions. In b) we consider a simple two qubit depolarizing error channel applied after each CNOT gate, since each gate requires one MS gate. If $\epsilon_1$ is the error of a single qubit gate (or an intra-ion gate), the CNOT gate on the right has an approximate error $p=14\epsilon_1$, while the three qubit depolarizing channels all have an error rate $11\epsilon_1$.  }
    \label{fig:rep_code_circuit}
\end{figure}

Here we elaborate on the error model used to simulate the bit-flip repetition code in Figure \ref{fig:threshold_repcode}. A circuit diagram showing a single round of error correction, along with a depiction of the error gates are shown in Figure \ref{fig:rep_code_circuit}.

We first consider the case when each ion has $n=1$ qubit, and a distance $d$ code then has $N=2d+1=L$ qubits and ions. A $\mathrm{CNOT}$ gate between two ions (suppose the control is the first qubit/ion) can be implemented using the decomposition $\mathrm{CNOT}=$$ R_{1, 0   1 }( \frac{\pi}{4} , \frac{\pi}{2})$ $\mathrm{MS}_{\{01\}\{01\}}( \frac{\pi}{4})$  $ R_{1, 0   1 }( -\frac{\pi}{4}, 0)$ $ R_{1, 0   1 }( -\frac{\pi}{4}, \frac{\pi}{2})$ $ R_{2, 0   1 }( -\frac{\pi}{4}, 0)$. One way to estimate the error after the application of a single qubit gate is to assume an independent error model such that if the error in a single qubit gate is $\epsilon_1$ and the error in the MS gate is $\epsilon_2$, then the total error after the CNOT gate is approximately $\epsilon_{E_3} = 1-(1-\epsilon_1)^4(1-\epsilon_2) \approx 4\epsilon_1+\epsilon_2$. One way to incorporate this in the error channel $\mathrm{E}_2$ applied on the two qubits after the application of the CNOT gate is to apply a depolarizing channel with $\lambda=4\epsilon_1+\epsilon_2$. Since it is easier to simulate pure state dynamics with Pauli gates, it can be equivalently applied as a random two qubit Pauli product gate with the probability $p=\lambda/4$.

Let's compare this to the next case when we incorporate the data qubits as virtual qubits inside an ion. Specifically when $n=2$ qubits per ion is considered, with auxiliary ions for stabilizer measurement each having a single qubit, the CNOT gates require fewer native gates as illustrated in Figures \ref{fig:BV} and \ref{fig:stab_meas} in the main text, and in Table \ref{tab:nv2_nv1_all}. There are then two types of CNOT gates, one which implement the $Z_1Z_2$ stabilizer measurement for the two qubits within the same ion (the first layer in Figure \ref{fig:rep_code_circuit}), and the other which implements a $Z_1Z_2$ stabilizer measurement for two qubits on two different ions. In either case, errors can potentially affect three qubits (two within an ion, and one in measurement ion), and are illustrated as error gates $E_1$ and $E_2$, respectively in Figure \ref{fig:rep_code_circuit}. A simple way to estimate the overall error is to again use an independent error model, implying that $E_1$ should have an error $\epsilon_{E_1} \approx \epsilon_1+\epsilon_2 $ and correspondingly for $E_2$, $\epsilon_{E_2} \approx \epsilon_1+\epsilon_2 $, using native gates in the decomposition in Table \ref{tab:nv2_nv1_all}.

To compare logical error rate, we make the assumption $\epsilon_2=10\epsilon_1$ for all MS and intra-ion gates, which means $\epsilon_{E_1} \approx 11\epsilon_1 $, $\epsilon_{E_2} \approx 11\epsilon_1 $ and $\epsilon_{E_3} \approx 14\epsilon_1 $. Using $p=14\epsilon_1$ on the x-axis, we plot the logical error rates after $d$ cycles for $n=1$ ion chain, and after $d/2$ cycles for the $n=2$ ion chain, such that for the same length of ion chains, the same number of rounds are performed. 

So, for the depolarizing error channel, we find slightly larger thresholds and smaller logical error rates for the $n=2$ ion chain. There are two aspects that may change the logical error rates in reality. The first one is that the bit-flip repetition code ignores phase ($Z$) errors, which are nevertheless included in the depolarizing channel. The second aspect is that we have assumed that the performance of all MS and intra-ion gates remains the same for the same number of ions, which may not be true in reality.\\

\section{Selecting the atomic states for the computational manifold}\label{sec:sup_cost_function}
As pointed out in Sec \ref{sec:guidelines_comp_manifold}, operations on $n>1$ qubits per ion are more susceptible to certain errors than $n=1$ in a way that depends on the specific states that make up the computational manifold. As such, the choice of states that make up the manifold is one that will directly impact the performance of the system. We can use the expected error per single-qubit gate as a numerical metric to benchmark different choices. While accurately estimating the errors is a difficult task that depends sensitively on experimental parameters (e.g. the magnetic field noise spectrum or the specific pulse shape used to drive two-state rotations) we can estimate an error metric that will give an indication of how well an encoding should perform. Because we also include some experimental details in this estimate (the value of the quantisation field and the polarisation of the driving fields) this metric can be used to estimate optimal parameters as well. 

The primary errors that differ between computational manifolds are off-resonant coupling errors (\textit{crosstalk} errors) and dephasing errors induced by noisy external magnetic fields (\textit{memory} errors). Additionally, if the is stored in the metastable level, the probability of a decay event to the ground level will depend upon the time taken to execute logical gates, which depends upon the connectivity and Rabi frequencies of the encoding. However, for long-lived levels such as those in $\mathrm{Ba}$ and $\mathrm{Yt}$, we expect these errors to be small ($\sim \ 10^{-5}$ for a 100 $\mu$s logical gate) and thus we neglect these in the analysis. 

We define the error metric as the sum of the expected crosstalk and memory errors per single-qubit gate, where these quantities are defined in Section \ref{sec:sup_crosstalk} and \ref{sec:sup_memory}, respectively. We obtain the encoding in Figure \ref{fig:Ba_level} by  evaluating the ten encodings with the lowest error metric in the \dlevel\ level and manually selecting one with a good compromise between Rabi frequencies and expected error. Relevant properties and error metrics are shown in Tables \ref{tab:sup_encoding_properties} and \ref{tab:sup_error_table} respectively, and the experimental parameters used in the simulation are listed in \ref{tab:sup_encoding_sim_params}. Simulations made use of the \texttt{networkx} \cite{networkx} and our in-house \texttt{atomic-physics} \cite{atomic-physics} Python packages.

\subsection{Constructing a computational manifold}\label{sec:sup_constructing_encoding}

We can think of the computational manifold of $n$ qubits as a graph, whose nodes are the $d=2^n$ states selected from the atomic manifold and whose edges are allowed transitions between them. Usually, transitions are deemed allowed or not depending on the low-field selection rules between hyperfine states for the gate-driving mechanism (either two-photon Raman or magnetic dipole transitions). However, the hyperfine splitting in the metastable level of many ions used is small enough that even with small quantisation fields, states labelled by $\ket{F, m_F}$ become mixed (see Figure \ref{fig:Ba_level}). This significantly modifies the selection rules, necessitating an alternative definition for an allowed transition.

We define a transition $T$ between states $\ket{\alpha}$ and $\ket{\beta}$ to be allowed if the absolute value of the matrix element $M_{T}$ connecting them (via the gate-driving mechanism employed) is larger than a minimum threshold value. We include the polarisation of the driving field as part of the matrix element, which makes it proportional to the Rabi frequency of the transition by factors that depend only on the driving field intensity (and the detuning of the beams from the intermediate excited state in the case of Raman transitions). Additionally, we impose that all allowed transitions must be well-resolved, either by frequency or polarisation. We say that a transition $T$ between states in the manifold is allowed if there is no transition $\Tilde{T}$ such that the expected off-resonant excitation of $\Tilde{T}$, $\Omega_{\Tilde{T}}^2 / (\omega_T - \omega_{\Tilde{T}})^2$, is greater than 0.05. This expression characterises the crosstalk error (see Section \ref{sec:sup_crosstalk}) where $\Omega_{\Tilde{T}}$ is the Rabi frequency of $\Tilde{T}$ and $\omega_{T, \Tilde{T}}$ are the angular frequencies of $T, \Tilde{T}$. Finally, it may only be experimentally possible to drive transitions in a certain bandwidth, further restricting allowed transitions.

\subsubsection{Raman transitions}\label{sec:app_raman}
The transition $T$ between states $\ket{\alpha}$ and $\ket{\beta}$ separated by energy $\hbar\omega_{T}$ can be driven by shining two laser beams that are far detuned by $\Delta$ from a dipole transition to a separate manifold, and whose frequency difference is equal to $\omega_T$. Driving this transition on-resonance with two beams whose electric fields at the ion are $\mathbf{E}_{1,2}=E_{1,2}\mathbf{e}_{1,2}$ results in a Rabi frequency of magnitude \cite{wineland2003quantum}

\begin{table}[h!]
\begin{center}
\begin{tabular}{ |c|c|c|c| }
\hline
\rule{0pt}{3ex}
\textit{Transition} & \textit{Frequency (MHz)} & \textit{Magnetic field sensitivity (kHz/mG)} & \textit{Rabi frequency (kHz)} \\ [0.5ex] 
\hline
\hline

\rule{0pt}{2.5ex}
$\ket{F=1, m=0} \leftrightarrow \ket{F=1, m=-1}$ & 56.8 & 2.468 & 94.99 \\ [0.5ex] 
\hline
\rule{0pt}{2.5ex}
$\ket{F=1, m=0} \leftrightarrow \ket{F=2, m=-2}$ & 137.1 & 3.263 & 48.80 \\ [0.5ex] 
\hline
\rule{0pt}{2.5ex}
$\ket{F=1, m=-1} \leftrightarrow \ket{F=2, m=-2}$ & 80.3 & 0.795 & 88.35 \\ [0.5ex] 
\hline
\rule{0pt}{2.5ex}
$\ket{F=2, m=-2} \leftrightarrow \ket{F=3, m=-3}$ & 63.1 & 0.618 & 71.36 \\ [0.5ex] 
\hline
\rule{0pt}{2.5ex}
$\ket{F=1, m=-1} \leftrightarrow \ket{F=3, m=-3}$ & 143.4 & 1.413 & 29.72 \\ [0.5ex] 
\hline
\hline
\end{tabular}
\end{center}
\caption{The properties of the optimal encoding presented in the text.}
\label{tab:sup_encoding_properties}
\end{table}

\begin{table}[h!]
\begin{center}
\begin{tabular}{ |c|c| }

\hline
\rule{0pt}{3ex}
\textit{Error type} & \textit{Error metric per gate} \\[0.5ex] 
\hline
\hline
\rule{0pt}{2.5ex}
Total error ($\varepsilon_M + \varepsilon_{\mathrm{Int}} + \kappa \varepsilon_{\mathrm{Spect}}$) & $2.988 \times 10^{-3}$ \\ [0.5ex]
\hline
\rule{0pt}{2.5ex}
Memory error ($\varepsilon_M$) & $1.168 \times 10^{-4}$ \\ [0.5ex]
\hline
\rule{0pt}{2.5ex}
Intra-encoding crosstalk error ($\varepsilon_{\mathrm{Int}}$) & $4.117 \times 10^{-4}$ \\ [0.5ex]
\hline
\rule{0pt}{2.5ex}
Spectator crosstalk error ($\varepsilon_{\mathrm{Spect}}$) & $4.920 \times 10^{-3}$ \\ [0.5ex]
\hline
\rule{0pt}{2.5ex}
$\kappa$ & 0.5\\[0.5ex] 
\hline

\end{tabular}
\end{center}
\caption{Error metric per single-qubit intra-ion gate for the encoding in the text. The memory error metric is defined in \ref{eq:sup_mem_error} and the crosstalk error metrics are defined in \ref{eq:sup_scat_error}.}
\label{tab:sup_error_table}
\end{table}

\begin{table}[h!]
\begin{center}
\begin{tabular}{ |c|c| }

\hline
\rule{0pt}{3ex}
\textit{Parameter} & \textit{Value} \\ 
\hline
\hline
\rule{0pt}{2.5ex}
$D$ & $1.018 \times 10^{64}$ Hz/m\textsuperscript{2} \\ [0.5ex] 
\hline
\rule{0pt}{2.5ex}
$\mathbf{e}_1$ & $1 / \sqrt{3} \left( 1, 1, 1 \right)$ \\ [0.5ex] 
\hline
\rule{0pt}{2.5ex}
$\mathbf{e}_2$ & $1 / \sqrt{3} \left( 1, 1, 1 \right)$ \\ [0.5ex] 
\hline
\rule{0pt}{2.5ex}
$B_Q$ & 20 G \\ [0.5ex] 
\hline
\rule{0pt}{2.5ex}
$\Delta B_{RMS}$ & 50 $\mu$G \\ [0.5ex] 
\hline

\end{tabular}
\end{center}
\caption{Parameters for the simulation used. $D$ is the ratio of Rabi frequency to Raman matrix element $\mathbf{e}_{1,2}$ are the polarisation vectors for the two Raman beams in the form $(\sigma_-, \pi, \sigma_+)$, $B_Q$ is the quantisation field used and $\Delta B_{RMS}$ is the RMS of the magnetic field.}
\label{tab:sup_encoding_sim_params}
\end{table}

\begin{equation}
\Omega_{T} = \frac{e^2 E_1 E_2}{4\hbar^2}\sum_k \left( \frac{\bra{j}\hat{\mathbf{r}} \cdot \mathbf{e}_1 \ket{k}\bra{k} \hat{\mathbf{r}} \cdot \mathbf{e}_2\ket{i}}{\Delta_k} \right) \ 
\approx \frac{e E_1 E_2}{4\hbar^2 \Delta} \sum_k \bra{j}\hat{\mathbf{r}} \cdot \mathbf{e}_1 \ket{k}\bra{k} \hat{\mathbf{r}} \cdot \mathbf{e}_2\ket{i} \ 
= \frac{e E_1 E_2}{4\hbar^2 \Delta} M_{T}
\end{equation}
where $e$ is the elementary charge and $\ket{k}$ are the states in the separate manifold via which the transition is driven. With this, we define the threshold matrix element for an allowed transition to be the matrix element that would result in a Rabi frequency of 10 kHz when the electric field at the centre of two $1$ mW beams beams at $532$ nm, with a waist radius of 1 $\mu$m, are used to drive a transition within the \dlevel\ level in \Ba\ via the transition at $614$ nm connecting it to the $P_{3/2}$ level (which sets $\Delta = (\omega_{532} - \omega_{614})$).

Since the constant of proportionality between $\Omega_T$ and $M_T$ is independent of $T$, it is constant for all Raman transitions within the encoding manifold. Thus, specifying the ratio $D=\Omega_T / M_T$ for just one transition (e.g. the strongest transition in the manifold, for the chosen polarisations) is sufficient to determine the Rabi frequency of all transitions.

\subsubsection{Magnetic dipole transitions}\label{sec:app_magnetic}
In a similar way to the Raman transitions, a magnetic field $\mathbf{B}=B\mathbf{e}$ drives the transition $T$ between states $\ket{i}$ and $\ket{j}$ with Rabi frequency \cite{harty_thesis}

\begin{equation}
\Omega_T=-\frac{B}{\hbar} \bra{j} \hat{\mathbf{\mu}} \cdot \mathbf{e} \ket{i} \ 
= -\frac{B}{\hbar} M_T
\end{equation}

with $\hat{\mathbf{\mu}} = -\mu_B g_J \hat{\mathbf{J}} + \mu_N g_I \hat{\mathbf{I}}$ and $M_T$ the magnetic dipole matrix element of the transition $T$. As before, the ratio $D=B/\hbar$ between Rabi frequency and matrix element is constant for all transitions.

\subsection{Estimating off-resonant crosstalk errors}\label{sec:sup_crosstalk}

When driving a two-state rotation within the computational manifold, other transitions with similar frequency may be off-resonantly driven. This results in errors in logical gates. In general, each gate in the universal set will exhibit different crosstalk errors due to its particular decomposition to two-state rotations. However, we can estimate an average error per gate by assuming that, on average, all rotations appear equally often in the decompositions. As such, the average number of times a particular transition will be driven will be the average number of two-state rotations per gate divided by the number of allowed transitions. This depends on the encoding's connectivity in a complex way and for an encoding with $d=2^n$ states, goes roughly as $d^{(2-x)}$, where 
\begin{equation}
x = \frac{A_{\textrm{max}} - A}{A_{\textrm{max}} - A_{\textrm{min}}}
\end{equation}
parametrises the connectivity of the encoding. $A$ is the number of allowed transitions in the encoding and $A_{\textrm{max, min}}$ are the maximum and minimum possible number of transitions in a manifold of $n$ qubits.

We make the further approximation that each rotation in the decomposition is a $\pi$-rotation to simplify the simulation. As such, the probability of transferring population between two states connected by a transition with Rabi frequency $\Omega$ that is detuned by $\delta$ from the driving field is approximately $\Omega^2/\delta^2$ in the limit $\delta\gg\Omega$.

Two types of crosstalk errors may occur. The off-resonant transition may connect a state in the computational to some other state in the computational manifold (\textit{intra-encoding crosstalk}) or to a spectator state in the atomic level (\textit{spectator crosstalk}). While the latter exists in single-qubit encodings and results in easily detectable leakage errors, the former only occurs when $n>1$, and results in an amplitude error in the stored state. As discussed in the main body, leakage errors can be corrected using erasure conversion \cite{sotirova2024high, wu2022erasure, kang2023quantum} whereas amplitude errors must be corrected via error-correction protocols that may not be feasible for small scale, near term devices. Thus, we treat the two error sources separately in the error metric and decrease the relative contribution of spectator crosstalk to the overall error metric, since its effect on fidelity should be smaller.

Both these effects can also occur due to off-resonantly driving a motional transition. Here, we only consider coupling to the first order sideband of one mode with frequency $\omega_M$. 

We define the cross-talk error metric as

\begin{gather}
\varepsilon_C = d^{2-x}(\varepsilon_{\mathrm{Int}} + \kappa \varepsilon_{\mathrm{Spect}}) \label{eq:sup_scat_error}
\\
\varepsilon_{i} = \frac{D^2}{N_{T_{\mathrm{Int}}}} \sum_{\{T\}_{\mathrm{Int}}} \sum_{\Tilde{T}_i} \left[ \
\frac{M_{\Tilde{T}}^2}{(\omega_T - \omega_{\Tilde{T}})^2} + \ 
\frac{\eta^2 M_{\Tilde{T}}^2}{(|\omega_T - \omega_{\Tilde{T}}| - \omega_M)^2} \right]~;~~~ 
\{\Tilde{{T}}\}_i=
\begin{cases}
    \{T\}_{\mathrm{Int}}, \Tilde{T} \neq T & i = \mathrm{Int}\\
    \{T\}_{\mathrm{Spect}} & i = \mathrm{Spect}
\end{cases}
\end{gather}
Here, $\{T\}_{\mathrm{Int}}$ is the set of allowed transitions within the encoding, $\{T\}_{\mathrm{Spect}}$ is the set of allowed transitions between states in the encoding and spectators, $\eta$ is the Lamb-Dicke parameter of the coupling between the driving field and the ion's motion, and $\kappa$ is a parameter that sets the relative importance of intra-encoding and spectator errors. We note that the off-resonant transition $\Tilde{T}$ does not need to involve either of the states connected by $T$ to cause errors, since all states in the encoding will in general be populated, so the inner sum is over all transitions in the set.

Because of the relatively small frequency difference between transitions in the \dlevel\ level in \Ba, spectator errors can be quite large. Setting $D$ to give Rabi frequencies of order $10-100$ kHz, we calculate single-qubit gate errors of $\sim 0.1$ \%. These can be brought down by further reducing the Rabi frequency (at the expense of longer gate times) or by shaping the envelope of the gate-driving pulses \cite{bauer1984Gaussian} to reduce the amplitude of frequency components away from the pulse's centre frequency.

\subsection{Estimating memory errors}\label{sec:sup_memory}

Here, we estimate the average error acquired by a random intra-ion $n$-qubit state due to a randomly fluctuating magnetic field $B(t) = B_0 + \Delta B(t)$. The evolution of the system is given by the Hamiltonian

\begin{equation}
    H(t) = H_0 + \Delta B(t) H_N
\end{equation}
where $H_0$ contains the atomic Hamiltonian and the contribution from the static field $B_0$, and $H_N = \sum_{i=1}^d \partial\omega_i / \partial B \ket{i}\bra{i}$, where $\omega_i$ is the rate of phase accrual in state $\ket{i}$ in the encoding. In the following, we will move to the interaction picture with respect to $H_0$. If the magnitude of the noise is small, a system that is initially in the pure state $\rho(0)=\ket{\psi}\bra{\psi}, ~ \ket{\psi}=\sum_{i=1}^d c_i\ket{i}$, will evolve according to \cite{harty_thesis}

\begin{equation}\label{eq:sup_noisy_evolution}
    \rho(t) = \rho(0) - \frac{1}{\hbar^2}\int_0^t dt_1 \int_0^{t_1}dt_2 G(t_1-t_2) [H_N, [H_N, \rho(0)]]
\end{equation}
where $G(\tau)$ is the autocorrelation
\begin{equation}\label{eq:sup_autocorrelation}
    G(\tau) = \langle \Delta B(\tau) \Delta B(0) \rangle
\end{equation}
It is straightforward to show that $[H_N, [H_N, \rho(0)]] = \sum_{i,j=1}^d \left(\tfrac{\partial \omega_{i,j}}{\partial B}\right)^2 c_i c^*_j \ket{i}\bra{j}$, where $\tfrac{\partial \omega_{i,j}}{\partial B}$ is the magnetic-field sensitivity of the transition between states $\ket{i}$ and $\ket{j}$ and $d=2^n$ is the number of states in the computational manifold of $n$ qubits.

For simplicity, we assume that the noisy field $\Delta B(t)$ is a stationary Gaussian process characterised by an exponentially decaying autocorrelation $G(\tau) = \langle \Delta B^2 \rangle e^{-\gamma \tau}$ with by a decay time $\tau_c = 1/\gamma$ \cite{monz_thesis}. In this case, the density matrix at time $t$ is given by 

\begin{equation}
    \rho(t) = \rho(0) - \langle \Delta B^2 \rangle\frac{e^{-\gamma t} + \gamma t - 1}{\gamma^2} \sum_{i,j=1}^d \left( \frac{\partial \omega_{i,j}}{\partial B} \right)^2 c_i c_j^* \ket
    {i}\bra{j}
\end{equation}
As a metric for the error accumulated over this time, we use the fidelity with respect to the original state $\mathcal{F}_{\ket{\psi}} = \mathrm{Tr}(\rho(0)\rho(t))$ \cite{harty_thesis} (since $\rho(0)$ is a pure state). This is given by 

\begin{equation}
    \mathcal{F}_{\ket{\psi}}(t) = 1 - \langle \Delta B^2 \rangle\frac{e^{-\gamma t} + \gamma t - 1}{\gamma^2} \sum_{i,j=1}^d \left( \frac{\partial \omega_{i,j}}{\partial B} \right)^2 |c_i|^2 |c_j|^2 
\end{equation}

We're interested in finding the loss in fidelity over some characteristic time, which we consider to be the average time taken to execute one of the gates in the intra-ion single-qubit universal gate set, $t_G$. This is approximately equal to the mean single-qubit rotation time multiplied by the number of rotations per gate, 

\begin{equation}
    t_G = d^{2-x} \frac{\sum_{T} \pi / \Omega_T}{N_{T_{\mathrm{int}}}} = \ 
    d^{2-x} \frac{\pi}{N_{T_{\mathrm{int}}}D} \sum_T \frac{1}{M_T} \ \triangleq d^{2-x}t_R
\end{equation}
Here $t_R$ is the mean two-state rotation time, $N_{T_{\mathrm{int}}}$ is the number of transitions in the encoding and $D, M_T$ are as defined in Sections \ref{sec:app_raman}, \ref{sec:app_magnetic}.

We assume that the correlation time $\tau_c$ is much longer than the single-qubit gate time $t_G$. In this limit, the power spectral density of the fluctuations exhibits a roughly $1/f$ dependence, which is often found when active \cite{hughes_thesis} or passive \cite{ruster2016b_field} measures are taken to reduce the fluctuations. In such cases, RMS fluctuations of $\sqrt{\langle \Delta B^2 \rangle}=2.7$ pT have been demonstrated; we conservatively assume RMS fluctuations of $50$ nT, similar to those found in \cite{hughes_thesis}.

In this regime, the fidelity becomes
\begin{equation}
    \mathcal{F}_{\ket{\psi}}(t_G) = 1 - \frac{1}{2} \langle \Delta B^2 \rangle t_G^2 \sum_{i,j=1}^d \left( \frac{\partial \omega_{i,j}}{\partial B} \right)^2 |c_i|^2 |c_j|^2 
\end{equation}

However, to fairly compare the performance of two choices of computational manifolds, we should consider the average fidelity over all possible states $\ket{\psi}$ in the $d$-dimensional Hilbert space. For this, we parametrise the state coefficients $c_i$ using hyperspherical coordinates \cite{hedemann2013hyperspherical}:

\begin{equation}\label{eq:hyperspherical}
    c_i(r, \{\theta_k\}, \phi_i) = r e^{i\phi_i} \times
    \begin{cases}
        \prod_{k=1}^{i-1}\sin\theta_k \cos\theta_i & i < d \\
        \prod_{k=1}^{d-1}\sin\theta_k & i=d    
    \end{cases}
\end{equation}

The $n-1$ angles $\theta_k \in [0, \pi/2]$ define the amplitude of the states, and the $n$ angles $\phi_k \in [0, 2\pi]$ define the phase. Enforcing normalisation of the state implies that the radius $r=1$ and we set $\phi_1\triangleq0$ to remove the global phase variable. The usual Bloch sphere is recovered in the case of a qubit ($d=2$) with the modification $\theta_1 \rightarrow \theta_1/2, \theta_1\in[0, \pi]$.

Since $\mathcal{F}_{\ket{\psi}}$ does not depend on the coordinates $\phi_k$, we only need to average the fidelity over the ${\theta_k}$ coordinates, which is equivalent to averaging over the first quadrant of the unit $d$-dimensional hypersphere:

\begin{align}\label{eq:fidelity_integral}
    \Bar{\mathcal{F}} & = \frac{1}{S_d}\int_{S_d}\mathcal{F}_{\psi} d\uline{\theta} \\
    & = 1 - \frac{1}{S_d} \sum_{i, j}\xi_{i, j} \int_{S_d}|c_i|^2|c_j|^2d\uline{\theta}
\end{align}
with $\xi_{i,j}=\tfrac{\langle \Delta B^2 \rangle t^2}{2} \left(\tfrac{\partial \omega_{i,j}}{\partial B}\right)^2$ and integration metric $d\uline{\theta} = \prod_{k=1}^{d-1}(\sin\theta_k)^{d-k-1} d\theta_k$. From this metric, the surface area $S_d$ is

\begin{align}
    S_d & = \prod_{k=1}^{d-1} \int_{0}^{\frac{\pi}{2}}(\sin\theta_k)^{d-k-1} d\theta_k.
\end{align}
This can be solved straightforwardly with the following identities:

\begin{gather}\label{eq:integral_identities}
    I_n \triangleq \int_0^{\pi/2}\sin^n \theta d\theta = \frac{(n-1)!!}{n!!} \times
    \begin{cases}
        \pi/2 & d ~ \mathrm{even} \\
        1 & d ~ \mathrm{odd} 
    \end{cases} \\
    I_nI_{n-1} = \frac{\pi/2}{n}.
\end{gather}
where the double factorial $x!! = x(x-2)(x-4)\cdots$. This gives

\begin{equation}
    S_d = \prod_{k=1}^{d-1}I_{(d-k-1)} = \frac{\left( \pi/2 \right)^{\left \lceil \frac{d}{2}\right\rceil}}{(n-2)!!}
\end{equation}

Now we can solve the integral in Eq. \ref{eq:fidelity_integral}. In the case $i>j,~i<d$, from Eq. \ref{eq:hyperspherical},

\begin{equation}\label{eq:expansion_of_integrand}
    |c_i|^2|c_j|^2 = \left( \prod_{k=1}^{j-1}  \sin^4\theta_k \right) \
    \left( \prod_{k=j}^{i-1} \sin^2\theta_k \right) \ 
    \left( 1 + \sin^2\theta_i\sin^2\theta_j - \sin^2\theta_i - \sin^2\theta_j \right)
\end{equation}
Similar expressions can be found for the cases $i=j$ and $i=n$. The four terms in the brackets can be integrated separately. The first one becomes
\begin{align}
    \prod_{k=1}^{j-1} I_{d-k+3} \ 
    \times \prod_{k=j}^{i-1} I_{d-k+1} \
    \times \prod_{k=i}^{d-1} I_{d-k-1} & = \ 
    \frac{\prod_{k=1}^{(d+4)-1} I_{(d+4)-k-1}}{ 
    I_{(d-j+3)} I_{(d-j+2)} I_{(d-i+1)} I_{(d-i)}} \\
    & = S_{(d+4)} \frac{(d-j+3)(d-i+1)}{(\pi/2)^2}
\end{align}
The remaining terms in Eq. \ref{eq:expansion_of_integrand} can be obtained by appropriately setting ${i,j} \rightarrow {i+1, j+1}$. From this, and using $S_{d+4} / S_d = (\pi/2)^2 / (d(d+2))$, we find

\begin{equation}
    \frac{1}{S_d} \int_{S_d} |c_i|^2|c_j|^2 d\uline{\theta} = \frac{1}{d(d+2)} \times \ 
    \begin{cases}
        1 & i \neq j \\
        3 & i = j
    \end{cases}
\end{equation}
As such, since $\xi_{i,i}=0$, Eq. \ref{eq:fidelity_integral} simplifies to 

\begin{equation}
    \mathcal{F}(t_G) = 1 - \frac{t_G^2 \langle\Delta B^2\rangle}{4d(d+2)} \sum_{T} \left( \frac{\partial \omega_{T}}{\partial B} \right)^2
\end{equation}
where we take the sum over all transitions $T$ rather than the states $i,j$.

We thus define the memory error per gate as the loss in fidelity of an encoding during the time it takes to execute an average gate, $1 - \mathcal{F}(t_G)$:

\begin{equation}\label{eq:sup_mem_error}
    \varepsilon_{M} = d^{4-2x} \frac{t_R^2 \langle\Delta B^2\rangle}{4d(d+2)} \sum_T\left( \frac{\partial \omega_T}{\partial B} \right)
\end{equation}

\subsection{Manual selection of the computational manifold}\label{sec:sup_manual_selection}
\begin{figure}[t]
    \centering
    \includegraphics[width=0.8\linewidth,trim = 0cm 0cm 0cm 0cm]{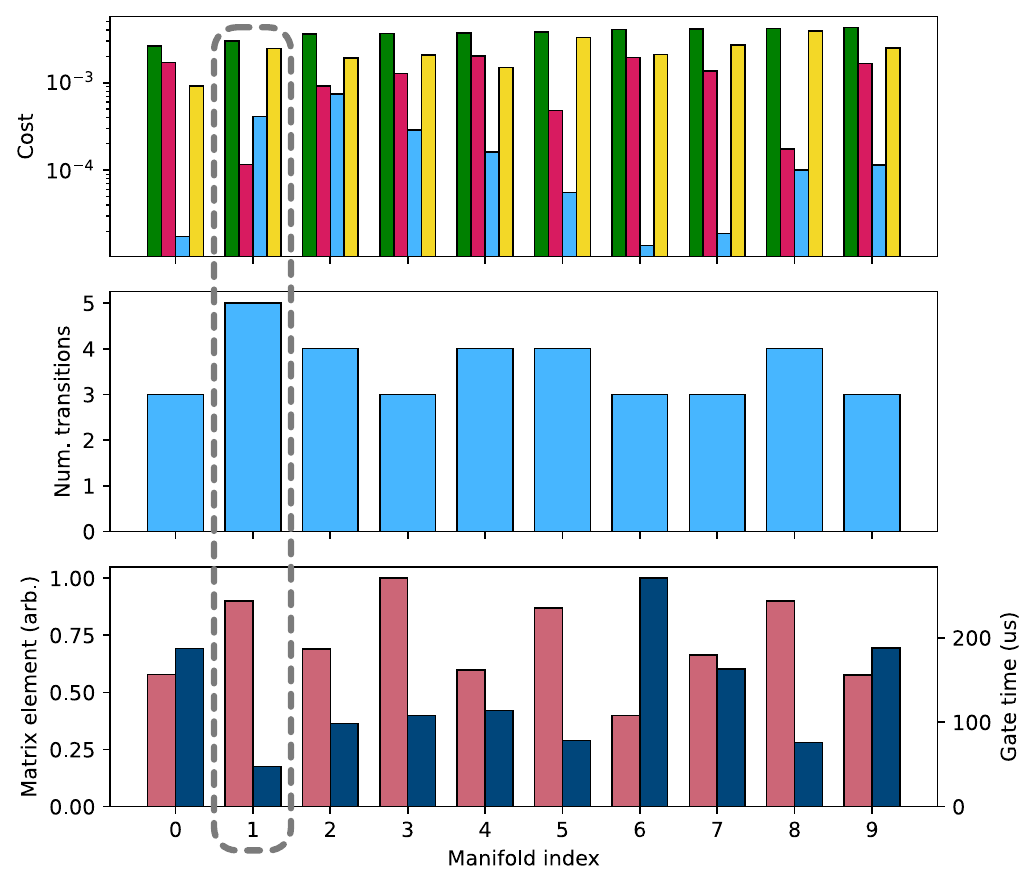}
\caption{Properties of the ten manifolds with the lowest cost for the simulation parameters in Table \ref{tab:sup_encoding_sim_params}. Encodings are labelled by increasing total cost. The properties of the encoding in the text are marked with a gray, dashed box. Top: total cost (green), cost associated to memory errors (red), cost associated to internal scattering errors (blue) and cost associated to spectator scattering errors, with a weighting $\kappa=0.5$ (yellow). Middle: number of allowed transitions in each encoding.R Bottom: average matrix element magnitude over the transitions in the encoding (salmon, left axis) and mean intra-ion gate time (blue, right axis).}
    \label{fig:sup_manifold_choices}
\end{figure}

Once the cost associated to all possible computational manifolds in the metastable level is evaluated, the ten manifolds with the lowest cost are found. The most desirable manifold for a given implementation is selected manually from this set. 

Relevant properties of the manifolds in the reduced set used to make the choice in the main text are shown in Figure \ref{fig:sup_manifold_choices}. The chosen encoding (marked by a dashed box) offers the best compromise between low associated cost, a large number of allowed transitions and a short intra-ion gate time. We note that the mean matrix element and gate times are not inversely proportional, since the gate time depends on the connectivity of the manifold.


\end{document}